\documentclass[article]{article}

\usepackage{amsmath}
\usepackage{amssymb}
\usepackage{MnSymbol}

\usepackage{tikz}
\usetikzlibrary{arrows,decorations.pathmorphing,backgrounds,positioning,fit,petri,calc,automata}
\usetikzlibrary{calc,decorations.pathreplacing}

\usepackage{algorithm}
\usepackage{algorithmicx}
\usepackage[noend]{algpseudocode}

\usepackage{subfig}
\usepackage{graphics}
\usepackage{epsfig}
\usepackage{color}

\usepackage{soul}
\usepackage{color}
\usepackage{times}

\usepackage{multirow}
\usepackage{enumerate}
\usepackage{paralist}

\newtheorem{definition}{Definition}
\newtheorem{lemma}{Lemma}
\newtheorem{proposition}{Proposition}
\newtheorem{theorem}{Theorem}
\newtheorem{corollary}{Corollary}

\newtheorem{example}{Example}
\newtheorem{fact}{Fact}

\renewcommand{\iff}{\Leftrightarrow}

\renewcommand{\Call}[2]{\ensuremath{\textsc{#1}} ({#2})}

\DeclareMathOperator{\lcm}{lcm}

\newcommand{\x}[0]{\vec{x}}
\newcommand{\yy}[0]{\vec{y}}
\newcommand{\zz}[0]{\vec{z}}
\newcommand{\xk}[1]{\vec{x}^{\scriptscriptstyle (#1)}}

\newcommand{\xxki}[2]{x^{\scriptscriptstyle (#1)}_{#2}}
\newcommand{\y}[0]{\vec{y}}

\newcommand{\mi}[0]{-\!}

\newcommand{\aw}[0]{\overline{w}}

\newcommand{\bin}[1]{|{#1}|}

\newcommand{\topleft}[1] {\!\! {~}^{\scriptscriptstyle{\blacksquare}} {#1}}
\newcommand{\botleft}[1] {\!\! {~}_{\scriptscriptstyle{\blacksquare}} {#1}}
\newcommand{\topright}[1] {{#1}^{{\scriptscriptstyle{\blacksquare}}}}
\newcommand{\botright}[1] {{#1}_{\scriptscriptstyle{\blacksquare}}}

\newcommand{\minw}{\mathrm{minw}}

\newcommand{\xmod}{\!\!\!\mod\!}

\newcommand{\idm}[1]{\mathbf{1}_{\scriptscriptstyle{#1}}}

\newcommand{\abs}[1]{\mathrm{abs}(#1)}
\newcommand{\id}[1]{\mathrm{I}_{#1}}

\renewcommand{\vec}[1]{{\bf {#1}}}

\def\set#1{{\left\{ #1 \right\}}}
\def\tuple#1{{\langle #1 \rangle}}

\newcommand{\true}{\top}
\newcommand{\false}{\bot}

\newcommand{\len}[1]{{|{#1}|}}

\newcommand{\card}[1]{\parallel\!\!{#1}\!\!\parallel}

\newcommand{\lang}[1]{{\mathcal L}({#1})}

\newcommand{\arrow}[2]{\xrightarrow{{\scriptscriptstyle #1}}}

\newcommand{\nat}{{\bf \mathbb{N}}}
\newcommand{\zed}{{\bf \mathbb{Z}}}

\newcommand{\proj}{\!\!\downarrow\!\!}

\def\vr{\kern-\arraycolsep & \kern-\arraycolsep}
\def\VR{\kern-\arraycolsep\strut\vrule}

\newcommand{\dbm}{\mathrm{DB}}
\newcommand{\oct}{\mathrm{OCT}}

\newcommand{\reachflat}{\textsc{ReachFlat}}

\newcommand*\circled[1]{\tikz[baseline=(char.base)]{        
  \node[shape=circle,draw,inner sep=1pt] (char) {$#1$};}}

\newcommand*\doublecircled[1]{\tikz[baseline=(char.base)]{        
  \node[shape=circle,draw,inner sep=0.1pt] (char) {\circled{#1}};}}

\newif\ifLongVersion\LongVersiontrue

\newcommand{\proof}[1]{\noindent {\em Proof}: {#1}}
\newcommand{\qed}{\hfill $\Box$\bigskip}

\usepackage{arrayjobx}

\newcommand{\gridName}[2]{n#1n#2}

\tikzset{
  nborder/.style={circle,inner sep=0.75mm,minimum size=0mm},
  ngrid/.style={nborder,fill=black!30,draw=black},
  matchingsep/.style={shape=circle,fill=red!10,inner sep=0mm,minimum size=0mm},
  matching/.style  = {matchingsep,midway,left}
}

\tikzset{
  zrBorder/.style={circle,inner sep=0.5mm,minimum size=0mm},
  zrNode/.style={zrBorder,fill=black!40} 
}

\tikzset{
  nborder/.style={circle,inner sep=0.5mm,minimum size=0mm},
  ngrid/.style={nborder,fill=black!40},
  matchingsep/.style={shape=circle,fill=red!10,inner sep=0mm,minimum size=0mm},
  matching/.style  = {matchingsep,midway,left}
}

\newcommand{\TermGridGenNoCap}[8]
{
  \begin{scope}
    \def\xIndent{#7}
    \def\yIndent{#8}
    \def\yBase{#4+(#6-1)*#5}
    \foreach \x in {1,...,#3} {
      \pgfmathsetmacro{\aa}{\xIndent+#1+(\x-1)*#2}
      \foreach \y in {1,...,#6} {
        \pgfmathsetmacro{\bb}{\yBase-(\y-1)*#5}
        \def\nName{\gridName{\x}{\y}}
        \node (\nName) at (\aa cm,\bb cm) [ngrid] {};
      }
    }
  \end{scope}
}

\newcommand{\TermGridGenNoCapBoxed}[8]
{
    \TermGridGenNoCap{#1}{#2}{#3}{#4}{#5}{#6}{#7}{#8}

    \pgfmathsetmacro{\cXleft}{#1+(#7)-0.2}
    \pgfmathsetmacro{\cXright}{#1+(#2)*(#3-1)+#7+0.2}
    \pgfmathsetmacro{\cYbot}{#4-0.2}
    \pgfmathsetmacro{\cYtop}{#4+(#5)*(#6-1)+0.2}
    \draw [rounded corners=4pt,dotted,very thick] (\cXleft cm,\cYbot cm) rectangle (\cXright cm,\cYtop);
}

\newcommand{\TermGridGen}[9]
{

  \begin{scope}

    \def\xIndent{#7}
    \def\yIndent{#8}

    \pgfmathsetmacro{\aa}{#1}
    \foreach \y in {1,...,#6} {
      \pgfmathsetmacro{\bb}{#4+(\y-1)*#5}
      \def\nName{\gridName{0}{\y}}
      \pgfmathtruncatemacro\inx{#6-\y+1}
      \node (\nName) at (\aa cm,\bb cm) [nborder] {$x_{\inx}$};
    }
    \pgfmathsetmacro{\bb}{#4+(#6-1)*#5+\yIndent}
    \foreach \x in {1,...,#3} {
      \pgfmathsetmacro{\aa}{\xIndent+#1+(\x-1)*#2}
      \def\nName{\gridName{\x}{0}}

      \pgfmathtruncatemacro\inxvv{mod(\x-1,#3-1)}
      \pgfmathtruncatemacro\inxww{\x-1}
      \def\inx{\ifthenelse{\equal{#9}{1}}{\inxvv}{\inxww}}

      \node (\nName) at (\aa cm,\bb cm) [nborder] {$\vec{x}^{(\inx)}$};
    }

    \TermGridGenNoCap{#1}{#2}{#3}{#4}{#5}{#6}{#7}{#8}

  \end{scope}
}
\newcommand{\TermGridGenDifferentCap}[9]
{

  \begin{scope}

    \def\xIndent{#7}
    \def\yIndent{#8}

    \pgfmathsetmacro{\aa}{#1}
    \pgfmathsetmacro{\aap}{#1+(#2)*(#3-1)+2*#7}
    \foreach \y in {1,...,#6} {
      \pgfmathsetmacro{\bb}{#4+(\y-1)*#5}
      \def\nName{\gridName{0}{\y}}
      \pgfmathtruncatemacro\inx{#6-\y+1}
      \node (\nName) at (\aa cm,\bb cm) [nborder] {$x_{\inx}$};
      \node (\nName) at (\aap cm,\bb cm) [nborder] {$x'_{\inx}$};
    }

    \TermGridGenNoCap{#1}{#2}{#3}{#4}{#5}{#6}{#7}{#8}

  \end{scope}
}

\newcommand{\TermGridEdgeC}[6]
{
  \draw [->,thick] (\gridName{#1}{#2}) -- (\gridName{#3}{#4})
     node[midway,#6] {#5}; 
}

\tikzset{
  zrBorder/.style={circle,inner sep=0.5mm,minimum size=0mm},
  zrNode/.style={zrBorder,fill=black!40} 
}

\newcommand{\TermZrName}[2]{n#1n#2}
\newcommand{\TermZrNameP}[2]{np#1n#2}
\newcommand{\TermZrNameC}[2]{nc#1n#2}

\newcommand{\TermZrCapState}[6]
{
  \pgfmathsetmacro{\capStart}{#1*(#4+#5)+(#4/2)} 
  \pgfmathsetmacro{\capStep}{#4+#5}
  \pgfmathtruncatemacro\aSize{#2-#1+1}
  \foreach \y in {1,...,\aSize} {
     \pgfmathsetmacro{\capPos}{\capStart+(\y-1)*\capStep}
     \node at (\capPos,-0.6	) {\aCaption(\y)};
  }
}
\newcommand{\TermZrCapGraph}[6]
{
  \pgfmathsetmacro{\capStart}{(#1-1)*(#4+#5)+#4+(#5/2)} 
  \pgfmathsetmacro{\capStep}{#4+#5}
  \pgfmathsetmacro{\capHeight}{(#3-1)*#6+0.5}
  \pgfmathtruncatemacro\aSize{#2-#1+1}
  \foreach \y in {1,...,\aSize} {
     \pgfmathsetmacro{\capPos}{\capStart+(\y-1)*\capStep}
     \node at (\capPos,\capHeight) {\aCaption(\y)};
  }
}

\newcommand{\TermZrBase}[9]
{
  \begin{scope}
    \pgfmathtruncatemacro\xStart{#1-1}
    \foreach \x in {#1,...,#2} {
      \pgfmathsetmacro{\aa}{#4+(\x-1)*(#4+#5)}
      \pgfmathsetmacro{\aaP}{\aa+#5}
      \foreach \y in {1,...,#3} {
        \pgfmathsetmacro{\bb}{(#3-\y)*#6}
        \node (\TermZrName{\x}{\y}) at (\aa cm,\bb cm) [zrNode] {};  
        \node (\TermZrNameP{\x}{\y}) at (\aaP cm,\bb cm) [zrNode] {}; 
      }
      \pgfmathsetmacro{\cXleft}{\aa-0.1}
      \pgfmathsetmacro{\cXright}{\aaP+0.1}
      \pgfmathsetmacro{\cYbot}{-0.2}
      \pgfmathsetmacro{\cYtop}{(#3-1)*#6+0.2}
      \draw [rounded corners=4pt,dotted,very thick] (\cXleft cm,\cYbot cm) rectangle (\cXright cm,\cYtop);
    }

    \def\TermZrXGap{0.2}
    \def\TermZrYExtra{0.3}
    \pgfmathsetmacro{\yBase}{(#3-1)*#6} 
    \foreach \x in {\xStart,...,#2} {
      \pgfmathsetmacro{\aaL}{(\x)*(#4+#5)+\TermZrXGap} 
      \pgfmathsetmacro{\aaR}{\aaL+#4-(2*\TermZrXGap)} 
      \pgfmathsetmacro{\bbB}{0-\TermZrYExtra} 
      \pgfmathsetmacro{\bbT}{\yBase+\TermZrYExtra} 
      \draw [rounded corners=4pt] (\aaL cm,\bbB cm) rectangle (\aaR cm,\bbT cm);
    }
    \foreach \x in {\xStart,...,#2} {
      \pgfmathsetmacro{\aaC}{\x*(#4+#5)+(#4/2)} 
      \foreach \y in {1,...,#3} {
        \pgfmathsetmacro{\bb}{(#3-\y)*#6}
        \node (\TermZrNameC{\x}{\y}) at (\aaC cm,\bb cm) {}; 
      }
    }
    \foreach \x in {1,...,#9} {
      \pgfmathsetmacro{\aaL}{#7*(#4+#5)+(\x-1)*(#4+#5)*#8+(\TermZrXGap/2)} 
      \pgfmathsetmacro{\aaR}{\aaL+(#4+#5)*#8} 
      \pgfmathsetmacro{\bbB}{0-\TermZrYExtra-0.6} 
      \pgfmathsetmacro{\bbT}{\yBase+\TermZrYExtra+0.6} 
      \draw [rounded corners=4pt,dotted,very thick] (\aaL cm,\bbB cm) rectangle (\aaR cm,\bbT cm);
      \pgfmathsetmacro{\capX}{(\aaL+\aaR)/2}
      \pgfmathsetmacro{\capY}{\bbT+0.3}
      \node at (\capX,\capY) {};
    }

  \end{scope}
}

\newcommand{\TermZrStateElem}[3]
{
    \path let \p1 = (\TermZrNameC{#1}{#2}) in node at (\x1,\y1) [nborder] {$#3$};
}

\newcommand{\TermZrEdgeFWLab}[4]
{
  \draw [->,thick] (\TermZrName{#1}{#2}) -- (\TermZrNameP{#1}{#3}) node[midway,above] {#4};
}

\newcommand{\TermZrEdgeBWLab}[4]
{
  \draw [->,thick] (\TermZrNameP{#1}{#2}) -- (\TermZrName{#1}{#3}) node[midway,above] {#4};
}

\usepackage[draft]{commenting}

\declareauthor{ri}{Radu}{red}
\declareauthor{mb}{Marius}{green}

\title{The Complexity of Reachability Problems for Flat Counter
  Machines with Periodic Loops}

\author{$\text{Marius Bozga}^1$ \and $\text{Radu Iosif}^1$ \and $\text{Filip Kone\v{c}n\'{y}}^2$ \\\\
${}^1$ Verimag/CNRS/Universit\'e de Grenoble Alpes (France) \\
${}^2$ NetSuite Brno (Czech Republic)}

\begin{document}
\maketitle

\begin{abstract}
This paper proves the NP-completeness of the reachability problem for
the class of flat counter machines with difference bounds and, more
generally, octagonal relations, labeling the transitions on the
loops. The proof is based on the fact that the sequence of powers
$\{R^n\}_{n=0}^\infty$ of such relations can be encoded as a periodic
sequence of matrices, and that both the prefix and the period of this
sequence are simply exponential in the size of the binary
representation of a relation $R$. This result allows to characterize
the complexity of the reachability problem for one of the most studied
class of counter machines \cite{Cav10,ComonJurski98}, and has a
potential impact on other problems in program verification.
\end{abstract}

\section{Introduction}

Counter machines are powerful abstractions of programs, commonly used
in software verification. Due to their expressive power, counter
machines can simulate Turing machines \cite{Minsky67}, thus their
decision problems (reachability, termination) are undecidable. This
early negative result motivated researchers to define classes of
systems with decidable reachability problems, such as: vector addition
systems \cite{Mayr81,Kosaraju82,Leroux12}, reversal-bounded counter
machines \cite{Ibarra78}, Datalog programs with gap-order constraints
\cite{Revesz93}, and flat counter machines
\cite{BoigelotPhD,ComonJurski98,Cav10}. Despite the fact that the
reachability problem is, in principle, decidable for these classes,
few of these results are actually supported by tools, and used for
real-life verification purposes. The main reason is that the
complexity of the reachability problems for these systems is, in
general, prohibitive. As a practical consequence, most software
verifiers rely on incomplete algorithms, which, due to the loss of
precision, may raise large numbers of false alarms. Improving the
precision of these tools requires mixed techniques such as
combinations of model checking, static analysis and acceleration, that
rely on identifying subproblems for which the set of reachable states,
or the transitive closure of the transition relation, can be computed
precisely \cite{Atva12,MonniauxGawlitza12}, by cost-effective
algorithms.

In this paper, we study the complexity of the reachability problems
for a class of {\em flat counter machines}, whose control structure
forbids nested loops and the transitions occurring inside loops are
labeled with {\em octagons}, i.e.\ conjunctions of inequalities of the
form $\pm x \pm y \leq c$ where $x,y$ denote the current or next
values of the counters and $c$ is an integer constant. Our main result
states that the reachability problem for this class of counter
machines is \textsc{Np}-complete. This result is a direct
generalization of the \textsc{Np}-completness of the reachability
problem for the subclass of \emph{difference bounds} constraints,
which are finite conjunctions of inequalities of the form $x-y \leq
c$, with $c$ being an integer constant.

Due to the particular syntax of the octagonal constraints, in which
the variables are always multiplied by coefficients from the set
$\set{-1,0,1}$, such relations can be represented by square matrices
of a fixed dimension, called \emph{difference bounds matrices}
(DBM). 
%
%
The main idea of the \textsc{Np}-completness proof is that sequence of
DBMs corresponding to the sequence of relations
$\set{R^n}_{n=0}^\infty$ is \emph{periodic}, in the sense that the
matrices situated at equal distance in the sequence, beyond a certain
prefix, differ by equal quantities. If the prefix and the period of
this sequence are known, one can build a quantifier-free formula of
Presburger arithmetic that characterizes this sequence, and reduce the
reachability problem to an instance of the satisfiability problem in
the quantifier-free fragment of Presburger arithmetic, known to be
\textsc{Np}-complete \cite{VermaSeidlSchwentick05}.

The main technical difficulty is, given an octagonal constraint that
defines a relation $R$, building such a Presburger formula in
polynomial time. To this end, we show that the prefix $b$ and the
period $c$ of the sequence of DBMs representing
$\set{R^n}_{n=0}^\infty$ are \emph{simply exponential} in the size of
the octagon defining $R$. Using this argument, one
can \begin{inparaenum}[(i)]
\item guess the prefix $b$ and the period $c$ of the relation, 
\item compute the powers
$R^b$ and $R^c$, using exponentiation by squaring, 
\item verify the validity of the guess for $b$ and $c$, and 
\item build the needed Presburger constraint in polynomial time.
\end{inparaenum}

Proving the simply exponential bounds for the prefix and the period of
an octagonal relation uses insights from the theory of weighted graphs
and tropical algebra \cite{Gaubert94,DeSchutter00}. We use the classical
representation of DBMs by weighted constraint graphs, such that the
$n$-th power of a difference bounds relation $R$ is defined by the
minimal weight paths of a constraint graph of width $n$, called an
\emph{unfolding graph} \cite{ComonJurski98}. Then we define a weighted
automaton, called \emph{zigzag automaton}, that recognizes the set of
constraint paths in the unfolding graph. The minimal weight paths,
needed to define the $n$-times composition of $R$ with itself, are
given by the $n$-th power of the incidence matrix of the transition
table of the zigzag automaton, where matrix multiplication is defined
in the tropical semiring $\tuple{\zed_{\pm\infty}, \min, +, \infty,
  0}$. Since the sequence of tropical powers of any given matrix is
periodic, we obtain that the sequence of DBMs representing the
relations $\set{R^n}_{n=0}^\infty$ is periodic as well.

We first prove the existence of simply exponential bounds on the
prefix and the period of the sequence of DBMs that represent the
sequence of relations $\set{R^n}_{n=0}^\infty$, where $R$ is a
relation defined by a difference bounds constraint. These bounds are
then generalized to octagonal relations. The most technical part is
proving the bound on the period of such sequences, which requires an
insight on the particular structure of loops in the zigzag automaton
describing the powers (w.r.t. composition) of a difference bounds
relations. The crucial point is restricting the zigzag automaton to
recognize a subset of constraint paths, with a bounded number of
direction changes, whose set of weights is sufficient to define the
relation $R^n$. This idea originates in the work of Comon and Jurski
\cite{ComonJurski98} on defining transitive closures of difference
bounds relations by formulae of Presbuger arithmetic, in order to
prove decidability of the reachability problem for flat counter
machines with difference bounds relations.

\subsection{Related Work} 

The study of the computational complexity of the reachability problem
(and other related problems, such as coverability and boundedness) for
various classes of counter machines has recently received much
attention.

An important class of counter machines with decidable reachability
problems are Vector Addition Systems with States (VASS). This problem
has been shown to be \textsc{Expspace}-hard by Lipton \cite{Lipton76}
and currently no upper bound has been found. On the other hand, the
problems of coverability and boundedness for VASS are shown to be
\textsc{Expspace}-complete \cite{Rackoff78}. Because the transition
relation of VASS can be defined as a finite disjunction of difference
bounds constraints, these counter machines are, in principle, not
flat. However, when restricting the number of counters to two,
Hopcroft and Pansiot \cite{HopcroftPansiot79} have shown that the set
of reachable configurations of a VASS is semilinear, thus definable in
Presburger arithmetic. Along this line, Leroux and Sutre
\cite{LerouxSutre04} showed that it is possible to build a flat
counter machine, with the same transitions as the original 2-counter
VASS and same reachable set of configurations. A close analysis of
their construction revealed that reachability of 2-counter VASS
(mostly known as 2-dimensional VASS) is a \textsc{Pspace}-complete
problem \cite{BlondinFinkelGoellerHaaseMcKenzie05}.

In their work, Ibarra and Gurari \cite{IbarraGurari81} study the
reachability problem for counter machines with increment, decrement
and zero test, in the {\em reversal-bounded} case, where the counters
are allowed to switch between non-decreasing and non-increasing modes
a number of times, bounded by a constant. It is found that, when the
number of counters and reversals are fixed constants (i.e.\ not part
of the representation of the counter machines) the emptiness problem
is decidable in logarithmic space, and hence, in polynomial
time. Moreover, if the machines under consideration are all
deterministic, the emptiness problem is
\textsc{Nlogspace}-complete. On the other hand, if the number of
counters and reversals are part of the input, the emptiness problem is
in \textsc{Pspace}. Our model of computation is incomparable, since
flat counter machines with non-deterministic counter updates are not
reversal-bounded, in general. For instance, if the future value of a
counter $x$ is chosen to be $x-1 \leq x' \leq x+1$ within a loop, the
counter can switch any number of times between increasing and
decreasing modes.

The class of \emph{gap-order} constraints, initially introduced by
Revesz \cite{Revesz93}, consists of finite conjunctions of difference
bounds constraints $x-y \leq c$, where $c$ is a positive
constant. Counter machines with gap-order constraints have been
studied by Bozzelli and Pinchinat \cite{BozzelliPinchinat12} who
coined their reachability problem to \textsc{Pspace}-complete. Our
result is incomparable to \cite{BozzelliPinchinat12}, as we show
\textsc{Np}-completeness for flat counter machines with strictly more
general\footnote{The generalization of gap-order to difference bound
  constraints suffices to show undecidability of non-flat counter
  machines, hence the restriction to flat control structures is
  crucial.} octagonal relations on loops.

The results which are probably closest to ours are the ones in
\cite{DemriDharSangnier12,DemriDharSangnier13}, where flat counter
machines with linear affine guards and vector addition updates are
considered. In \cite{DemriDharSangnier12} it is shown that
model-checking for Linear Temporal Logic is \textsc{Np}-complete for
these systems, matching thus our complexity for reachability with
difference bounds constraints, while model-checking first-order logic
and linear $\mu$-calculus is \textsc{Pspace}-complete
\cite{DemriDharSangnier13}, matching the complexity of CTL* model
checking for gap-order constraints \cite{BozzelliPinchinat12}. These
results are again incomparable with ours,
since \begin{inparaenum}[(i)]
\item the linear affine guards are more general, while
\item the vector addition updates are more restrictive (e.g. the
  direct transfer of values $x'_i=x_j$ for $i \neq j$ is not allowed).
\end{inparaenum}

The result of this paper is a refinement of earlier decidability
proofs for the reachability problem concerning flat counter machines
with loops labeled by difference bounds
\cite{ComonJurski98,Fundamenta09} constraints. The first such result,
due to Comon and Jurski \cite{ComonJurski98} defines the sequence of
$n$-times compositions $\set{R^n}_{n=0}^\infty$ of a difference bounds
relation $R$ by a formula in Presburger arithmetic. The essence of
their proof is the definition, by a formula of Presburger arithmetic,
of a subset of paths, in the constraint graph representing $R^n$, that
encompasses the set of paths of minimal weight, relevant for the
definition of the relation $R^n$. They show that only certain paths,
that roughly go back and forth from one extremity of the graph to the
other, without changing direction in between, are important in the
definition of the closed form. The idea of considering only such
\emph{simple} paths is instrumental in our work, for establishing a
simply exponential upper bound on the period of these relations.

By exploring further the structure of these simple paths,
Kone\v{c}n\'{y} \cite{Konecny14} showed that the sequence of the
closed form of the power sequence of a difference bounds
(respectively, octagonal) relation can be defined by a quantifier-free
Presburger formula which, moreover, can be built in polynomial time by
a deterministic algorithm. As a result, the reachability problem for
flat counter machines can be proved to be in \textsc{Nptime} directly,
by polynomial reduction to the satisfiability of quantifier-free
Presburger arithmetic. Unlike the proof given in this paper,
Kone\v{c}n\'{y}'s proof \cite{Konecny14} does not use periodic
sequences, relying on an enumeration of polynomially many minimal
weight paths. Besides providing an alternative proof of
\textsc{Np}-completness to the reachability problem, the results in
this paper define closed forms using only finite disjunctions of
difference bounds constraints (respectively, octagons) whose
coefficients are parameterized by $n$. This characterization of the
closed forms is of particular interest for other problems, such as,
e.g.\ the complexity of the termination problem for periodic classes
of relations \cite{TerminationLmcs}, or extensions of the model of
flat counter machines with recursive calls \cite{GantyIosif15}.

\section{Preliminaries}

We denote by $\zed$, $\nat$ and $\nat_+$ the sets of integers,
positive (including zero) and strictly positive integers,
respectively. We define $\zed_\infty=\zed \cup \set{\infty}$ and
$\zed_{\pm\infty}=\zed_\infty \cup \set{-\infty}$. We write $[n]$ for
the interval $\{0,\ldots,n-1\}$, $\abs{n}$ for the absolute value of
the integer $n\in\zed$, and $\gcd(n_1,\ldots,n_k)$,
$\lcm(n_1,\ldots,n_k)$ for the greatest common divisor and least
common multiple of the natural numbers $n_1,\ldots,n_k \in \nat$,
respectively. The cardinality of a finite set $S$ is denoted by
$\card{S}$.

A \emph{weighted graph} is a tuple $G= \tuple{V, E, w}$, where $V$ is
a~set of vertices, $E \subseteq V \times V$ is a~set of edges, and $w:
E \rightarrow \zed$ is a~weight function. If $G$ is clear from the
context, we write $u \arrow{\alpha}{} v$ for $(u,v) \in E$ and $w(u,v)
= \alpha$. A {\em path} in $\mathcal{G}$ is a sequence of the form
$\pi ~:~ v_0 \arrow{w_1}{} v_1 ~\cdots~ v_{k-1} \arrow{w_k}{} v_k$. We
denote by $\mathit{src}(\pi)=v_0$, $\mathit{dst}(\pi)=v_k$ its source
and destination vertices, by $\len{\pi}=k$ and by $w(\pi) =
\sum_{i=1}^k w_i$ its weight. The path $\pi$ is said to
be~\begin{inparaenum}[(i)]
\item \emph{elementary} if $v_i=v_j$ only if $i=0$ and $j=k$, 
\item \emph{a cycle} if $\mathit{src}(\pi)=\mathit{dst}(\pi)$, and
\item \emph{minimal} if, for any path $\pi'$ such that
  $\mathit{src}(\pi')=\mathit{src}(\pi)$,
  $\mathit{dst}(\pi')=\mathit{dst}(\pi)$, 
  we have $w(\pi)\leq w(\pi')$.
\end{inparaenum}
We denote by $\mu(G) = \max(\set{\abs{\alpha} ~|~ u \arrow{\alpha}{}
  v} \cup \set{1})$ the maximum between the absolute values of the
weights of $G$ and $1$. 

The set of $n \times n$ square matrices with coefficients in
$\zed_\infty$ ($\zed_{\pm\infty}$) is denoted as $\zed^{n \times
  n}_\infty$ ($\zed^{n \times n}_{\pm\infty}$). Each matrix $M \in
\zed_\infty^{n \times n}$ is the \emph{incidence matrix} of a weighted
graph $G_M = \tuple{V_M,E_M,w_M}$, where $V_M = \set{1,\ldots,n}$,
$E_M = \set{(i,j) \mid M_{ij} < \infty}$ and $w(i,j) = M_{ij}$, for
all $i,j \in \set{1,\ldots,n}$. In this case, we also define
$\mu(M)=\mu(G_M)$.

A \emph{term} $t$ over a set of variables $\x = \{x_1,\ldots,x_N\}$ is
a linear combination $a_0 + a_1x_1 + \ldots a_Nx_N$, for some integer
constants $a_0, a_1, \ldots, a_N \in \zed$. An \emph{atomic
  proposition} is a predicate of the form $t \leq 0$ or $t \equiv_c
0$, where $t$ is a term, $c \in \nat_+$ is a constant, and $\equiv_c$
denotes equality modulo $c$. The boolean constants \emph{false} and
\emph{true} are denoted by $\false$ and $\true$, respectively.
\emph{Quantifier-free Presburger Arithmetic} (QFPA) is the set of
boolean combinations of atomic propositions of the above form. For a
QFPA formula $\phi$, let $Atom(\phi)$ denote the set of atomic
propositions in $\phi$, and $\phi[t/x]$ denote the formula obtained by
substituting the variable $x$ with the term $t$ in $\phi$. We assume
that all integers are encoded in binary and denote by $\bin{\phi}$ the
size of the binary encoding of a formula $\phi$.

Let $\x$ denote a nonempty set of integer variables. A
\emph{valuation} of $\x$ is a function $\smash{\nu : \x \arrow{}{}
  \zed}$. The set of valuations is denoted by $\zed^{\x}$. If $\nu \in
\zed^{\x}$ is a valuation, we denote by $\nu \models \varphi$ the fact
that the formula obtained from $\varphi$ by replacing each occurrence
of $x\in\vec{x}$ with the integer $\nu(x)$ is valid under the standard
interpretation of the first-order arithmetic. A formula $\varphi$ is
said to be \emph{consistent} if and only if there exists a valuation
$\nu$, such that $\nu \models \varphi$. The consistency problem (also
known as the satisfiability problem) for QFPA is NP-complete
\cite[Lemma 5]{VermaSeidlSchwentick05}.

Let $\x'$ denote the set $\{x' ~|~ x \in \x\}$ of \emph{primed}
variables. A formula $\phi(\x,\x')$ is evaluated with respect to two
valuations $\nu,\nu'\in\zed^\x$, by replacing each occurrence of
$x\in\vec{x}$ with $\nu(x)$ and each occurrence of $x'\in\vec{x'}$
with $\nu'(x)$ in $\phi$. We write $(\nu,\nu') \models \phi$ when the
formula obtained from these replacements is valid. A formula
$\phi(\x,\x')$ is said to define a relation $R \subseteq \zed^{\x}
\times \zed^{\x}$ whenever for all $\nu,\nu' \in \zed^{\x}$, we have
$(\nu,\nu')\in R$ iff $(\nu,\nu') \models \phi$. The empty relation is
denoted by $\emptyset$. The composition of two relations $R_1, R_2
\subseteq \zed^\x \times \zed^\x$ defined by formulae
$\varphi_1(\vec{x},\vec{x'})$ and $\varphi_2(\vec{x},\vec{x'})$,
respectively, is the relation $R_1 \circ R_2 \subseteq \zed^{\x}
\times \zed^{\x}$, defined by the formula $\exists \vec{y} ~.~
\varphi_1(\vec{x},\vec{y}) \wedge \varphi_2(\vec{y},
\vec{x'})$. 

The {\em identity} on $\x$ is the relation $\id{\x} \subseteq
\zed^{\x} \times \zed^{\x}$ defined by the formula
$\bigwedge_{x\in\vec{x}} x'=x$. For any relation $R \subseteq
\zed^{\x} \times \zed^{\x}$, we define $R^0 = \id{\x}$ and $R^{n+1} =
R^n \circ R = R \circ R^n$, for all $n \in \nat$. $R^n$ is called the
$n$-th \emph{power} of $R$ in the sequel. The infinite sequence of
relations $\set{R^n}_{n=0}^\infty$ is called the \emph{power sequence}
of $R$. With these notations, $R^+ = \bigcup_{n=1}^\infty R^n$ denotes
the \emph{transitive closure} of $R$, and $R^* = R^+ \cup \id{\x}$
denotes the \emph{reflexive and transitive closure} of $R$. A relation
$R$ is said to be \emph{$*$-consistent} if and only if $R^n \neq
\emptyset$, for all $n \in \nat_+$. If $R$ is not $*$-consistent,
there exists $b>0$ such that $R^n = \emptyset$, for all $n\geq b$.

\begin{definition}\label{def:rel-class}
A {\em class} of relations $\mathcal{R}$ is the union of all monoids
$\tuple{\mathcal{R}_{\x},\circ,\id{\x}}$, where $\mathcal{R}_{\x}
\subseteq 2^{\zed^{\x} \times \zed^{\x}}$ is a set of relations over
$\x$ closed under conjunction and composition, containing the
relations $\id{\x}$ and $\emptyset$.
\end{definition}
In this paper we will define classes of relations by a fragments of
QFPA. In fact, any fragment of QFPA that contains equality and is
closed under conjunction and quantifier elimination defines a class of
relations.

\section{The Reachability Problem for Flat Counter Machines}
\label{sec:reachability-flat-cm}

In this section we define {\em counter machines}, which are
essentially a generalization of Rabin-Scott finite nondeterministic
automata, extended with a set of integer counters, and transitions
described by quantifier-free Presburger formulae. Formally, a counter
machine (CM) is a tuple $M = \tuple{\x, \mathcal{L},
  \ell_{\mathrm{init}}, \ell_{\mathrm{fin}}, \Delta}$,
where: \begin{compactitem}
\item  $\x$ is a set of \emph{variables} (counters) ranging over $\zed$, 
\item $\mathcal{L}$ is a set of \emph{control locations}, 
\item $\ell_{\mathrm{init}},\ell_{\mathrm{fin}} \in \mathcal{L}$ are
  the \emph{initial} and \emph{final} control locations, respectively,
\item $\Delta$ is a set of \emph{transition rules} of the form $\ell
  \arrow{\phi(\x,\x')}{} \ell'$, where $\ell,\ell' \in \mathcal{L}$
  are control locations and $\phi(\x,\x')$ is a QFPA formula defining
  both \begin{inparaenum}[(i)]
\item the conditions on the current values $\x$ that enable the
  transition, and
\item the updates of the current values $\x$ to the next values $\x'$.
\end{inparaenum}
\end{compactitem}
The size of the binary representation of a counter machine is defined
as $\bin{M}=\sum_{\ell\arrow{\phi}{}\ell'} \bin{\phi}$, i.e.\ the sum
of the sizes of all formulae labeling the transition rules of $M$. 

A \emph{configuration} of $M$ is a pair $(\ell,\nu)$, where $\ell \in
\mathcal{L}$ is a control location, and $\nu \in \zed^{\x}$ is a
valuation of the variables. A \emph{run} of $M$ is a sequence of
configurations $(\ell_0,\nu_0), \ldots, (\ell_n,\nu_n)$, where $\ell_0
= \ell_{\mathrm{init}}$, $\ell_n = \ell_{\mathrm{fin}}$ and for each
$i = 0,\ldots,n-1$, there exists a transition rule $\ell_i
\arrow{\phi_i}{} \ell_{i+1}$ such that $(\nu_i,\nu_{i+1}) \models
\phi_i$. The \emph{reachability problem} asks, given a counter machine
$M$, does $M$ have a run?

Let us now define the flatness restriction on counter machines. The
\emph{control flow graph} of $M$ is the labeled graph whose vertices
are the control locations $\mathcal{L}$ and whose edges are the
transition rules in $\Delta$. A cycle in this graph is
\emph{elementary} if it does not contain another cycle. A counter
machine $M$ is {\em flat} if and only if every control location
belongs to at most one elementary cycle in its control flow graph.

For a set of relations $\mathcal{R}$, we denote by
$\reachflat(\mathcal{R})$ the class of reachability problems for all
flat counter machines $M$ where, for each transition rule $\ell
\arrow{\phi(\x,\x')}{} \ell'$ belonging to a cycle in the control flow
graph of $M$, the formula $\phi(\x,\x')$ defines a relation from
$\mathcal{R}$. The main result is that $\reachflat(\oct)$ is
\textsc{Np}-complete, where $\oct$ is the set of relations defined
below.

\begin{definition}\label{odbc}
  A~formula $\phi(\x)$ is an \emph{octagonal constraint} if it is
  a~finite conjunction of atomic propositions of the form $\pm x_i
  \leq \alpha_i$ or $\pm x_i \pm x_j \leq \beta_{ij}$, where
  $\alpha_i,\beta_{ij} \in \zed$, for all $1 \le i,j\le N$. We denote
  by $\oct$ the set of relations $R \subseteq \zed^\x \times \zed^\x$
  defined by octagonal constraints $\phi(\x,\x')$.
\end{definition}

\begin{figure}[htb]
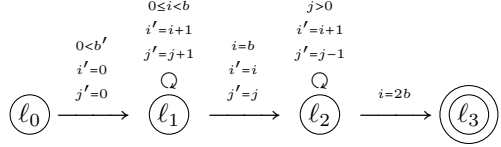

\[\circled{\ell_0} \arrow{\begin{array}{c}
    \scriptscriptstyle{0 < b'}\\[-1mm]
    \scriptscriptstyle{i'=0} \\[-1mm] 
    \scriptscriptstyle{j'=0} \end{array}}{} 
  \stackrel{\stackrel{\begin{array}{c}
      \scriptscriptstyle{0 \leq i < b} \\[-1mm]
      \scriptscriptstyle{i' = i+1} \\[-1mm]
      \scriptscriptstyle{j' = j+1} 
  \end{array}}{\lcirclearrowdown}}{\circled{\ell_1}}
  \arrow{\begin{array}{c}
      \scriptscriptstyle{i=b}\\[-1mm]
      \scriptscriptstyle{i'=i} \\[-1mm]
      \scriptscriptstyle{j'=j}
  \end{array}}{}
  \stackrel{\stackrel{\begin{array}{c}
      \scriptscriptstyle{j>0} \\[-1mm]
      \scriptscriptstyle{i' = i+1} \\[-1mm]
      \scriptscriptstyle{j' = j-1}
  \end{array}}{\lcirclearrowdown}}{\circled{\ell_2}}
  \arrow{\begin{array}{c}
      \scriptscriptstyle{i = 2b}
  \end{array}}{}
  \doublecircled{\ell_3}\]
\caption{A flat counter machine}
\label{fig:flat-cm}
\end{figure}

\begin{example}\label{ex:flat-cm}
Fig. \ref{fig:flat-cm} shows a flat counter machine $M =
\tuple{\set{i,j,b},\set{\ell_0,\ell_1,\ell_2,\ell_3},\ell_0,\ell_3,\Delta}$. The
machine increments both counters $i$ and $j$ by executing the
self-loop on state $\ell_1$ a number of times equal to the value of
$b$, that was guessed on the transition $\ell_0 \arrow{}{} \ell_1$,
then it will move to $\ell_2$ and will increment $i$, while
decrementing $j$, until $j=0$. Finally, it moves to its final state if
$i=2b$. Observe that all transition rules, except for $\ell_2
\arrow{i=2b}{} \ell_3$, are labeled with octagonal
relations. \hfill$\blacksquare$
\end{example} 

In the rest of this section we set the ground for the proof of the
\textsc{Np}-completness result by presenting necessary background
notions concerning octagonal constraints, starting with a simpler
class of formulae, called difference bounds constraints. In
particular, we show that the set of octagonal relations are closed
under compositions, thus $\oct$ is a class of relations in the sense
of Definition \ref{def:rel-class}.

\subsection{Difference Bounds Constraints}

Let $\vec{x}=\set{x_1, x_2, \ldots, x_N}$ be a~set of variables, for
some $N\in\nat_+$. Without losing generality, we consider only
formulae in which each atomic proposition involves exactly two
variables. Atomic propositions $x_i \leq \alpha_i$, $x_i \geq
\alpha_i$, for $\alpha_i\in\zed$, are replaced by $x_i - \zeta \leq
\alpha_i$, $\zeta - x_i \leq -\alpha_i$, respectively, for an extra
variable $\zeta$, with the implicit assumption $\zeta=0$.

\begin{definition}\label{dbc}
A \emph{difference bounds constraint} $\phi(\vec{x})$ is a finite
conjunction of atomic propositions of the form $x_i-x_j \le
\alpha_{ij},~ 1 \le i,j \le N$, where $\alpha_{ij}\in\zed$. A relation
$R \subseteq \zed^\x \times \zed^\x$ is a~{\em difference bounds
  relation} if it is defined by a~difference bounds constraint
$\phi_R(\x,\x')$.
\end{definition}

\begin{figure}
\begin{center}
\begin{tabular}{cc}
\mbox{\scalebox{0.8}{\begin{minipage}{0.50\textwidth}
\begin{center}
\begin{picture}(42, 28)(0,-12)
\put(0,0){\bordermatrix{~ & x_1 & x_2 & ~ & x_1' & x_2' \cr
                  x_1 & 0 & \infty & \VR & 1 & -1 \cr
                  x_2 & \infty & 0 & \VR & -2 & 2 \cr                       
                  \cline{2-6}
                  x_1' & \infty & \infty & \VR & 0 & \infty \cr
                  x_2' & \infty & \infty & \VR & \infty & 0 }}
\end{picture}
\end{center}
\end{minipage}
}}
& \mbox{\begin{minipage}{0.50\textwidth}
  \begin{center}
  \begin{tikzpicture}
    \scriptsize
    \tikzset{
      sState/.style={draw=black,circle,inner sep=1.5pt,semithick}
    }
    \node[sState] (x2) at (0mm,0mm) {$x_2$};
    \node[sState] (x2p) at (10mm,0mm) {$x_2'$};
    \node[sState] (x1) at (0mm,12mm) {$x_1$};
    \node[sState] (x1p) at (10mm,12mm) {$x_1'$};
    \path[->] 
       (x1) edge node [above] {$1$} (x1p)
            edge [bend right,bend angle=80] node [right] {$-1$} (x2p)
       (x2) edge node [below] {$2$} (x2p)
            edge [bend left,bend angle=80] node [right] {$-2$} (x1p);
  \end{tikzpicture}
  \end{center}
\end{minipage}}
\\\\
(a) $M_R^*$ &
(b) $\mathcal{G}_R$ 
\end{tabular}
\end{center}
\caption{Let $\phi(x_1,x_2,x_1',x_2') \equiv x_1-x_1' \leq 1 \wedge
  x_1-x_2' \leq -1 \wedge x_2-x_1' \leq -2 \wedge x_2-x_2' \leq 2$ be
  a difference bounds constraint. \textbf{(a)} shows the closed DBM of
  $M_\phi^*$ and \textbf{(b)} shows the constraint graph
  $\mathcal{G}_\phi$.}
\label{fig:dbm:ex}
\end{figure}
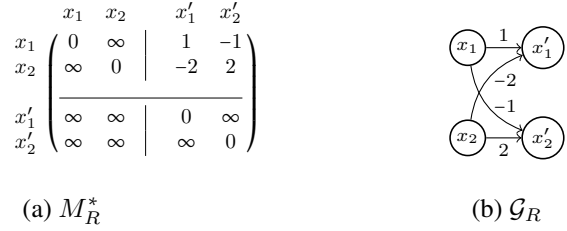

If $\phi(\vec{x})$ is a~difference bounds constraint, the~{\em
  difference bounds matrix} (DBM) representing $\phi$ is the matrix
$M_\phi \in \zed_\infty^{N \times N}$, where $(M_\phi)_{ij} =
\alpha_{ij}$ if $x_i-x_j\le \alpha_{ij} \in Atom(\phi)$, and
$(M_\phi)_{ij} = \infty$, otherwise (see Fig. \ref{fig:dbm:ex} (a) for
an example). In particular, any inconsistent difference bounds
constraint is represented by the $N \times N$ matrix $[-\infty]_N$,
whose coefficients are all $-\infty$. Dually, a matrix $M \in
\zed_\infty^{N \times N}$ corresponds to the difference bounds
constraint $\Phi(M) \equiv \bigwedge_{M_{ij} < \infty} x_i - x_j <
M_{ij}$. $M$ is \emph{consistent} if $\Phi(M)$ is a consistent
formula. The {\em constraint graph} $\mathcal{G}_\phi$ of a difference
bounds constraint $\phi$ is the weighted graph whose incidence matrix
is $M_\phi$ (see Fig. \ref{fig:dbm:ex} (b) for an example).

\begin{definition}\label{def:closed-dbm}
A consistent DBM $M \in \zed_\infty^{N \times N}$ is
said to be {\em closed} if and only if:
\begin{compactenum}
\item $M_{ii} = 0$, for all $1 \leq i \leq N$, and
\item all triangle inequalities $M_{ik}
\leq M_{ij} + M_{jk}$ hold, for all $1 \leq i,j,k \leq N$. 
\end{compactenum}
\end{definition}
Given a~consistent DBM $M$, the unique closed DBM which is logically
equivalent to $M$ is denoted by $M^*$. If $M$ is an inconsistent DBM,
we denote $M^*=[-\infty]_N$, by convention. Observe that the closed
DBM is a canonical (unique) representation of a difference bounds
constraint. Moreover, this canonical representation of a DBM can be
computed in cubic time, using the classical Floyd-Warshall shortest
path algorithm.

It is known that quantifier elimination for difference bounds
constraints takes cubic time in the size of the binary representation
of the constraint\footnote{To eliminate $\exists x . \phi(x)$, one
  computes the closed DBM $M_\phi^*$ in cubic time in the binary size
  of $\phi$ and eliminates the row and column corresponding to $x$
  from it.}. Then the set of relations $R \subseteq \zed^\x \times
\zed^\x$ defined by difference bounds constraints $\phi_R(\x,\x')$ is
closed under relational composition. Since the identity relation
$\id{\x}$ is definable by a difference constraint and the empty
relation $\emptyset$ is definable by any inconsistent constraint the
set of difference bounds relations forms a class (Definition
\ref{def:rel-class}), denoted as $\dbm$ in the following.

Because any difference bounds relation $R \subseteq \zed^\x \times
\zed^\x$, defined by a formula $\phi(\x,\x')$, is uniquely represented
by the difference bounds constraint $\Phi({M_\phi^*})$, we define the
size of its binary representation as $\bin{R}=\bin{\Phi(M_\phi^*)}$,
independently of the choice of $\phi$. In principle, any algorithm
that takes as input a difference bounds relation $R$ can be considered
w.l.o.g. to work directly on its canonical representation, because the
time needed to compute the canonical representation of $R$ is
$\mathcal{O}(\bin{R}^3)$, thus any super-polynomial bound derived with
this assumption carry over to the general case.

\subsection{Octagonal Constraints}

Given a set of variables $\x=\set{x_1,\ldots,x_N}$, an octagonal
constraint $\phi(\x)$ (Definition \ref{odbc}) is usually represented
by a difference bounds constraints $\overline{\phi}(\y)$, where
$\yy=\set{y_1,\ldots,y_{2N}}$, $y_{2i-1}$ stands for $+x_i$ and
$y_{2i}$ stands for $-x_i$, with the implicit requirement that
$y_{2i-1}=-y_{2i}$, for each $1\leq i\leq N$. Observe that the latter
condition cannot be directly represented as a difference bounds
constraint. Formally, we have:
\[ \begin{array}{lcl}
  (x_i-x_j\leq c)\in Atom(\phi) & \iff & (y_{2i-1}-y_{2j-1}\leq c), (y_{2j}-y_{2i}\leq c)\in Atom(\overline{\phi}) \\
  (-x_i+x_j\leq c)\in Atom(\phi) & \iff & (y_{2j-1}-y_{2i-1}\leq c), (y_{2i}-y_{2j}\leq c)\in Atom(\overline{\phi}) \\
  (-x_i-x_j\leq c)\in Atom(\phi) & \iff & (y_{2i}-y_{2j-1}\leq c), (y_{2j}-y_{2i-1}\leq c)\in Atom(\overline{\phi}) \\
  (x_i+x_j\leq c)\in Atom(\phi) & \iff & (y_{2i-1}-y_{2j}\leq c), (y_{2j-1}-y_{2i}\leq c)\in Atom(\overline{\phi}) 
\end{array} \]
In order to handle the $\yy$ variables in the following, we define
$\bar{\imath} = i-1$, if $i$ is even, and $\bar{\imath}=i+1$ if $i$ is
odd. Obviously, we have $\bar{\bar{\imath}} = i$, for all $i \in
\zed,~ i \geq 1$. 

An octagonal constraint $\phi(\x)$ is represented by the matrix
$M_{\overline{\phi}} \in \zed_\infty^{2N \times 2N}$, corresponding to
$\overline{\phi}(\y)$. A matrix $M \in \zed_\infty^{2N \times 2N}$ is
\emph{coherent} if $M_{ij} = M_{\bar{\jmath}\bar{\imath}}$ for all $1
\leq i,j \leq 2N$. This property is needed because an atomic
proposition $x_i - x_j \leq c$ can be represented as both $y_{2i-1} -
y_{2j-1} \leq c$ and $y_{2j} - y_{2i} \leq c$. Dually, a~coherent
matrix $M \in \zed_\infty^{2N \times 2N}$ corresponds to the following
octagonal constraint:
\[
\Omega(M) \equiv \!\!\!\! \bigwedge_{\scriptscriptstyle 1 \leq i,j \leq N}\!\!\!\!
  x_i - x_j \leq M_{\scriptscriptstyle 2i-1,2j-1} 
  \wedge \!\!\!\!\bigwedge_{\scriptscriptstyle 1 \leq i,j \leq N}\!\!\!\!
  x_i + x_j \leq M_{\scriptscriptstyle 2i-1,2j} 
  \wedge \bigwedge_{\scriptscriptstyle 1 \leq i,j \leq N}
  -x_i-x_j \leq M_{\scriptscriptstyle 2i,2j-1}
\]
A coherent DBM $M$ is said to be \emph{octagonal-consistent} if
$\Omega(M)$ is consistent.

\begin{definition}\label{def:tclose}
An octagonal-consistent coherent DBM $M \in \zed_\infty^{2N \times 2N}$
is said to be \emph{tightly closed} if and only if it is closed and
$M_{ij} \le \lfloor \frac{M_{i\bar{\imath}}}{2} \rfloor + \lfloor
\frac{M_{\bar{\jmath}j}}{2} \rfloor$, for all $1 \leq i,j \leq N$.
\end{definition}

The last condition from Definition \ref{def:tclose} ensures that the
knowledge induced by the implicit conditions $y_i + y_{\bar\imath} =
0$, which cannot be represented as difference constraints, has been
propagated through the DBM. Since $2y_i = y_i - y_{\bar\imath} \leq
M_{i\bar\imath}$ and $-2y_j = y_{\bar\jmath} - y_j \leq M_{\bar\jmath
  j}$, we have $y_i \leq \lfloor\frac{M_{i\bar\imath}}{2}\rfloor$ and
$-y_j \leq \lfloor\frac{M_{\bar\jmath j}}{2}\rfloor$, which implies
$y_i - y_j \leq \lfloor\frac{M_{i\bar\imath}}{2}\rfloor +
\lfloor\frac{M_{\bar\jmath j}}{2}\rfloor$, thus $M_{ij} \leq
\lfloor\frac{M_{i\bar\imath}}{2}\rfloor + \lfloor\frac{M_{\bar\jmath
    j}}{2}\rfloor$ must hold, if $M$ is supposed to be the most
precise DBM representation of an octagonal constraint. If $j =
\bar\imath$ in the previous, we obtain $M_{i\bar\imath} \leq 2 \lfloor
\frac{M_{i\bar\imath}}{2} \rfloor$, implying that $M_{i\bar\imath}$ is
necessarily even, if $M$ is tightly closed.

The following theorem \cite{BagnaraHillZaffanella08} provides an
effective way of testing octagonal-consistency and computing the tight
closure of a~coherent DBM. Moreover, it shows that the tight closure
of a~given DBM is unique and can also be computed within the same
cubic time upper bound, as the DBM closure:

\begin{theorem}\label{thm:bhz}
Let $M \in \zed_\infty^{2N \times 2N}$ be a~coherent DBM. Then $M$ is
octagonal-consistent iff $M$ is consistent and $\lfloor
\frac{M^*_{i\bar{\imath}}}{2} \rfloor + \lfloor
\frac{M^*_{\bar{\imath}i}}{2} \rfloor \geq 0$, for all $1 \leq i \leq
2N$. Moreover, if $M$ is octagonal-consistent, the tight closure of
$M$ is the DBM $M^t \in \zed_\infty^{2N \times 2N}$ defined
as: $$M^t_{ij}=\min\left\{M^*_{ij}, \left\lfloor
\frac{M^*_{i\bar{\imath}}}{2} \right\rfloor + \left\lfloor
\frac{M^*_{\bar{\jmath}j}}{2} \right\rfloor\right\}$$ for all $1 \leq
i,j \leq 2N$, where $M^* \in \zed_\infty^{2N \times 2N}$ is the closure
of $M$.
\end{theorem}
\proof{\cite[Theorems 2 and 3]{BagnaraHillZaffanella08}.\qed}

The tight closure of DBMs is needed for checking entailment between
octagonal constraints and for quantifier elimination, as shown by the
following proposition.

\begin{proposition}\label{prop:oct-entl-qe}
  Let $\phi(\x)$ and $\psi(\x)$ be two consistent octagonal
  constraints. Then, the following hold:
  \begin{compactenum}
  \item\label{it:oct-entl-qe1} $\phi \Rightarrow \psi$ if and only if
    $\left(M^t_{\overline{\phi}}\right)_{ij} \leq
    \left(M^t_{\overline{\psi}}\right)_{ij}$, for all $1 \leq i,j \leq
    2N$.
  \item\label{it:oct-entl-qe2} $\exists x_k.\phi(\x) \iff
    \Omega(M')$, where $M'$ is the DBM obtained by eliminating the
    lines and columns $2k$ and $2k+1$ from $M^t_{\overline{\phi}}$.
  \end{compactenum}
\end{proposition}
\proof{Point (\ref{it:oct-entl-qe1}) is by \cite[Theorem
    4.4.1]{mine-PHD04} and point (\ref{it:oct-entl-qe2}) is by
  \cite[Theorem 2]{tacas09}. \qed}

Since octagonal constraints have quantifier elimination, by
Proposition \ref{prop:oct-entl-qe} (\ref{it:oct-entl-qe2}), the set
$\oct$ of octagonal relations forms a class, in the sense of
Definition \ref{def:rel-class}. Moreover, since a tightly closed DBM
is a canonical representation of an octagonal relation, we can define
w.l.o.g. the size of the binary representation of a relation $R \in
\oct$ as $\bin{R}=\bin{\Omega(M^t_{\overline{\phi}})}$, where $\phi$
is any octagonal constraint that defines $R$. Again, this definition
has no impact on the computational complexity of the decision problems
involving octagonal relations, because the canonical representation of
any $R \in \oct$ can be computed in time $\mathcal{O}(\bin{R}^3)$. 

\begin{figure}
\begin{tabular}{c@{\extracolsep{0mm}}c}
\mbox{\begin{minipage}{6.0cm}
  \begin{tikzpicture}
    \tiny
    \tikzset{
      sState/.style={draw=black,circle,inner sep=1.5pt,semithick}
    }

    \node[sState] (xpm) at (0mm,0mm) {$y_2'$};
    \node[sState] (xm) at (15mm,0mm) {$y_2$};
    \node[sState] (ym) at (30mm,0mm) {$y_4$};
    \node[sState] (ypm) at (45mm,0mm) {$y_4'$};

    \node[sState] (xpp) at (0mm,15mm) {$y_1'$};
    \node[sState] (xp) at (15mm,15mm) {$y_1$};
    \node[sState] (yp) at (30mm,15mm) {$y_3$};
    \node[sState] (ypp) at (45mm,15mm) {$y_3'$};

    \path[->,bend angle=5] 
       (xpp) edge [bend left] node [below] {$-2$} (xp)
       (xm) edge [bend left] node [above] {$-2$} (xpm)
       (ypp) edge [bend right] node [below] {$-3$} (yp)
       (ym) edge [bend right] node [above] {$-3$} (ypm)
       (xp) edge [bend right] node {} (ym)
       (yp) edge [bend left] node {} (xm)
       (ypp) edge [bend angle=25,bend right] node [below] {$1$} (xpp)
       (xpm) edge [bend angle=25,bend right] node [above] {$1$} (ypm)
       (ypp) edge [bend left,dashed] node [left] {\shortstack{\st{$1$} \\ $0$ }} (ypm);

    \node at (20mm,11mm) {$5$};
    \node at (25mm,11mm) {$5$};

  \end{tikzpicture}
\end{minipage}}
&
\mbox{\begin{minipage}{6.5cm}
\scalebox{0.85}{\bordermatrix{
   ~ & y_1 & y_2 & y_3 & y_4 & y_1' & y_2' & y_3' & y_4' \cr
   y_1 & 0 & \infty & \infty & 5 & \infty & \infty & \infty & 2 \cr
   y_2 & \infty & 0 & \infty & \infty & \infty & -2 & \infty & -1 \cr
   y_3 & \infty & 5 & 0 & \infty & \infty & 3 & \infty & 4 \cr
   y_4 & \infty & \infty & \infty & 0 & \infty & \infty & \infty & -3 \cr
   y_1' & -2 & \infty & \infty & 3 & 0 & \infty & \infty & 0 \cr
   y_2' & \infty & \infty & \infty & \infty & \infty & 0 & \infty & 1 \cr
   y_3' & -1 & 2 & -3 & 4 & 1 & 0 & 0 & 0 \cr
   y_4' & \infty & \infty & \infty & \infty & \infty & \infty & \infty & 0 
}}
\end{minipage}}
\\[2mm] $\mathcal{G}_{\overline{\phi}}$ & $M_{\overline{\phi}}^t$
\end{tabular}
\caption{Graph and matrix representation of the difference bounds
  representation $\overline{\phi}(\y,\y')$ of an octagonal relation defined by
  $\phi(\x,\x') \equiv x_1+x_2 \leq 5 ~\wedge~ x_1'-x_1 \leq -2 ~\wedge~
  x_2'-x_2 \leq -3 ~\wedge~ x_2'-x_1' \leq 1$.}
\label{fig:oct:ex}
\end{figure}
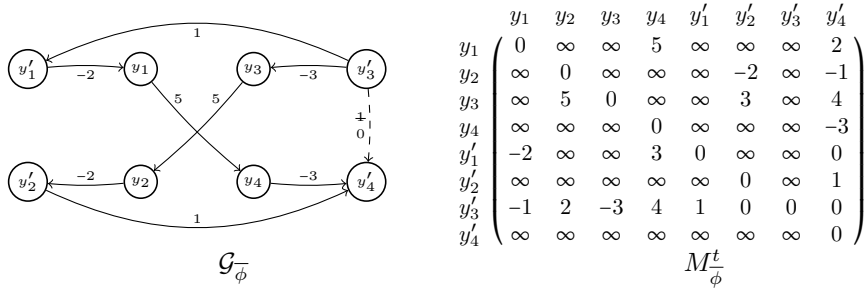

\begin{example}
Consider the octagonal relation defined by $\phi(x_1,x_2,x_1',x_2') \equiv x_1+x_2 \leq 5
\wedge x_1'-x_1 \leq -2 \wedge x_2'-x_2 \leq -3 \wedge x_2'-x_1' \leq
1$.  Its difference bounds representation is $\overline{\phi}(\y,\y')
\iff y_1-y_4 \leq 5 \wedge y_3-y_2 \leq 5 \wedge y_1'-y_1 \leq -2
\wedge y_2-y_2' \leq -2 \wedge y_3'-y_3 \leq -3 \wedge y_4-y_4' \leq
-3 \wedge y_3'-y_1' \leq 1 \wedge y_2'-y_4' \leq 1$, where
$\y=\{y_1,\dots,y_4\}$. Figure \ref{fig:oct:ex}(a) shows the graph
representation $\mathcal{G}_{\overline\phi}$. Note that the implicit constraint
$y_3'-y_4' \leq 1$, represented by a~dashed edge in Figure
\ref{fig:oct:ex}(a), is not tight. The tightening step replaces the bound
$1$, crossed in Figure \ref{fig:oct:ex}(a), with $0$. Figure
\ref{fig:oct:ex}(b) shows the tightly closed DBM representation of $R$,
denoted $M^t_{\overline{\phi}}$. \hfill$\blacksquare$
\end{example}

\subsection{Periodic Relations}

As shown in the previous, both difference bounds and octagonal
relations are closed under compositions and have canonical matrix
representations. When studying the complexity of the reachability
problem $\reachflat(\oct)$, a crucial point concerns the behavior of
the power sequence $\set{R^k}_{k=0}^\infty$, for a relation $R \in
\oct$. A first observation is that, for any $k\geq0$, we have that
$R^k \in \oct$. If we denote by $\sigma(R)$ the canonical DBM
representation of $R$, we show that the sequence of DBMs
$\{\sigma(R^k)\}_{k=0}^\infty$ is \emph{periodic}, in a sense that we
define next. This fact is important for the definition of a
nondeterministic algorithm that solves the above reachability
problem.

We consider infinite sequences $\set{s_k}_{k=0}^\infty$ in
$\zed_{\pm\infty}$, with the following extension of
addition: \begin{inparaenum}[(i)]
\item for all $x\in\zed_\infty$, $x+\infty=\infty+x=\infty$, and
\item for all $x\in\zed_{\pm\infty}$, $x+(-\infty)=-\infty+x=-\infty$.
\end{inparaenum} 
A sequence $\set{s_k}_{k=0}^\infty$ is an \emph{arithmetic
  progression} if there exists a constant
$\lambda\in\zed_{\pm\infty}$, called \emph{rate}, such that $s_{k+1} =
s_k + \lambda$, for all $k\geq0$. A generalization of this notion are
\emph{periodic} sequences, defined below.

\begin{definition}\label{def:periodic-sequence}
  An infinite sequence $\set{s_k}_{k=0}^\infty$, where $s_k \in
  \zed_{\pm\infty}$, for all $k\geq0$, is said to be \emph{periodic}
  if and only if there exist integer constants $b\geq0$, $c>0$ and
  $\lambda_0,\ldots,\lambda_{c-1}\in \zed_{\pm\infty}$ such that
  $s_{b+(k+1)c+i}=s_{b+kc+i}+\lambda_i$, for all $k \geq 0$ and all
  $i\in[c]$. The smallest $b$, $c$ and $\lambda_i$ are called the
  \emph{prefix}, \emph{period} and \emph{rates} of the sequence.
\end{definition}
Note that an arithmetic progression is a periodic sequence with prefix
$0$ and period $1$.

In the following, we consider sequences of square matrices and say
that an infinite sequence $\set{M_k}_{k=0}^\infty$ of matrices
$M_k\in\zed_{\pm\infty}^{n\times n}$ is \emph{periodic} if every
sequence $\{\left(M_k\right)_{ij}\}_{k=0}^\infty$ is periodic, for all
$i,j \in [n]$. The next lemma provides a characterization of
periodicity for a sequence of matrices, with an estimation of its
prefix and period.

\begin{lemma}\label{lemma:periodic-matrix-sequence}
  A a sequence of $\zed_{\pm\infty}^{n \times n}$ matrices
  $\set{M_k}_{k=0}^\infty$ is periodic iff there exist integers
  $b\geq0$, $c>0$ and matrices $\Lambda_0,\ldots,\Lambda_{c-1} \in
  \zed_{\pm\infty}^{n \times n}$ such that: \[\forall k\geq0 \forall
  i\in[c] ~.~ M_{b+(k+1)c+i}=M_{b+kc+i}+\Lambda_i\enspace.\] If,
  moreover, $b_{ij}$ and $c_{ij}$ are the prefix and period of the
  sequence $\{\left(M_k\right)_{ij}\}_{k=0}^\infty$, then
  $b=\max_{1 \leq i,j\leq n} (b_{ij})$, $c=\lcm_{1\leq i,j\leq n} (c_{ij})$ are the
  smallest such integers.
\end{lemma}
\proof{ ``$\Rightarrow$'' Suppose that the sequence
  $\set{M_k}_{k=0}^\infty$ is periodic. Then, for each $i,j\in[n]$,
  the sequence $\set{(M_k)_{ij}}_{k=0}^\infty$ is periodic, and let
  $\lambda^{ij}_0,\ldots,\lambda^{ij}_{c_{ij}-1}$ be the rates of this
  sequence. For all $\ell\in[c]$ and all $i,j\in[n]$, define
  $\left(\Lambda_\ell\right)_{ij}=\frac{c}{c_{ij}}\cdot\left(\lambda^{ij}_{(b-b_{ij}+\ell)
    \xmod c_{ij}}\right)$. The check
  $M_{b+(k+1)c+\ell}=M_{b+kc+\ell}+\Lambda_\ell$, for all $k\geq0$ and
  $\ell\in[c]$ is straightforward. ``$\Leftarrow$'' For each
  $i,j\in[n]$, the sequence $\set{\left(M_k\right)_{ij}}_{k=0}^\infty$
  is periodic: $\left(M_{b+(k+1)c+\ell}\right)_{ij} =
  \left(M_{b+kc+\ell}\right)_{ij} + \left(\Lambda_\ell\right)_{ij}$,
  for all $k\geq0$ and $\ell\in[c]$.

  For the last point, suppose first, by contradiction, that there
  exists $b' < \max_{1\leq i,j\leq n} (b_{ij})$ such that
  $M_{b'+(k+1)c+\ell}=M_{b'+kc+\ell}+\Lambda_\ell$, for all $k\geq0$
  and $\ell\in[c]$. Let $1 \leq s,t\leq n$ be such that
  $b_{st}=\max_{1 \leq i,j\leq n} (b_{ij})$. Then the sequence
  $\set{\left(M_k\right)_{st}}_{k=0}^\infty$ is periodic with prefix
  $b'<b_{st}$, contradiction. Second, suppose by contradiction, that
  there exists $c' < \lcm_{1\leq i,j\leq n} (c_{ij})$ such that
  $M_{b+(k+1)c'+\ell}=M_{b+kc'+\ell}+\Lambda_\ell$, for all $k\geq0$
  and $\ell\in[c']$. Then there exists $1\leq s,t\leq n$ such that
  $c_{st}$ does not divide $c'$, which contradicts the fact that the
  sequence $\set{\left(M_k\right)_{st}}_{k=0}^\infty$ is periodic with
  period $c_{st}$. \qed}

Let us focus now on sequences of matrices that represent the power
sequences $\{R^k\}_{k=0}^\infty$, where $R \in \oct$. Given a set of
variables $\x=\set{x_1,\ldots, x_N}$, we denote by $\oct_\x$ the class
of octagonal relations $R \subseteq \zed^\x \times \zed^\x$. Formally,
let $\sigma : \oct_\x \rightarrow \zed_\infty^{4N \times 4N} \cup
\set{[-\infty]_{4N}}$ be the bijection that maps each consistent
relation $R$ into its canonical DBM $\sigma(R) \in \zed_\infty^{4N
  \times 4N}$ and the inconsistent relation into $\sigma(\emptyset) =
[-\infty]_{4N}$. Then $R$ is said to be \emph{periodic} if the matrix
sequence $\{\sigma(R^k)\}_{k=0}^\infty$ is periodic. If every
relation in a certain class is periodic, we call that class periodic
as well.

\begin{example}\label{ex:periodic-rel}
Consider the octagonal relation $R \subseteq \zed^{\set{x,y}} \times
\zed^{\set{x,y}}$ defined by the formula $x'=y+1 \wedge y'=x$, where
for all $\ell\in\nat$:
\[
\resizebox{\hsize}{!}{$\sigma(R^{2\ell+1}) = \begin{array}{c|cccc}
& x & y & x' & y' \\
\hline
x  & 0      & \infty & \infty & \ell    \\
y &  \infty & 0      & -\ell-1   & \infty \\
x' & \infty & \ell+1    & 0      & \infty \\
y' & -\ell      & \infty & \infty & 0
\end{array}
\hspace*{4mm}
\sigma(R^{2\ell+2}) = \begin{array}{c|cccc}
& x & y & x' & y' \\
\hline
x & 0      & \infty & -\ell-1   & \infty \\
y & \infty & 0      & \infty & -\ell-1   \\
x' & \ell+1    & \infty & 0      & \infty \\
y' & \infty & \ell+1    & \infty & 0     
\end{array}$}
\]
The sequence $\{\sigma(R^k)\}_{k=0}^\infty$ is periodic with prefix
$b=1$ and period $c=2$, where:
\[
\resizebox{0.7\hsize}{!}{$
\Lambda_0 = \begin{array}{|cccc|}
0 & 0 & 0 & 1 \\
0 & 0 & -1 & 0 \\
0 & 1 & 0 & 0 \\
-1 & 0 & 0 & 0
\end{array}
\hspace*{2cm}
\Lambda_1 = \begin{array}{|cccc|}
0 & 0 & -1 & 0 \\
0 & 0 & 0 & -1 \\
1 & 0 & 0 & 0 \\
0 & 1 & 0 & 0
\end{array}$}\] 
\hfill$\blacksquare$
\end{example}

One of the results in this paper is that the class $\oct$ is
periodic. The proof of this fact relies essentially on the fact that
the class $\dbm$ is periodic, and uses (a variant of) Theorem
\ref{thm:bhz} to generalize this result from difference bounds to
octagonal relations. In the next section we give a generic
nondeterministic decision procedure for the problem
$\reachflat(\mathcal{R})$, where $\mathcal{R}$ is a periodic class of
relations. Moreover, we identify certain conditions under which each
branch of the procedure terminates in polynomial time, which provides
an $\textsc{Np}$ upper bound for the $\reachflat(\mathcal{R})$
problem.

\section{An Algorithm for the Reachability Problem}

In general, the decision procedures for the reachability problem for
flat counter machines rely on \emph{acceleration}
\cite{BoigelotPhD,LerouxFinkel02}, which is defining the transitive
closure of the relations that occur on the cycles of these machines by
formulae from the quantifier-free fragment of Presburger
arithmetic. To show that these reachability problems belong to the
class \textsc{Np}, it is essential to build these QFPA formulae in
polynomial time.

For the sake of simplicity, we explain the idea of a nondeterministic
algorithm (Algorithm \ref{alg:ndtm}) for the reachability problem on
the flat counter machine below:
\begin{equation}\label{eq:one-loop-cm}
\ell_{\mathrm{init}} \arrow{I(\x)}{}
\stackrel{\stackrel{\phi(\x,\x')}{\lcirclearrowdown}}{\ell} 
\arrow{F(\x)}{} \ell_{\mathrm{fin}}
\end{equation}
where $I(\x)$ and $F(\x)$ are QFPA formulae and $\phi(\x,\x')$ is an
octagonal constraint defining a relation $R \in \oct_\x$, for a given
set of variables $\x=\set{x_1,\ldots,x_N}$.

Let us assume for now that this relation is periodic. The algorithm
guesses candidate values for the prefix $b\geq0$ and period $c>0$ of
$R$ (line \ref{ln:prefix-period-guess}), computes a candidate rate
$\Lambda$ (line \ref{ln:rate-guess}), and checks if $b$, $c$ and
$\Lambda$ satisfy the following condition (line \ref{ln:ind}):
\begin{equation}\label{eq:ind}
\textsc{Ind}(B,C,\Lambda):\enspace\forall n\geq0 ~.~ 
\sigma\left(\sigma^{-1}(B+n \cdot \Lambda) \circ \sigma^{-1}(C)\right)
= B+(n+1) \cdot \Lambda
\end{equation}
where $B, C$ and $\Lambda$ are square matrices of equal dimension, in
our case $B=\sigma(R^b)$, $C=\sigma(R^c)$ and $\Lambda$ is such that
$\sigma(R^b)+\Lambda=\sigma(R^{b+c})$. Intuitively, this means that
$b$, $c$ and $\Lambda$ are valid choices for the prefix, period and
rate of the sequence of matrices $\set{\sigma(R^k)}_{k=0}^\infty$, in
the sense of Lemma \ref{lemma:periodic-matrix-sequence}. 

In case the reachability problem for $M$ has a positive answer,
i.e.\ there exists a run from $\ell_{\mathrm{init}}$ to
$\ell_{\mathrm{fin}}$, two cases are possible. Either the number of
iterations of the loop is
\begin{inparaenum}[(i)]
\item strictly smaller than $b$, or
\item between $b+nc$ and $b+(n+1)c$, for some $n\geq0$.
\end{inparaenum}
The first case is captured by the QFPA formula $\phi^{<b}$ (line
\ref{ln:lt-b}), where $\Omega(\sigma(R^i))$ is the canonical octagonal
constraint representing the relation $R^i$. 

The second case is encoded by the QFPA formula $\phi^{\geq b}$ (line
\ref{ln:gt-b}). Here $k\not\in \x$ is a parameter variable and by
$\zed[k]_\infty$ we denote the set of univariate linear terms of the
form $a \cdot k + b$, with $a,b\in\zed_\infty$. Also
$\zed[k]_\infty^{m \times m}$ denotes the set of $m\times m$ square
matrices of such terms. With these notations, $\varsigma$ is a mapping
of matrices $M[k] \in \zed[k]_\infty^{4N \times 4N}$ into parametric
octagonal constraints consisting of atomic propositions of the form
$\pm x \pm y \leq a \cdot k + b$, defined in the same way as the
octagonal constraint $\Omega(M)$ is defined for a matrix $M \in
\zed_\infty^{4N \times 4N}$. Moreover, $\varsigma$ satisfies the
following condition:
\begin{equation}
\forall M \in \zed[k]_\infty^{4N \times 4N} \forall n \in \nat ~.~
\varsigma(M)(n) = \sigma^{-1}(M[n/k])
\end{equation}
The final step is checking the satisfiability of the disjunction
$\phi^{<b} \vee \phi^{\geq b}$ (line \ref{ln:disj}). If the formula
produced by a nondeterministic branch of the algorithm is
satisfiable, the reachability question has a positive
answer. Otherwise, if no branch produces a satisfiable formula, the
reachability question has a negative answer.

\begin{algorithm}[t]
\begin{algorithmic}[0]
\State {\bf input:} 
$M=\tuple{\x,\set{\ell_{\mathrm{init}},\ell,\ell_{\mathrm{fin}}},\ell_{\mathrm{init}},\ell_{\mathrm{fin}},\Delta}$
of the form (\ref{eq:one-loop-cm}), where $\x=\set{x_1,\ldots,x_N}$
\State {\bf output:} \textsc{Yes} if and only if $M$ has a
run from $\ell_{\mathrm{init}}$ to $\ell_{\mathrm{fin}}$
\end{algorithmic}
\begin{algorithmic}[1]
  \State {\bf let} $R$ be the relation defined by $\phi(\x,\x')$
  \State {\bf chose} $b\geq0$ and $c>0$
  \label{ln:prefix-period-guess}
  \State {\bf let} $\Lambda \in \zed_\infty^{4N \times 4N}$ be a matrix such that 
  $\sigma(R^{b}) + \Lambda = \sigma(R^{b+c})$
  \label{ln:rate-guess}
  \If{$\textsc{Ind}(\sigma(R^b),\sigma(R^c),\Lambda)$}
  \label{ln:ind}
  \State {\bf chose} $i \in [b]$ 
  \State $\phi^{<b} \leftarrow I(\x) \wedge \Omega(\sigma(R^i)) \wedge F(\x')$
  \label{ln:lt-b}
  \State {\bf chose} $j \in [c]$ 
  \State $\phi^{\geq b} \leftarrow k \geq 0 \wedge I(\x) \wedge
  \varsigma(\sigma(R^{b+j}) + k \cdot \Lambda) \wedge F(\x')$
  \label{ln:gt-b}
  \If{$\phi^{<b} \vee \phi^{\geq b}$ is satisfiable}
  \label{ln:disj}
  \State {\bf return} \textsc{Yes}
  \EndIf
  \EndIf
  \State {\bf fail}
  \label{ln:fail}
\end{algorithmic}
\caption{nondeterministic algorithm for the reachability problem 
  (\ref{eq:one-loop-cm}) \label{alg:ndtm}}
\end{algorithm}

To prove that the class of reachability problems $\reachflat(\oct)$ is
in \textsc{Np}, it is enough to show that, for any machine $M$ of the
form (\ref{eq:one-loop-cm}), each branch of Algorithm \ref{alg:ndtm}
terminates in $\textsc{Ptime}(\bin{M})$. For this, the matrices
$\sigma(R^c)$, $\sigma(R^i)$ and $\sigma(R^{b+j})$ must be computable
in $\textsc{Ptime}(\bin{R})$, for all $i = 0,\ldots,b$ and
$j=0,\ldots,c$ and, moreover, the condition
$\textsc{Ind}(\sigma(R^b),\sigma(R^c),\Lambda)$ (\ref{eq:ind}) must be
decidable in $\textsc{Nptime}(\bin{R})$. Under these conditions, the
QFPA formulae $\phi^{<b}$ and $\phi^{\geq b}$ are of polynomial size
in $\bin{M}$, and the satisfiability of their disjunction is decidable
in $\textsc{Nptime}(\bin{M})$.

The following theorem generalizes this argument to arbitrary flat
counter machines by giving sufficient conditions under which the class
$\reachflat(\oct)$ is \textsc{Np}-complete.

\begin{theorem}\label{thm:flat-cm-np}
$\reachflat(\oct)$ is \textsc{Np}-complete if there exists a constant
  $d$, such that the following hold, for each relation $R \in \oct$: \begin{compactenum}
  \item\label{it1:flat-cm-np} $\len{R^n}
    = \mathcal{O}((\len{R} \cdot \log n)^d)$, for all $n>0$,
  \item\label{it2:flat-cm-np} $R$ is periodic with prefix and period
    of the order of $2^{\mathcal{O}(\len{R}^d)}$.
  \end{compactenum} 
\end{theorem}

Before giving the proof of Theorem \ref{thm:flat-cm-np}, we show that
the condition $\textsc{Ind}(B,C,\Lambda)$ is decidable in
nondeterministic polynomial time, by reduction to the satisfiability
of a QFPA formula. The proof relies on a symbolic tight closure
algorithm (Algorithm \ref{alg:symb-tc}), which builds such a formula
using a cubic number of steps.

\begin{algorithm}[htb]
\begin{algorithmic}[0]
\State {\bf input:} a matrix $M \in \zed_\infty[k]^{m
  \times m}$ of univariate linear terms
\State {\bf output:} a matrix $T \in \zed_\infty[k]^{m \times m}$ of
univariate terms over $\min,+$, and $\lfloor\frac{.}{2}\rfloor$ 
\end{algorithmic}
\begin{algorithmic}[1]
  \Function{SymbFW}{$M$}
  \For{$i=1,\ldots,m$}
  \For{$j=1,\ldots,m$}
  \State $P_{ij} \leftarrow M_{ij}$
  \EndFor
  \EndFor
  \For{$k=1,\ldots,m$}
  \For{$i=1,\ldots,m$}
  \For{$j=1,\ldots,m$}
  \State $P_{ij} \leftarrow \min\left(P_{ij}, P_{ik} + P_{kj}\right)$
  \label{ln:min}
  \EndFor
  \EndFor
  \EndFor
  \State {\bf return} $P$
  \EndFunction
\end{algorithmic}
\begin{algorithmic}[1]
  \State $T \leftarrow \Call{SymbFW}{M}$
  \For{$i=1,\ldots,m$}
  \For{$j=1,\ldots,m$}
  \State $T_{ij} \leftarrow \min(T_{ij}, 
  \lfloor\frac{T_{i\bar\imath}}{2}\rfloor + 
  \lfloor\frac{T_{\bar\jmath j}}{2}\rfloor)$
  \EndFor
  \EndFor
  \State {\bf return} $T$
\end{algorithmic}
\caption{Symbolic Tight Closure algorithm}
\label{alg:symb-tc}
\end{algorithm}

\begin{lemma}\label{lemma:ind-np}
  Given $N>0$ and matrices $B,C,\Lambda \in \zed_\infty^{4N \times
    4N}$, the condition $\textsc{Ind}(B,C,\Lambda)$ is decidable in
  nondeterministic polynomial time.
\end{lemma}
\proof{The condition $\textsc{Ind}(B,C,\Lambda)$ checks the validity
  of the equivalence:
 \[\forall k\geq0 ~.~ \sigma(\sigma^{-1}(B+k\cdot\Lambda) \circ \sigma^{-1}(C)) = B + (k+1)\cdot\Lambda
 \enspace.\] For a matrix $M \in \zed_\infty[k]^{4N \times 4N}$ of
 univariate linear terms in $k$, we define the labeled graph
 $\mathcal{H}_M = \tuple{\y\cup\y',\rightarrow}$, where
 $\y=\set{y_1,\ldots,y_{2N}}$ are the variables used in the difference
 bounds encoding of an octagonal relation with variables
 $\set{x_1,\ldots,x_N}$, and whose labeled edges are \(y_i
 \arrow{M_{i,j}}{} y_j,~ y_i \arrow{M_{i,j+2N}}{} y'_j,~ y'_i
 \arrow{M_{i+2N,j}}{} y_j,~ y'_i \arrow{M_{i+2N,j+2N}}{} y'_j,\) for
 all $1 \leq i,j \leq 2N$.
The left-hand side of the equivalence $\textsc{Ind}(B,C,\Lambda)$
corresponds to the graph $\mathcal{H}_{\mathrm{lhs}}$ with vertices
$\y^{(0)}\cup\y^{(1)}\cup\y^{(2)}$, such that: \begin{compactitem}
 \item $\mathcal{H}_{\mathrm{lhs}}\proj_{\y^{(0)}\cup\y{(1)}}$ is the
   graph $\mathcal{H}_{B+k\cdot\Lambda}$, in which the vertices
   $\y^{(0)}$ and $\y^{(1)}$ replace $\y$ and $\y'$, and
 \item $\mathcal{H}_{\mathrm{lhs}}\proj_{\y^{(1)}\cup\y{(2)}}$ is
   the constraint graph $\mathcal{G}_{C}$ representing the difference
   bounds constraint $\Phi(C)$, in which $\y^{(1)}$ and
   $\y^{(2)}$ replace the vertices $\y$ and $\y'$ of
   $\mathcal{G}_C$, respectively.
 \end{compactitem}
 The right-hand side of the equivalence $\textsc{Ind}(B,C,\Lambda)$ is
 represented, in a similar way, by the graph
 $\mathcal{H}_{\mathrm{rhs}}$, which equals the graph
 $\mathcal{H}_{B+(k+1)\cdot\Lambda}$, with vertices $\y^{(0)}$ and
 $\y^{(2)}$ replacing $\y$ and $\y'$, respectively.
 
Since both graphs denote (parametric) octagonal constraints, by
Proposition \ref{prop:oct-entl-qe} (\ref{it:oct-entl-qe1}) we need to
prove that, for all $k\geq0$ the path labels corresponding to the
minimal paths within the \emph{tight closures} of the incidence
matrices of $\mathcal{H}_{\mathrm{lhs}}$ and
$\mathcal{H}_{\mathrm{rhs}}$ are equal, for all $k\geq0$. These tight
closures can be expressed by univariate terms with variable $k$, built
in time $\mathcal{O}(N^3)$, from constants $c\in\zed$ and the
functions $\min,+$ and $\lfloor\frac{.}{2}\rfloor$, by Algorithm
\ref{alg:symb-tc}, which implements the result of Theorem
\ref{thm:bhz}. Notice that the size of each such term is bounded by
the time needed to build it, i.e.\ $\mathcal{O}(N^3)$. Finally, each
term can be encoded in QFPA, because all constituent functions,
i.e.\ $\min,+$ and $\lfloor\frac{.}{2}\rfloor$ are QFPA-definable. As
a direct consequence, the condition $\textsc{Ind}(B,C,\Lambda)$ is
decidable in \textsc{Nptime}. \qed}

\paragraph{\bf Proof of Theorem \ref{thm:flat-cm-np}} \textsc{Np}-hardness 
is by reduction from the satisfiability problem for QFPA, and the fact
that any transition rule of a flat CM, that is not part of a cycle,
can be labeled by an arbitrary QFPA formula. Given an instance
$\phi(\x)$ of the QFPA satisfiability problem, we consider the CM
\(\ell_{\mathrm{init}} \arrow{\phi}{} \ell_{\mathrm{fin}}\). The
reachability problem has a positive answer iff $\phi$ has a satisfying
assignment.

To prove that $\reachflat(\oct)$ is contained in \textsc{Np}, let \(M
= \tuple{\x, \mathcal{L}, \ell_1, \ell_n, \Delta}\) be a flat CM,
where $\mathcal{L} = \set{\ell_1, \ldots, \ell_n}$. First, we reduce
the control flow graph $\tuple{\mathcal{L}, \Delta}$ of $M$ to a dag
and several self-loops, by replacing each non-trivial cycle:
\[\ell_{i_0} \arrow{\phi_0}{} \ell_{i_1}
\arrow{\phi_1}{} \ell_{i_2} ~\ldots~ \ell_{i_{k-2}}
\arrow{\phi_{k-2}}{} \ell_{i_{k-1}} \arrow{\phi_{k-1}}{}
\ell_{i_0}\] where $k > 1$, with the following sequence:
\begin{equation}\label{eq:self-loops}
  \stackrel{\stackrel{\lambda_0(\x,\x')}{\lcirclearrowdown}}{\ell_{i_0}} 
  \arrow{\phi_0}{}
  \stackrel{\stackrel{\lambda_1(\x,\x')}{\lcirclearrowdown}}{\ell_{i_1}} 
  ~\ldots~ 
  \stackrel{\stackrel{\lambda_{k-1}(\x,\x')}{\lcirclearrowdown}}{\ell_{i_{k-1}}}
  \arrow{\phi_{k-1}}{}
  \ell^\prime_{i_0} \arrow{\lambda_0(\x,\x')}{}  \ldots 
  \arrow{\phi_{k-1}}{} \ell^\prime_{i_{k-1}}
\end{equation}
where $\lambda_{j}(\x,\x') = \exists \x_1,\ldots,\x_{j+1} ~.~
\phi_{j}(\x,\x_{j+1}) \wedge \ldots \wedge \phi_{k-1}(\x_{k-1},\x_k)
\wedge \phi_0(\x_k,\x_1) \wedge \ldots \wedge
\phi_{j-1}(\x_{j-1},\x')$, $\ell^\prime_{i_1}, \ldots,
\ell^\prime_{i_{k-1}}$ are fresh control locations not in
$\mathcal{L}$, and for each rule $\ell_{i_j} \arrow{\phi}{} \ell_m$ of
$M$, where $m \neq i_{(j+1)\!\!\mod k}$, we add a rule
$\ell_{i_j}^\prime \arrow{\phi}{} \ell_m$, for each
$j=0,\ldots,k-1$. This step doubles at most the number of control
locations in $\mathcal{L}$. W.l.o.g., we can consider henceforth that
each control location $\ell_i$ belongs to at most one self loop
labeled by a formula $\phi_i$, for $i = 1,\ldots,2n$.

For each cycle $\ell_i \arrow{\phi_i}{} \ell_i$, where $\phi_i$
defines a relation $R_i \in \oct_\x$, for each $i=1,\ldots,2n$, the
nondeterministic algorithm performs the steps of Algorithm
\ref{alg:ndtm}, namely: \begin{compactenum}
\item Guess values $b_i\geq0$ and $c_i>0$, of the order of
  $2^{\mathcal{O}(\bin{R_i}^d)}$, compute the powers $R_i^{b_i}$,
  $R_i^{c_i}$ and $R_i^{b_i+c_i}$ and find $\Lambda_i$ such that
  $\sigma(R_i^{b_i+c_i}) = \sigma(R_i^{b_i}) + \Lambda_i$. By the
  hypotheses (\ref{it1:flat-cm-np}) and (\ref{it2:flat-cm-np}) this
  computation is possible in $\textsc{Ptime}(\bin{R_i})$, using
  exponentiation by squaring.
\item Check the validity of the condition
  $\textsc{Ind}(\sigma(R_i^{b_i}),\sigma(R_i^{c_i}),\Lambda_i)$ in
  $\textsc{Nptime}(\bin{R_i})$, which is possible by Lemma
  \ref{lemma:ind-np}. 
\item Build a QFPA formula $\phi_i(k,\x,\x') = \phi_i^{<b_i}(\x,\x')
  \vee \phi_i^{\geq b_i}(k,\x,\x')$ in $\textsc{Ptime}(\bin{R_i})$.
  \end{compactenum}
  
The second step uses a breadth-first dag traversal to label each
control location in $\ell_i$, for $i=1,\ldots,2n$, with a QFPA formula
$\theta_i(\vec{x},\vec{x'})$ that captures the summary (effect) of the
set of executions of $M$ from the initial state $\ell_1$ to
$\ell_i$. We assume w.l.o.g. that \begin{inparaenum}[(i)]
\item for every location $\ell_i$, $i=2,\ldots,2n$, there exists a
  path in $M$ from $\ell_1$ to $\ell_i$, and
\item there is no self-loop involving $\ell_1$ in $M$.
\end{inparaenum}
We consider the sets of variables $\vec{k}=\set{k_1, \ldots, k_{2n}}$
and $\x^t_i = \set{x^t_{i,\ell} \mid \ell = 1,\ldots,N}$, where
$i=1,\ldots,2n$ and $t\in\set{\mathrm{in}, \mathrm{out}}$. We define
$\theta_1 = \true$, and for all $j = 2,\ldots,2n$:
\[\theta_j(\x^{\mathrm{in}}_{1,\ldots,2n},\x^{\mathrm{out}}_{1,\ldots,2n}) = 
\bigvee_{\ell_i \stackrel{R_{ij}}{\Rightarrow} \ell_j}
\theta_i(\x^{\mathrm{in}}_{1,\ldots,2n},\x^{\mathrm{out}}_{1,\ldots,2n})
\wedge \phi_{ij}(\x^{\mathrm{out}}_i,\x^{\mathrm{in}}_j) \wedge
\phi_j(k_j,\x^{\mathrm{in}}_j,\x^{\mathrm{out}}_j)\] where
$\phi_{ij}$ is the formula defining $R_{ij}$. It is not difficult to
prove that, for all $i=1,\ldots,2n$ and $\nu,\nu'\in\zed^\x$:
\((\nu,\nu') \models \theta_i \iff \mbox{$M$ has a run
  $(\ell_1,\nu), \ldots, (\ell_i,\nu')$}\enspace.\)

Since for every location $\ell_i$, $i=2,\ldots,2n$, there exists a
control path from $\ell_1$ to it, the breadth-first traversal
guarantees that each predecessor $\ell_i$ of a location $\ell_j$ is
labeled with the summary $\theta_i$ before $\ell_j$ is visited by the
algorithm, ensuring that the definition above is correct. Moreover,
the dag structure (excepting the self-loops) of the CM guarantees that
it is sufficient to visit each locations only once in order to label
it with a summary. Thus, the labeling takes polynomial time and,
consequently, $\bin{\theta_i}$ is polynomially bounded by $\bin{M}$,
for each $i=1,\ldots,2n$. Since the size of the sumary labeling the
final location is polynomial in $\bin{M}$, and the satisfiability
problem is in $\textsc{Nptime}(\bin{M})$, it follows that
$\reachflat(\oct)$ is contained in $\textsc{Np}$. \qed

In the next section we prove the point (\ref{it1:flat-cm-np}) from
Theorem \ref{thm:flat-cm-np}. The rest of the paper is dedicated to
proving point (\ref{it2:flat-cm-np}), which requires a more complex
technical argument.

\subsection{The First Ingredients of the Proof}

In order to apply Theorem \ref{thm:flat-cm-np} we start by proving
that the first assumption from its statement holds for each octagonal
relation $R \in \oct$, namely that the size of the binary
representation of the $n$-th power $R^n$ is bounded by a polynomial
function with arguments $\bin{R}$ and $\log n$. In this case, the
binary size of an exponentially large power $R^n$, where
$n=2^{\mathcal{O}(\bin{R}^d)}$ and $d$ is a constant, is bounded by a
polynomial in $\bin{R}$. Moreover, such powers can be computed in a
polynomial number of steps, using exponentiation by squaring. This is
essential in proving that each branch of the nondeterministic
Algorithm \ref{alg:ndtm} terminates in polynomial time, and also in
the generalization of this reasoning to arbitrary flat CM with cycles
labeled by octagonal constraints (Theorem \ref{thm:flat-cm-np}).

As in most proofs in the rest of this paper, it is useful to prove the
statement first for the simpler class of difference bounds relations,
and use Theorem \ref{thm:bhz} (or a variant thereof) to relate
difference bounds with octagonal relations. Since difference bounds
relations can be represented by weighted graphs (see
Fig. \ref{fig:dbm:ex} (b) for an example), we use a weighted graph to
represent the $n$-th power of a relation $R \in \dbm$. 

We write $\mathcal{G}_R$ for the weighted graph
$\mathcal{G}_{\sigma(R)}$, in which each vertex $1 \leq i \leq N$ is
replaced by the variable $x_i$, and each vertex $N < i \leq 2N$ is
replaced by $x'_i$. For a matrix $M \in \zed_\infty^{2N\times 2N}$, we
denote its top-left, top-right, bottom-left and bottom-rigth $N\times
N$ corners as $\topleft{M}, \topright{M},\botleft{M}$, and
$\botright{M}$, respectively (see Fig. \ref{fig:dbm:ex} (a) for an
example).

\begin{definition} \label{def:unfolding}
  Let $R \in \mathrm{DB}_\x$ be a relation and $n\in\nat_+$ a
  constant. Let \(\mathcal{G}_R^n = \tuple{\bigcup_{k=0}^n \xk{k},
    \arrow{}{}, w}\) be a weighted graph, where
  $\xk{k}=\set{\xxki{k}{i} \mid 1 \leq i \leq N}$ and for all
  $k\in[n]$: \begin{compactitem}
  \item $\xxki{k}{i} \arrow{c}{} \xxki{k}{j}$ if and only if $x_i
    \arrow{c}{} x_j$ is an edge of $\mathcal{G}_R$, 
  \item $\xxki{k}{i} \arrow{c}{} \xxki{k+1}{j}$ if and only if $x_i
    \arrow{c}{} x'_j$ is an edge of $\mathcal{G}_R$, 
  \item $\xxki{k+1}{i} \arrow{c}{} \xxki{k}{j}$ if and only if $x'_i
    \arrow{c}{} x_j$ is an edge of $\mathcal{G}_R$,
  \item $\xxki{k+1}{i} \arrow{c}{} \xxki{k+1}{j}$ if and only if $x'_i
    \arrow{c}{} x'_j$ is an edge of $\mathcal{G}_R$.
  \end{compactitem}
\end{definition}
The constraint graph $\mathcal{G}_R^n$ is said to be an {\em
  unfolding} of the constraint graph $\mathcal{G}_R$. The key
observation relating the power $R^n$ of $R$ and the unfolding graph
$\mathcal{G}_R^n$ is the following: each difference constraint
defining $R^n$ is given by a minimal path between the extremal
vertices (from the set $\xk{0}\cup\xk{n}$) in
$\mathcal{G}_R^n$. Formally, for all $i,j \in \set{1,\ldots,N}$, the
power $R^n$ is defined by the conjunction of the following
constraints:
\begin{equation}\label{eq:dbm-min-paths}
\begin{array}{rcl}
x_i - x_j & \leq & \left(\topleft{\sigma(R^n)}\right)_{ij} = \minw_{\mathcal{G}_R^n}(\xxki{0}{i},\xxki{0}{j}) \\
x_i - x_j' & \leq & \left(\topright{\sigma(R^n)}\right)_{ij} = \minw_{\mathcal{G}_R^n}(\xxki{0}{i},\xxki{n}{j}) \\
x'_i - x_j & \leq & \left(\botleft{\sigma(R^n)}\right)_{ij} = \minw_{\mathcal{G}_R^n}(\xxki{n}{i},\xxki{0}{j}) \\
x'_i - x'_j & \leq & \left(\botright{\sigma(R^n)}\right)_{ij} = \minw_{\mathcal{G}_R^n}(\xxki{n}{i},\xxki{n}{j}) 
\end{array}
\end{equation}
where $\minw_{\mathcal{G}_R^n}(\xxki{p}{i},\xxki{q}{j}) =
\min_{\ell\in\nat}\set{\minw_{\mathcal{G}_R^n}(\xxki{p}{i},\xxki{q}{j},\ell)}$,
and $\minw_{\mathcal{G}_R^n}(\xxki{p}{i},\xxki{q}{j},\ell)$ is the
minimal weight among all paths of length $\ell$ between $\xxki{p}{i}$
and $\xxki{q}{j}$ in $\mathcal{G}_R^n$, or $\infty$, if no such path
exists. When the length is not important, we denote a path between
$\xxki{p}{i}$ and $\xxki{q}{j}$ as $\xxki{p}{i} \arrow{*}{}
\xxki{q}{j}$.

\newcommand{\symbolGone}[0]{
      \scalebox{0.7}{\begin{tikzpicture}
        \TermGridGenNoCapBoxed{0.0}{0.7}{2}{0.0}{0.4}{4}{0.4}{0.4}
        \foreach \ii in {1,...,1} {
          \pgfmathtruncatemacro\jj{\ii+1}
          \TermGridEdgeC{\ii}{1}{\jj}{3}{\tiny$0$}{above}
          \TermGridEdgeC{\jj}{4}{\ii}{4}{\tiny$0$}{above}
        }
      \end{tikzpicture}}
}
\newcommand{\symbolGtwo}[0]{
      \scalebox{0.7}{\begin{tikzpicture}
        \TermGridGenNoCapBoxed{0.0}{0.7}{2}{0.0}{0.4}{4}{0.4}{0.4}
        \foreach \ii in {1,...,1} {
          \pgfmathtruncatemacro\jj{\ii+1}
          \TermGridEdgeC{\ii}{3}{\jj}{2}{\tiny$0$}{above}
          \TermGridEdgeC{\jj}{4}{\ii}{4}{\tiny$0$}{above}
        }
      \end{tikzpicture}}
}
\newcommand{\symbolGthree}[0]{
      \scalebox{0.7}{\begin{tikzpicture}
        \TermGridGenNoCapBoxed{0.0}{0.7}{2}{0.0}{0.4}{4}{0.4}{0.4}
        \foreach \ii in {1,...,1} {
          \pgfmathtruncatemacro\jj{\ii+1}
          \TermGridEdgeC{\ii}{2}{\jj}{1}{\tiny$-\!1$}{above}
          \TermGridEdgeC{\jj}{4}{\ii}{4}{\tiny$0$}{above}
        }
      \end{tikzpicture}}
}
\newcommand{\symbolGfour}[0]{
      \scalebox{0.7}{\begin{tikzpicture}
        \TermGridGenNoCapBoxed{0.0}{0.7}{2}{0.0}{0.4}{4}{0.4}{0.4}
        \foreach \ii in {1,...,1} {
          \pgfmathtruncatemacro\jj{\ii+1}
          \TermGridEdgeC{\ii}{1}{\jj}{3}{\tiny$0$}{above}
          \TermGridEdgeC{\jj}{3}{\ii}{4}{\tiny$0$}{above}
        }
      \end{tikzpicture}}
}
\newcommand{\symbolGeps}[0]{
      \scalebox{0.7}{\begin{tikzpicture}
        \TermGridGenNoCapBoxed{0.0}{0.7}{2}{0.0}{0.4}{4}{0.4}{0.4}
        \foreach \ii in {1,...,1} {
          \pgfmathtruncatemacro\jj{\ii+1}
        }
      \end{tikzpicture}}
}
\newcommand{\symbolGfive}[0]{
      \scalebox{0.7}{\begin{tikzpicture}
        \TermGridGenNoCapBoxed{0.0}{0.7}{2}{0.0}{0.4}{4}{0.4}{0.4}
        \foreach \ii in {1,...,1} {
          \pgfmathtruncatemacro\jj{\ii+1}
          \TermGridEdgeC{\ii}{3}{\jj}{2}{\tiny$0$}{above}
          \TermGridEdgeC{\jj}{3}{\ii}{4}{\tiny$0$}{above}
        }
      \end{tikzpicture}}
}
\newcommand{\symbolGsix}[0]{
      \scalebox{0.7}{\begin{tikzpicture}
        \TermGridGenNoCapBoxed{0.0}{0.7}{2}{0.0}{0.4}{4}{0.4}{0.4}
        \foreach \ii in {1,...,1} {
          \pgfmathtruncatemacro\jj{\ii+1}
          \TermGridEdgeC{\ii}{2}{\jj}{1}{\tiny$-\!1$}{above}
          \TermGridEdgeC{\jj}{3}{\ii}{4}{\tiny$0$}{above}
        }
      \end{tikzpicture}}
}

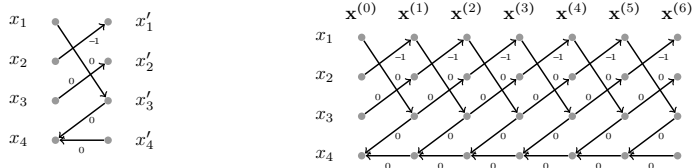
\begin{figure}[htb]
\centering{
\begin{tabular}{c}
 \begin{tabular}{cc}
  \mbox{\begin{minipage}{2.5cm}
      \scalebox{0.7}{\begin{tikzpicture}
        \TermGridGenDifferentCap{0.0}{1.0}{2}{0.0}{0.75}{4}{0.7}{0.5}{0}
        \foreach \ii in {1,...,1} {
          \pgfmathtruncatemacro\jj{\ii+1}
          \TermGridEdgeC{\ii}{2}{\jj}{1}{\tiny$-\!1$}{right}
          \TermGridEdgeC{\ii}{3}{\jj}{2}{\tiny$0$}{left}
          \TermGridEdgeC{\ii}{1}{\jj}{3}{\tiny$0$}{right}
          \TermGridEdgeC{\jj}{4}{\ii}{4}{\tiny$0$}{below}
          \TermGridEdgeC{\jj}{3}{\ii}{4}{\tiny$0$}{right}
        }
      \end{tikzpicture}}
  \end{minipage}} & 
  \mbox{\begin{minipage}{8cm}
      \scalebox{0.7}{\begin{tikzpicture}
        \TermGridGen{0.0}{1.0}{7}{0.0}{0.75}{4}{0.7}{0.5}{0}
        \foreach \ii in {1,...,6} {
          \pgfmathtruncatemacro\jj{\ii+1}
          \TermGridEdgeC{\ii}{2}{\jj}{1}{\tiny$-\!1$}{right}
          \TermGridEdgeC{\ii}{3}{\jj}{2}{\tiny$0$}{left}
          \TermGridEdgeC{\ii}{1}{\jj}{3}{\tiny$0$}{right}
          \TermGridEdgeC{\jj}{4}{\ii}{4}{\tiny$0$}{below}
          \TermGridEdgeC{\jj}{3}{\ii}{4}{\tiny$0$}{right}
        }
      \end{tikzpicture}}
  \end{minipage}}
  \\
  (a) $\mathcal{G}_R$ -- the constraint graph of $R$ & 
  (b) $\mathcal{G}_R^6$ -- the $6$-times unfolding of $\mathcal{G}_R$ 
  \end{tabular}
\end{tabular}}
\caption{Constraint graphs for the DB relation $R \equiv x_2\mi
  x'_1\leq -1 \wedge x_3\mi x'_2\leq 0 \wedge x_1\mi x'_3\leq 0 \wedge
  x'_4\mi x_4\leq 0 \wedge x'_3\mi x_4\leq 0$. }\label{fig:unfolded}
\end{figure}

\begin{example}\label{ex:unfold-dbm}
Consider the difference bounds relation $R$ defined by the formula
$\phi \equiv x_2\mi x'_1\leq -1 \wedge x_3\mi x'_2\leq 0 \wedge x_1\mi
x'_3\leq 0 \wedge x'_4\mi x_4\leq 0 \wedge x'_3\mi x_4\leq
0$. Fig.~\ref{fig:unfolded} (a) shows the constraint graph
$\mathcal{G}_R$ and Fig.~\ref{fig:unfolded} (b) depicts the unfolding
$\mathcal{G}^6_R$ of $\mathcal{G}_R$. \hfill$\blacksquare$
\end{example}

We are now ready to prove that the first condition of Theorem
\ref{thm:flat-cm-np} holds for any difference bounds relation. 

\begin{lemma}\label{lemma:dbm-poly-log}
  There exists a constant $d>0$ such that, for every relation $R \in
  \dbm$, we have $\bin{R^n} = \mathcal{O}((\bin{R} \cdot \log n)^d)$,
  for all $n > 0$.
\end{lemma}
\proof{We assume w.l.o.g. that $R^n$ is consistent, otherwise $R^m =
  \emptyset$ for all $m\geq n$ and $\bin{R^m}=\bin{R^n}$. Then the
  unfolding $\mathcal{G}_R^n$ does not contain cycles of negative
  weight, thus any minimal path in $\mathcal{G}_R^n$ does not contain
  a cycle. Since $\mathcal{G}_R^n$ has $(n+1)\cdot N$ nodes, any
  minimal path has weight at most $\mu \cdot (n+1) \cdot N$, where
  $\mu$ is the maximal value among the labels of
  $\mathcal{G}_R$. Since there are at most $4N^2$ such paths in the
  definition of $R^n$ (\ref{eq:dbm-min-paths}), we compute:
  \[\begin{array}{rclcl}
  \bin{R^n} & \leq & 4N^2 \log(\mu \cdot (n+1) \cdot N) 
  & \leq & 4\bin{R}^2 \log(\bin{R} \cdot (n+1) \cdot \bin{R}) \\
  & \leq & 4\bin{R}^2 (2\log\bin{R} + \log n + 1) 
  & \leq & 16\bin{R}^3 \cdot \log n \leq 16(\bin{R}\cdot\log n)^3\enspace.\enspace\text{ \qed}
  \end{array}\]
}

In order to establish a similar result for the more general class
$\oct$, we must first relate the powers of an octagonal relation $R
\in \oct$, defined by a constraint $\phi(\x,\x')$ with the powers of
the difference bounds relation $\overline{R} \in \dbm$, defined by the
constraint $\overline{\phi}(\y,\y')$, obtained from $\phi$ by doubling
the number of variables. The following lemma establishes the needed
correspondence between $R^n$ and $\overline{R}^n$. We recall that a
relation is said to be $*$-consistent if each of its powers is
consistent.

\begin{lemma}\label{lemma:oct-dbm-powers}
  For any $*$-consistent relation $R \in \oct$, the following holds,
  for any $n\geq0$:
  \[\sigma(R^n)_{ij} = \min\left(\sigma({\overline{R}}^n)_{ij},  
  \left\lfloor \frac{\sigma({\overline{R}}^n)_{i\bar{\imath}}}{2} \right\rfloor + 
  \left\lfloor \frac{\sigma({\overline{R}}^n)_{\bar{\jmath}j}}{2} \right\rfloor\right)\enspace.\]
\end{lemma}
\proof{\cite[Lemma 4.30]{TerminationLmcs}. \qed} 

The following lemma proves the validity of the first condition of
Theorem \ref{thm:flat-cm-np}, for every octagonal relation. 

\begin{lemma}\label{lemma:oct-poly-log}
  There exists a constant $d>0$ such that, for every relation $R \in
  \oct$, we have $\bin{R^n} = \mathcal{O}((\bin{R} \cdot \log n)^d)$,
  for all $n > 0$.
\end{lemma}
\proof{ We assume w.l.o.g. that $R$ is $*$-consistent, otherwise there
  exists $n > 0$ such that $R^m = \emptyset$, thus $\bin{R^m} =
  \bin{R^n}$, for all $m \geq n$. By Lemma \ref{lemma:oct-dbm-powers},
  we infer, for any $n>0$:
\[\bin{R^n} = \sum_{i,j=1}^{4N} \log\left(\sigma(R^n)\right)_{ij} \leq 
  \sum_{i,j=1}^{4N} \log\left(\sigma(\overline{R}^n)\right)_{ij} =
  \bin{\overline{R}^n}
  \enspace.\] By Lemma \ref{lemma:dbm-poly-log}, we have
  $\bin{\overline{R}^n}= \mathcal{O}((\bin{\overline{R}} \log n)^d)$
  for a constant $d>0$, not depending on $R$. Observe that $\bin{R}
  \leq 2\bin{\overline{R}}$, since every constraint of $R$ is encoded
  by two constraints of $\overline{R}$. We obtain thus $\bin{R^n} =
  \mathcal{O}((\bin{R} \log n)^d)$, for any $n>0$.  \qed}

\section{The Periodicity of Octagonal Relations}

In this section we prove that the class of octagonal relations is
periodic, that is, for each relation $R \in \oct$, the sequence of
matrices $\{\sigma(R^k)\}_{k=0}^\infty$ is periodic. Moreover, we
show that the prefix and the period of this sequence are of the order
of $2^{\mathcal{O}(\bin{R}^d)}$, for a constant $d>0$ that does not depend on the
choice of $R$. By Theorem \ref{thm:flat-cm-np}, a consequence is that
the class of problems $\reachflat(\oct)$ is \textsc{Np}-complete.

The core of the proof is showing periodicity of difference bounds
relations and establishing the upper bounds for the prefix and period
of sequence $\{\sigma(R^k)\}_{k=0}^\infty$, where $R \in \dbm$. The
main idea is that the coefficients of any matrix $\sigma(R^k)$ can be
derived from the $k$-th power of a larger matrix $\mathcal{M}_R$,
where the matrix product is defined using $\min$ as addition and $+$
as multiplication. Intuitively, a sequence
$\{\mathcal{M}^k_R\}_{k=0}^\infty$ gives the minimal weights of the
paths of length $k=0,1,\ldots$ in a weighted graph $\mathcal{A}_R$,
whose incidence matrix is $\mathcal{M}_R$. To obtain the simply
exponential bounds on the period and prefix of a sequence
$\{\sigma(R^k)\}_{k=0}^\infty$, we develop this periodicity result
further, by exploiting the structure of the strongly connected
components of $\mathcal{A}_R$.

In a nutshell, the set of paths from an unfolding $\mathcal{G}_R^k$ of
the constraint graph $\mathcal{G}_R$ that represents the relation $R
\in \dbm$ (Definition \ref{def:unfolding}) is the language, consisting
of words of length $k$, recognized by a weighted automaton
$\mathcal{A}_R$ (called \emph{zigzag automaton} in the following). We
use an idea of Comon and Jurski \cite{ComonJurski98} that show that
the set of minimal weights of these paths can be captured by a subset
of paths, in which only a bounded number of direction changes may
occur. Based on this fact, we define $\mathcal{A}_R$ to recognize only
these simple paths from $\mathcal{G}^k_R$, by considering a
\emph{saturated} relation $R_\mathrm{sat}$, with the same periodic
behavior as $R$. The simply exponential upper bound on the period of
the sequence $\{\sigma(R^k)\}_{k=0}^k$ follows by a proof of the fact
that, in each strongly connected component of $\mathcal{A}_R$, there
is an elementary cycle of minimal weight/length ratio, whose length is
of the order of $2^{\mathcal{O}(N)}$, where $N$ is the number of
variables from $R$.

\subsection{Saturation of Difference Bounds Relations}

We start by proving that the periodic behavior of a sequence of powers
$\{R^k\}_{k=0}^\infty$ of a relation $R \in \dbm$ can be analyzed by
considering a \emph{saturated} version of $R$, denoted as
$R_{\mathrm{sat}}$. First we show that such a relation exists for each
$R \in \dbm$ and that all powers of $R$, beyond a certain threshold,
can be computed by a function taking as arguments powers of
$R_{\mathrm{sat}}$ instead. As a consequence, $R$ is periodic if
$R_{\mathrm{sat}}$ is periodic and the period of $R$ is bounded by the
period of $R_{\mathrm{sat}}$.  The salient property of a saturated
difference bounds relation is that, every power $R^k_{\mathrm{sat}}$
is defined by the weights of the minimal paths in the unfolding
$\mathcal{G}^k_R$ of the constraint graph defining $R$, with a bounded
number of direction changes. This detail is instrumental in providing
an accurate upper bound on the period of difference bounds relations.

Let $\x=\set{x_1,\ldots,x_N}$ be a set of variables. We recall first
the notion of \emph{folded graph} introduced by Comon and Jurski
\cite{ComonJurski98}. Given a relation $R \in \dbm_\x$, we consider
the weighted graph \(\mathcal{G}^f_R = \tuple{\x, \rightarrow_f,
  w_f}\) which has an edge $x_i \arrow{\alpha}{}_f x_j$ for each edge
$x_i \arrow{\alpha}{} x_j$, $x_i \arrow{\alpha}{} x'_j$, $x'_i
\arrow{\alpha}{} x_j$, or $x'_i \arrow{\alpha}{} x'_j$ in
$\mathcal{G}_R$. In other words, the folded graph $\mathcal{G}^f_R$ is
obtained from the weighted graph $\mathcal{G}_R$ by merging all
vertices $x_i$ and $x'_i$, respectively.

\begin{figure}[htb]
  \centering{
    \begin{tabular}{ccccc}
   \mbox{\begin{minipage}{2.5cm}
      \scalebox{0.7}{\begin{tikzpicture}
        \TermGridGenDifferentCap{0.0}{1.0}{2}{0.0}{0.75}{4}{0.7}{0.5}{0}
        \foreach \ii in {1,...,1} {
          \pgfmathtruncatemacro\jj{\ii+1}
          \TermGridEdgeC{\ii}{2}{\jj}{1}{\tiny$-\!1$}{right}
          \TermGridEdgeC{\ii}{3}{\jj}{2}{\tiny$0$}{left}
          \TermGridEdgeC{\ii}{1}{\jj}{3}{\tiny$0$}{right}
          \TermGridEdgeC{\jj}{4}{\ii}{4}{\tiny$0$}{below}
          \TermGridEdgeC{\jj}{3}{\ii}{4}{\tiny$0$}{right}
        }
      \end{tikzpicture}}
  \end{minipage}}
   & \hspace*{2mm} & 
  \mbox{\begin{minipage}{1.5cm}
      \scalebox{1.0}{\begin{tikzpicture}
          \scriptsize
    \tikzset{
      sState/.style={draw=black,circle,inner sep=1.5pt,semithick}
    }
    \node[sState] (x4) at (0mm,0mm) {$x_4$};
    \node[sState] (x3) at (0mm,8mm) {$x_3$};
    \node[sState] (x2) at (0mm,16mm) {$x_2$};
    \node[sState] (x1) at (0mm,24mm) {$x_1$};
    \path[->] 
       (x1) edge [bend left,bend angle=5] node[right]{0} (x3)
       (x3) edge [bend left,bend angle=5] node[left]{0} (x4) 
            edge [bend left,bend angle=5] node[left]{0} (x2)
       (x2) edge [bend left,bend angle=5] node[left]{-1} (x1)
       (x4) edge [loop right,looseness=7] node[right]{0} (x4);
        \end{tikzpicture}
      }
  \end{minipage}}
  && 
  \mbox{\begin{minipage}{8cm}
      \scalebox{1.0}{\begin{tikzpicture}
          \scriptsize
          \TermGridGen{0.0}{0.8}{6}{0.0}{0.4}{4}{0.7}{0.5}{0}
          \TermGridEdgeC{1}{2}{2}{1}{$-\!1$}{above}
          \TermGridEdgeC{2}{1}{3}{3}{$0$}{above}
          \TermGridEdgeC{3}{3}{4}{2}{$0$}{above}
          \TermGridEdgeC{4}{2}{5}{1}{$-\!1$}{above}
          \TermGridEdgeC{5}{1}{6}{3}{$0$}{above}
          \TermGridEdgeC{6}{3}{5}{4}{$0$}{above}
          \TermGridEdgeC{5}{4}{4}{4}{$0$}{above}
          \TermGridEdgeC{4}{4}{3}{4}{$0$}{above}
          \TermGridEdgeC{3}{4}{2}{4}{$0$}{above}
          \TermGridEdgeC{2}{4}{1}{4}{$0$}{above}
      \end{tikzpicture}}
      \end{minipage}
  }
\\\\
(a) && (b) && (c)
\end{tabular}}
\caption{(a) Constraint graph $\mathcal{G}_R$, (b) folded graph
  $\mathcal{G}_R^f$ and (c) zigzag paths in the $\mathcal{G}^5_R$
  unfolding of the difference bounds relation $R$, defined by $x_2\mi
  x'_1\leq -1 \wedge~ x_3\mi x'_2\leq 0 \wedge x_1\mi x'_3\leq 0
  \wedge x'_4\mi x_4\leq 0 \wedge x'_3\mi x_4\leq 0$. }
\label{fig:folded-graph}
\end{figure}
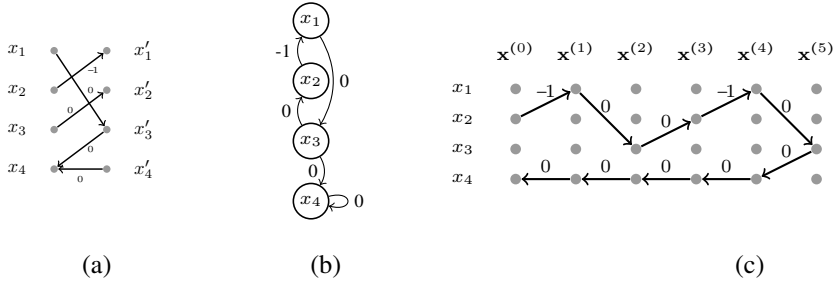

\begin{example}\label{ex:folded-graph}
For example, Fig.~\ref{fig:folded-graph} (b) shows the folded graph
for the relation defined by the formula $x_2\mi x'_1\leq -1 \wedge~
x_3\mi x'_2\leq 0 \wedge x_1\mi x'_3\leq 0 \wedge x'_4\mi x_4\leq 0
\wedge x'_3\mi x_4\leq 0$, whose constraint graph is given in
Fig.~\ref{fig:folded-graph} (a). \hfill$\blacksquare$
\end{example}

The folded graph induces the following equivalence relation on indices
of variables: $i \sim_R j$ iff $x_i$ and $x_j$ belong to the same
strongly connected component of $\mathcal{G}^f_R$. For example, the
equivalence classes of the $\sim_R$ relation, induced by the folded
graph in Fig.~\ref{fig:folded-graph} (b), are $\set{1,2,3}$ and
$\set{4}$.

\begin{figure}[h!]
\begin{center}
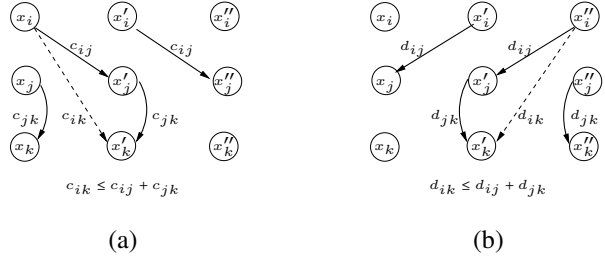
\end{center}
\vspace*{-5mm}
\caption{Corner inequalities}\label{fig:corner-inequalities}
\end{figure}

The following \emph{corner inequalities} are generalized triangle
inequalities that occur in an unfolding of size two of the constraint
graph $\mathcal{G}_R$ of a relation $R \in \dbm$ (see Fig.
\ref{fig:corner-inequalities}). 

\begin{definition}\label{def:saturation}
  A relation $R \in \dbm_\x$ is \emph{saturated} if, for all $1
  \leq i,j,k \leq N$, such that $i \sim_R j \sim_R k$, the following hold:
  \[\begin{array}{rcl}
  \left(\topright{\sigma(R)}\right)_{ik} & \leq & 
  \left(\topright{\sigma(R)}\right)_{ij}+\left(\topleft{\sigma(R)}\right)_{jk} 
  \\
  \left(\botleft{\sigma(R)}\right)_{ik} & \leq & 
  \left(\botleft{\sigma(R)}\right)_{ij} + \left(\botright{\sigma(R)}\right)_{jk}
  \end{array}\]
\end{definition}
The first inequality is depicted in Fig. \ref{fig:corner-inequalities}
(a) and the second case in Fig. \ref{fig:corner-inequalities} (b).
The importance of the above definition lies in the fact that the
powers $R^n$ of a saturated relation $R \in \dbm$ can be defined using
only a subset of the minimal paths between the (extremal) vertices in
the unfolding $\mathcal{G}_R^n$ of the constraint graph of $R$ (see
the constraints (\ref{eq:dbm-min-paths}) for a definition of $R^n$
using the minimal paths of $\mathcal{G}_R^n$). Essentially, these are
the minimal paths that, moreover, do not change direction while
traversing variables from the same equivalence class of the $\sim_R$
relation.

\begin{definition}\label{def:saturated-paths}
  Let $\mathcal{G}_R^n$ be the $n$-th unfolding of the constraint
  graph $\mathcal{G}_R$ of a relation $R \in \dbm$. An edge
  $\xxki{p}{i} \arrow{}{} \xxki{q}{j}$ of $\mathcal{G}_G^n$ is
  \emph{forward} (\emph{backward}) if $q=p+1$ ($q=p-1$). A path is
  \emph{forward} (\emph{backward}) if it consists only of forward
  (backward) edges. A path $\pi$ is \emph{saturated} if for each
  forward (backward) subpath $\rho : \xxki{p}{i} \arrow{*}{}
  \xxki{q}{j}$ of $\pi$, of length $\len{\rho} > 1$, we have $i \sim_R
  j$.
\end{definition}
We show that the $n$-th power of a saturated relation can be defined
by considering only the saturated minimal paths in the $n$-th
unfolding of its constraint graph. For instance, the path in
Fig. \ref{fig:folded-graph} (c) is saturated.

\begin{lemma}\label{lemma:sat-min-paths}
  Given a saturated relation $R \in \dbm$, for each minimal path $\pi
  : \xxki{p}{i} \arrow{*}{} \xxki{q}{j}$ in the unfolding
  $\mathcal{G}^n_R$ of its constraint graph $\mathcal{G}_R$, there
  exists a saturated path $\pi_{\mathrm{sat}} : \xxki{p}{i}
  \arrow{*}{} \xxki{q}{j}$ such that $w(\pi)=w(\pi_{\mathrm{sat}})$.
\end{lemma}
\proof{We define a sequence $\pi_0, \pi_1, \ldots$ of paths such that
  $\pi_0=\pi$ and for each $t \geq 0$, $\pi_{t+1}$ is obtained from
  $\pi_t$ as follows. Because $R$ is saturated: 
  \begin{compactitem}
  \item for every two adjacent edges $\xxki{p}{i} \arrow{\alpha}{}
    \xxki{p+1}{j} \arrow{\beta}{} \xxki{p+1}{\ell}$ on $\pi_t$, where
    $i \sim_R \ell$, there exists an edge $\xxki{p}{i} \arrow{\gamma}{}
    \xxki{p+1}{\ell}$, where $\gamma \leq \alpha+\beta$, and
  \item for every two adjacent edges $\xxki{p+1}{i} \arrow{\alpha}{}
    \xxki{p}{j} \arrow{\beta}{} \xxki{p}{\ell}$, where $i \sim_R \ell$,
    there exists an edge $\xxki{p+1}{i} \arrow{\gamma}{}
    \xxki{p}{\ell}$, where $\gamma\leq\alpha+\beta$.
  \end{compactitem}
  $\pi_{t+1}$ is obtained by replacing all pairs $\xxki{p}{i}
  \arrow{\alpha}{} \xxki{p+1}{j} \arrow{\beta}{} \xxki{p+1}{\ell}$
  with $\xxki{p}{i} \arrow{\gamma}{} \xxki{p+1}{\ell}$ and all pairs
  $\xxki{p+1}{i} \arrow{\alpha}{} \xxki{p}{j} \arrow{\beta}{}
  \xxki{p}{\ell}$, where $i \sim_R \ell$, with the edges $\xxki{p}{i}
  \arrow{\gamma}{} \xxki{p+1}{\ell}$ and $\xxki{p}{i} \arrow{\gamma}{}
  \xxki{p+1}{\ell}$, respectively. Clearly, $\pi_{t+1}$ is a path in
  $\mathcal{G}_R^n$ and, moreover, we have that $\len{\pi_{t+1}} <
  \len{\pi_t}$ and $w(\pi_{t+1}) \leq w(\pi_t)$. The sequence is
  finite, because $\len{\pi_t}$ decreases at each step, and the last
  path in the sequence is saturated. If, moreover, $\pi$ is minimal,
  all paths in the sequence are minimal as well. \qed}

In the rest of this section, we prove that for each difference bounds
relation $R$ it is possible to find a saturated relation
$R_{\mathrm{sat}}$ with the same periodic behavior as $R$. The problem
of periodicity and the evaluation of the upper bounds for the prefix
and period of $R$ is carried out considering $R_{\mathrm{sat}}$
instead. This step is instrumental in proving a simply exponential
upper bound on the period of $R$.

At this point, it is useful to distinguish between relations
that are $*$-consistent ($R^k\neq\emptyset$ for all $k\geq0$) and the
ones that are not. In the latter case, the period of the power
sequence $\{R^k\}_{k=0}^\infty$ is $1$ and the prefix is bounded by
the cut-off result below.

\begin{lemma}\label{lemma:spiral}
  For any relation $R \in \dbm_\x$, the following hold: \begin{compactenum}
  \item\label{it1:spiral} $R$ is $*$-consistent only if for every
    $n\in\nat_+$ and $1 \leq i < j \leq N$, such that $i \sim_R j$,
    for any paths $\pi_i : \xxki{0}{i} \arrow{*}{} \xxki{k}{i}$ and
    $\pi_j : \xxki{k}{j} \arrow{*}{} \xxki{0}{j}$ in $\mathcal{G}_R^n$
    with $0 < k \leq n$, we have $w(\pi_i)+w(\pi_j)\geq0$.
  \item\label{it2:spiral} $R$ is not $*$-consistent only if
    $R^n=\emptyset$, for all $n \geq 6N^7\cdot\mu(\mathcal{G}_R)$.
  \end{compactenum}
\end{lemma}
\proof{In the following, the \emph{travel} of a path $\pi :
  \xxki{n_1}{i_1} \arrow{}{} \ldots \arrow{}{} \xxki{n_k}{i_k}$ in the
  unfolding $\mathcal{G}^n_R$ (Definition \ref{def:unfolding}) of the
  constraint graph $\mathcal{G}_R$ is defined by $\tau(\pi) =
  \max_{j=1}^{k}\set{n_j}-\min_{j=1}^{k-1}\set{n_j}$.

  \noindent
  (\ref{it1:spiral}) By contradiction, suppose that $R$ is
  $*$-consistent and let $\pi_i : \xxki{0}{i} \arrow{*}{} \xxki{k}{i}$
  and $\pi_j : \xxki{k}{j} \arrow{*}{} \xxki{0}{j}$ in
  $\mathcal{G}_R^n$, such that $i \sim j$ and
  $w(\pi_i)+w(\pi_j)<0$. Because $i\sim j$, there exist paths $\zeta :
  \xxki{d}{j} \rightarrow \ldots \rightarrow \xxki{d+p}{i}$ and $\xi :
  \xxki{d}{i} \rightarrow \ldots \rightarrow \xxki{d+q}{j}$ in
  $\mathcal{G}_R^n$, for some $-N \leq p,q \leq N$ and $d\geq
  \max(\abs{p},\abs{q})$. For any $m\in\nat$ we build the following
  path:
  \[\zeta.\pi_i^m.\xi = \xxki{d}{j} \rightarrow \ldots \rightarrow 
  \xxki{d+p}{i} \rightarrow \ldots \rightarrow \xxki{d+p+mk}{i}
  \rightarrow \ldots \rightarrow \xxki{d+p+q+mk}{j}\enspace.\] For
  $m\geq\lceil\frac{-(p+q)}{k}\rceil$, we have $p+q+mk>0$, i.e.\ the
  above path has a positive travel. Let us now repeat this path $k$
  times, and concatenate it with the path \[\pi_j^{p+q+km} :
  \xxki{k(d+p+q+mk)}{j} \rightarrow \ldots \rightarrow
  \xxki{d}{j}\enspace.\] We obtain the cycle
  $(\zeta.\pi_i^m.\xi)^k.\pi_j^{p+q+km}$, starting and ending in
  $\xxki{d}{j}$. Observe that the unfolding $\mathcal{G}_R^n$ has to
  be sufficiently large to accomodate this cycle. Since $n$ can be
  taken arbitrarily large, this is not a restriction. The weight of
  this cycle is:
  \[\begin{array}{rcl}
  k\cdot(w(\zeta)+w(\xi)+m\cdot w(\pi_i))+(p+q+km)\cdot w(\pi_j) & = & \\
  km\cdot(w(\pi_i)+w(\pi_j))+k\cdot(w(\zeta)+w(\xi))+(p+q)\cdot w(\pi_j) \enspace.
  \end{array}\]
  For a sufficiently large $m\geq\lceil\frac{-(p+q)}{k}\rceil$, the
  weight of this cycle is negative, which contradicts with the
  assumption that $R$ is $*$-consistent. It follows that
  $w(\pi_i)+w(\pi_j)\geq0$.

  \vspace*{\baselineskip}\noindent (\ref{it2:spiral}) If $R$ is not
  $*$-consistent, there exists $n\in\nat_+$ such that $R^n=\emptyset$,
  thus there exists a cycle $\gamma$ of negative weight in
  $\mathcal{G}_R^n$. Let $\gamma : \xxki{n_1}{i_1} \rightarrow \ldots
  \rightarrow \xxki{n_{k-1}}{i_{k-1}} \rightarrow \xxki{n_1}{i_1}$ be
  a negative cycle of minimal travel in $\mathcal{G}_R^n$. If
  $\tau(\gamma) \leq N^2$, the cycle is present also in
  $\mathcal{G}_R^{N^2}$. Hence $R^n=\emptyset$, for every $n\geq
  N^2$. We consider thus that $\tau(\gamma) > N^2$. By the pigeonhole
  principle, there exist a pair of variables $x_i,x_j$ such that the
  pairs of vertices $\xxki{k}{i},\xxki{k}{j}$ and
  $\xxki{\ell}{i},\xxki{\ell}{j}$ occur on the cycle, for some $0 \leq
  k < \ell \leq n$. Thus there exist paths $\pi_i : \xxki{k}{i}
  \rightarrow \ldots \rightarrow \xxki{\ell}{i}$ and $\pi_j :
  \xxki{\ell}{j} \rightarrow \ldots \rightarrow \xxki{k}{j}$. Observe
  that we can always chose the pairs $\xxki{k}{i},\xxki{k}{j}$ and
  $\xxki{\ell}{i},\xxki{\ell}{j}$ such that $\ell-k \leq
  N^2$. Clearly, we have $i \sim j$, since $x_i$ and $x_j$ occur on
  the cycle $\gamma$. Suppose now that $w(\pi_i)+w(\pi_j)\geq0$. Since
  $\ell-k>0$, we obtain a cycle of smaller travel, by eliminating
  $\pi_i$ and $\pi_j$ from $\gamma$, and concatenating the path
  $\xxki{k}{j} \rightarrow \ldots \rightarrow \xxki{k}{i}$ with
  $\xxki{\ell-k}{i} \rightarrow \ldots \xxki{\ell-k}{j}$. But then we
  obtain a cycle $\gamma'$ of weight $w(\gamma') \leq w(\gamma) < 0$
  and travel $\tau(\gamma') < \tau(\gamma)$. This would contradict the
  assumption that $\tau(\gamma)$ is the minimal travel of all negative
  weight cycles in $\mathcal{G}_R^n$. Hence it must be the case that
  $w(\pi_i)+w(\pi_j)<0$.

  Given $\pi_i$ and $\pi_j$ such that $i \sim j$ and
  $w(\pi_i)+w(\pi_j)<0$, we apply the construction of point
  (\ref{it1:spiral}) to obtain a cycle
  $(\zeta.\pi_i^m.\xi)^{\ell-k}.\pi_j^{p+q+(\ell-k)m}$ of weight:
  \[m\cdot(\ell-k)\cdot(w(\pi_i)+w(\pi_j))+(\ell-k)\cdot(w(\zeta)+w(\xi))+(p+q)\cdot w(\pi_j)\]
  where $\zeta$, $\xi$, $p$ and $q$ are the ones from point
  (\ref{it1:spiral}). To obtain a negative cycle, it is thus
  sufficient to chose $m = (\ell-k)\cdot(w(\zeta)+w(\xi))+(p+q)\cdot
  w(\pi_j)$. Since $\zeta$ and $\pi$ are elementary paths, we have
  $w(\zeta) + w(\xi) \leq 2N\cdot\mu(\mathcal{G}_R)$ and $p+q\leq
  2N$. Moreover, since $\ell-k\leq N^2$, we have $w(\pi_j) \leq N^2
  \cdot \mu(\mathcal{G}_R)$. Then it is sufficient to take $m =
  4N^3\cdot\mu(\mathcal{G}_R)$. The travel of the cycle thus
  constructed is at most: \[\begin{array}{rcl}
  (\ell-k)\cdot(p+q+(\ell-k)\cdot m) & \leq & N^2 \cdot (2N + N^2
  \cdot 4N^3 \cdot \mu(\mathcal{G}_R)) \\ & \leq & 4N^7 \cdot
  \mu(\mathcal{G}_R) + 2N^3 \leq 6N^7 \cdot
  \mu(\mathcal{G}_R) \enspace.
  \end{array}\] 
  The last inequality is obtained from the following observation:
  since $w(\pi_i)+w(\pi_j)<0$, there must exist at least an edge of
  non-zero weight in $\mathcal{G}_R$, hence
  $\mu(\mathcal{G}_R)>0$.
\qed}

In the light of the previous lemma, we consider, from now on, that $R
\in \dbm_\x$ is a $*$-consistent relation. We are now ready to define
a saturated relation $R_{\mathrm{sat}}$, which is periodic if and only
if $R$ is periodic, in which case the prefix and the period of
$R_{\mathrm{sat}}$ bound the prefix and the period of $R$,
respectively. 

Let $\phi(\x,\x')$ be any difference bounds constraint (Definition
\ref{dbc}) that defines $R$ and $\widetilde{\phi}$ be the conjunction
of all atomic propositions from $\mathit{Atom}(\phi)$ involving
$\sim_R$-equivalent variables. We define the following sequence of
formulae:
\[\begin{array}{rclr}
\psi_0(\yy,\x,\x',\zz) & \equiv & \widetilde{\phi}(\yy,\x) \wedge
\widetilde{\phi}(\x,\x') \wedge \widetilde{\phi}(\x',\zz)
\\ \psi_{n+1}(\yy,\x,\x',\zz) & \equiv & \exists\yy'\exists\zz' ~.~
\widetilde{\phi}(\yy,\yy') \wedge \psi_n(\yy',\x,\x',\zz') \wedge
\widetilde{\phi}(\zz',\zz), &\mbox{for all $n\in\nat$}
\end{array}\]
where $\yy=\set{y_1,\ldots,y_N}$, $\zz=\set{z_1,\ldots,z_N}$. The
following lemma shows that the sequence $\{\exists \yy \exists \zz ~.~
\psi_n\}_{n=0}^\infty$ converges in at most $N^2$ steps and its limit
defines a saturated difference bounds relation.

\begin{lemma}\label{lemma:sat-lim}
For every $n \geq N^2$, we have
\(\exists\yy\exists\zz ~.~ \psi_{n+1}(\yy,\x,\x',\zz) \iff 
\exists\yy\exists\zz ~.~ \psi_n(\yy,\x,\x',\zz)\) and the formula
$\exists\yy\exists\zz~.~\psi_{N^2}(\yy,\x,\x',\zz) \wedge \phi$
defines a saturated relation $R_{\mathrm{sat}}$.
\end{lemma}
\proof{``$\Rightarrow$'' We have, for all $n\geq 0$:
    \[\begin{array}{rcl}
    \psi_{n+1}(\yy,\x,\x',\zz) & \iff &
    \exists\yy'\exists\zz' ~.~ 
    \widetilde\phi(\yy,\yy') \wedge \psi_n(\yy',\x,\x',\zz') \wedge \widetilde\phi(\zz',\zz) \\
    & \Rightarrow & \exists\yy\exists\zz~.~ \psi_n(\yy,\x,\x',\zz) \enspace.
    \end{array}\]
    ``$\Leftarrow$'' For each $n\geq0$, $\exists \yy \exists \zz ~.~
    \psi_n$ is equivalent to a difference bounds constraint, obtained
    by eliminating the existential quantifiers and let $M_n$ be the
    DBM of canonical the quantifier-free difference bounds constraint
    that is equivalent to $\exists \yy \exists \zz ~.~
    \psi_n$. Suppose, by contradiction, that there exists $n \geq N^2$
    such that $\exists \yy \exists \zz ~.~ \psi_n \not\Rightarrow
    \exists \yy \exists \zz ~.~ \psi_{n+1}$. Then there exist
    $k,\ell\in\set{1,\ldots,2N}$ such that $\left(M^*_n\right)_{k\ell}
    > \left(M^*_{n+1}\right)_{k\ell}$. There are four cases, namely
    $k\leq N$, $k>N$ and $\ell\leq N$, $\ell>N$. We prove the case
    $k,\ell\leq N$, the other three cases being symmetric.

  Because $\left(M^*_{n+1}\right)_{k\ell} < \infty$, there exists a
  path $\pi : \xxki{n}{k} \arrow{}{} \ldots \arrow{}{} \xxki{n}{\ell}$
  of weight $w(\pi) = \left(M^*_{n+1}\right)_{k\ell}$ in the unfolding
  graph $\mathcal{G}_{\widetilde{\phi}}^{2(n+1)+1}$. But the only
  paths in $\mathcal{G}_{\widetilde\phi}^{2(n+1)+1}$ are among
  $\sim_R$-equivalent variables, by the definition of
  $\widetilde\phi$, thus it must be the case that $k \sim_R
  \ell$. Moreover, $\pi$ is not a path in
  $\mathcal{G}_{\widetilde\phi}^{2n+1}$, or else we would have
  $w(\pi)\geq \left(M^*_{n}\right)_{k\ell}$, hence
  $\left(M^*_{n+1}\right)_{k\ell} \geq \left(M^*_{n}\right)_{k\ell}$,
  which contradicts our assumption. Since $n+1 > N^2$ and there
  are at most $N^2$ pairs of variables in $\x$, by the pigeonhole
  principle there exists a pair $i,j$ and two positions $0 \leq
  n_1 < n_2 \leq 2(n+1)+1$ such that $\pi$ can be factorized as:
  \[\begin{array}{cccccc}
  \xxki{n}{k} & \rightarrow \ldots \rightarrow & \xxki{n_1}{i} & 
  \underbrace{\rightarrow \ldots \rightarrow}_\xi & \xxki{n_2}{i} \\[-5mm]
  &&&& \downarrow \\
  &&&& \vdots \\
  &&&& \downarrow \\
  \xxki{n}{\ell} & \leftarrow \ldots \leftarrow & \xxki{n_1}{j} & 
  \underbrace{\leftarrow \ldots \leftarrow}_\zeta & \xxki{n_2}{j} 
  \end{array}\]
  Let $\pi'$ denote the path obtained from $\pi$ by replacing the
  segment $\xi:\xxki{n_1}{i} \rightarrow \ldots \rightarrow
  \xxki{n_1}{j}$ with the segment $\zeta:\xxki{n_2}{i} \rightarrow
  \ldots \rightarrow \xxki{n_2}{j}$, in which each position is shifted
  by $n_1-n_2>0$. Hence $\pi'$ is a path between $\xxki{n}{k}$ and
  $\xxki{n}{\ell}$ in $\mathcal{G}_R^{2n+1}$, thus $w(\pi') \geq
  \left(M_n^*\right)_{k\ell} > w(\pi)$. Since $w(\pi) = w(\pi') +
  w(\xi)+w(\zeta)$, we obtain that $w(\xi)+w(\zeta)< 0$, and because
  $k \sim_R \ell$, we have that $i \sim_R j$ as well. By Lemma
  \ref{lemma:spiral} (\ref{it1:spiral}), this contradicts the
  assumption that $R$ is $*$-consistent.

  For the second point,
  $\exists\yy\exists\zz~.~\psi^{N^2}_R(\yy,\x,\x',\zz) \wedge \phi_R
  \Rightarrow \phi_R$, hence $R_{\mathrm{sat}} \subseteq R$. Now
  suppose, by contradiction, that $R^s$ is not saturated. Then there
  exist three indices $i \sim_R j \sim_R k$ that violate one of the
  corner inequalities from Def. \ref{def:saturation}. Assume w.l.o.g
  that \(\left(\topright{\sigma(R)}\right)_{ik} >
  \left(\topright{\sigma(R)}\right)_{ij}+\left(\topleft{\sigma(R)}\right)_{jk}\)
  the other case being symmetric. Then we obtain that
  $\exists\yy\exists\zz~.~\psi_R^{N^2} \not\Rightarrow
  \exists\yy\exists\zz~.~\psi_R^{N^2+1}$, contradicting the first
  point of the Lemma. \qed}

Below we relate the powers of $R$ with those of the relation
$R_{\mathrm{sat}}$, defined in the statement of Lemma
\ref{lemma:sat-lim}, in the sense that a proof of periodicity for
$R_{\mathrm{sat}}$ constitutes a proof for the periodicity of
$R$. Moreover, the prefix and period of $R_{\mathrm{sat}}$ are used to
bound the prefix and period of $R$, respectively. We recall that, for
a difference bounds relation $R\in\dbm$, $\sigma(R)$ is the closed DBM
that defines $R$ and $\mu(\sigma(R))$ is the maximum between the
absolute values of the coefficients of $\sigma(R)$ and $1$.

\begin{lemma}\label{lemma:saturation-prefix-period}
  Given a $*$-consistent relation $R \in \dbm_\x$, where
  $\x=\set{x_1,\ldots,x_N}$, for every $n\geq 2N^2$, we have $R^n =
  R^{N^2} \circ R_{\mathrm{sat}}^{n-2N^2} \circ R^{N^2}$. Moreover,
  $R$ is periodic with prefix $b$ and period $c$ if $R_{\mathrm{sat}}$
  is periodic with prefix $b_{\mathrm{sat}}$ and period
  $c_{\mathrm{sat}}$, where:
  \begin{compactitem}
  \item $b = b_{\mathrm{sat}} + \mathcal{O}(2^{N\log N}) \cdot
    \max\left(\mu(\sigma(R^{N^2})),\max_{0 \leq i < c_{\mathrm{sat}}}
    \mu(\sigma(R_{\mathrm{sat}}^{b_{\mathrm{sat}}+i}))\right)$ and
  \item $c$ divides $c_{\mathrm{sat}}$.
  \end{compactitem}
\end{lemma}
\proof{Since $R_{\mathrm{sat}} \subseteq R$, by Lemma
  \ref{lemma:sat-lim}, we obtain that $R^n \supseteq R^{N^2} \circ
  R_{\mathrm{sat}}^{n-2N^2} \circ R^{N^2}$, for each $n\geq2N^2$. The
  dual inclusion follows by noticing that, for any difference bounds
  constraint $\phi$ that defines $R$, we have $\phi(\x,\x')
  \Rightarrow \widetilde{\phi}(\x,\x')$, because $\widetilde{\phi}$ is
  obtained by dropping several atomic propositions from $\phi$. Also,
  since $R$ is $*$-consistent, it must be the case that
  $R_{\mathrm{sat}}$ is $*$-consistent as well. 

  Assume that $R_{\mathrm{sat}}$ is periodic with prefix
  $b_{\mathrm{sat}}$ and period $c_{\mathrm{sat}}$, thus the sequence
  of matrices $\{\sigma(R_{\mathrm{sat}}^k\}_{k=0}^\infty$ is
  periodic. By Lemma \ref{lemma:periodic-matrix-sequence}, there exist
  matrices $\Lambda_i \in \zed_\infty^{2N \times 2N}$, such that
  \(\sigma(R_{\mathrm{sat}}^{b_{\mathrm{sat}}+kc_{\mathrm{sat}}+i}) =
  \sigma(R_{\mathrm{sat}}^{b_{\mathrm{sat}}+i})+k\cdot\Lambda_i\), for
  all $k\geq0$ and $i\in[c_{\mathrm{sat}}]$.  Since
  $R^{b_{\mathrm{sat}}+2N^2+\ell}=R^{N^2} \circ
  R^{b_{\mathrm{sat}}+\ell} \circ R^{N^2}$ for all $\ell\geq0$, it is
  sufficient to show that the sequence
  \(\{\sigma(R^{b_{\mathrm{sat}}+kc_{\mathrm{sat}}+i+2N^2})\}_{k=0}^\infty\)
  is periodic, for each $i \in [c_{\mathrm{sat}}]$. To this end, let
  us observe that, for each $i \in [c_{\mathrm{sat}}]$, the matrix
  $\sigma(R^{b_{\mathrm{sat}}+kc_{\mathrm{sat}}+i+2N^2}) \in
  \zed_\infty[k]^{2N \times 2N}$ is the incidence matrix of the
  labeled graph $\mathcal{G}^i = \tuple{\bigcup_{j=0}^3 \x^{(j)},
    \rightarrow^i, w^i}$, defined as follows: \begin{compactitem}
  \item $\mathcal{G}^i\proj_{\x^{(0)} \cup \x^{(1)}}$ is
    $\mathcal{G}_{R^{N^2}}$ with $\x^{(0)}$ and $\x^{(1)}$ replacing
    $\x$ and $\x'$, respectively.
  \item $\mathcal{G}^i\proj_{\x^{(2)} \cup \x^{(3)}}$ is
    $\mathcal{G}_{R^{N^2}}$ with $\x^{(2)}$ and $\x^{(3)}$ replacing
    $\x$ and $\x'$, respectively.
  \item $\mathcal{G}^i\proj_{\x^{(1)} \cup \x^{(2)}}$ is defined by
    the following edges labeled by univariate linear terms with
    variable $k$, for each $s,t = 1,\ldots,N$:
    \[{\scriptstyle\begin{array}{c}
      \xxki{1}{s} \arrow{\sigma(R^{b_{\mathrm{sat}}+i})_{\scriptscriptstyle s,t}~+~k\cdot{(\Lambda_i)}_{\scriptscriptstyle s,t}}{}  \xxki{1}{t} \\
      \xxki{1}{s} \arrow{\sigma(R^{b_{\mathrm{sat}}+i})_{\scriptscriptstyle s,t+N}~+~k\cdot{(\Lambda_i)}_{\scriptscriptstyle s,t+N}}{} \xxki{2}{t} 
      \end{array}
      \hspace*{2mm}
      \begin{array}{c}
        \xxki{2}{s} \arrow{\sigma(R^{b_{\mathrm{sat}}+i})_{\scriptscriptstyle s+N,t}~+~k\cdot{(\Lambda_i)}_{\scriptscriptstyle s+N,t}}{} \xxki{1}{t} \\
        \xxki{2}{s} \arrow{\sigma(R^{b_{\mathrm{sat}}+i})_{\scriptscriptstyle s+N,t+N}~+~k\cdot{(\Lambda_i)}_{\scriptscriptstyle s+N,t+N}}{} \xxki{2}{t} 
    \end{array}}\]    
  \end{compactitem}
In other words, the middle graph $\mathcal{G}^i\proj_{\x^{(1)} \cup
  \x^{(2)}}$ corresponds to the matrix $\sigma(R^{b_{\mathrm{sat}}+i})
+ k\cdot\Lambda_i \in \zed_\infty[k]^{2N \times 2N}$, whereas the
extremities $\mathcal{G}^i\proj_{\x^{(0)} \cup \x^{(1)}}$ and
$\mathcal{G}^i\proj_{\x^{(1)} \cup \x^{(2)}}$ are constraint graphs
defining the relation $R^{N^2}$.  Clearly, for each
$i\in[c_{\mathrm{sat}}]$ and $k\geq0$, the coefficients of the matrix
$\sigma(R^{b_{\mathrm{sat}}+kc_{\mathrm{sat}}+i+2N^2})$ are given by
the weights of the minimal paths from and to the vertices in the set
$\xk{0} \cup \xk{3}$ in $\mathcal{G}^i$. Since $R$ is $*$-consistent,
no cycle of negative weight can be found in $\mathcal{G}^i$, for any
$k\geq0$ and $i\in[c_{\mathrm{sat}}]$. Thus the minimal paths in
$\mathcal{G}^i$ are necessarily elementary, thus of length at most
$4N$, which is the number of vertices in
$\mathcal{G}^i$. Consequently, there exist at most $(4N)^N =
\mathcal{O}(2^{N\log N})$ such paths, and
\((\sigma(R^{b_{\mathrm{sat}}+kc_{\mathrm{sat}}+i+2N^2}))_{st} =
\min(w^{i,1}_{st}(k), \ldots, w^{i,L_{st}}_{st}(k))\), for all $s,t
\in \set{1,\ldots,2N}$, where $w^{i,1}_{st}(k), \ldots,
w^{i,L_{st}}_{st}(k)$ are the univariate linear terms labeling the
elementary paths from $\mathcal{G}^i$ of the form: \begin{compactitem}
\item $x_s^{(0)} \rightarrow^* x_t^{(0)}$ if $1 \leq s,t \leq N$, 
\item $x_s^{(0)} \rightarrow^* x_{t-N}^{(3)}$ if $1 \leq s \leq N$ and $N < t \leq 2N$,
\item $x_{s-N}^{(3)} \rightarrow^* x_t^{(0)}$ if $N < s \leq 2N$ and $1 \leq t \leq N$,
\item $x_{s-N}^{(3)} \rightarrow^* x_{t-N}^{(3)}$ if $N < s,t \leq 2N$, 
\end{compactitem} 
and $L_{st}=\mathcal{O}(2^{N\log N})$ is the number of such
paths. Moreover, because each term $w^{i,j}_{st}(k)$ denotes a path of
length at most $4N$ in $\mathcal{G}^i$, we have that:
\begin{equation}\label{eq:monomial}
\abs{w^{i,j}_{st}(0)} \leq 4N \cdot \max(\mu(\sigma(R^{N^2})),\max_{0 \leq i < c_{\mathrm{sat}}}
\mu(\sigma(R_{\mathrm{sat}}^{b_{\mathrm{sat}+i}})))
\end{equation}

The upper bound on the period of $\{\sigma(R^k)\}_{k=0}^\infty$ can be
found by the following argument. Observe that
\(\{(\sigma(R_{\mathrm{sat}}^{b_{\mathrm{sat}}+kc_{\mathrm{sat}}+i})_{st}\}_{k=0}^\infty\)
is an arithmetic progression, for all $1 \leq s,t \leq 2N$, thus each
sequence $\{w^\ell_{st}(k)\}_{k=0}^\infty$ is an arithmetic
progression. By Lemma \ref{lemma:polynomial-periodicity}, the sequence
\(\{\min(w^1_{st}(k), \ldots, w^{L_{st}}_{st}(k))\}_{k=0}^\infty\) has
period $1$ and by Lemma \ref{lemma:periodic-matrix-sequence}, the
sequence of matrices
\(\{\sigma(R^{b_{\mathrm{sat}}+kc_{\mathrm{sat}}+i+2N^2})\}_{k=0}^\infty\)
has period $1$, for each $i\in[c_{\mathrm{sat}}]$. The period of the
sequence $\{\sigma(R^k)\}_{k=0}^\infty$ is thus a divisor of
$c_{\mathrm{sat}}$. Considering that $L_{st} = \mathcal{O}(2^{N \log
  N})$, the upper bound on the prefix of $R$ is obtained as follows:
\[\begin{array}{rcll}
b & \leq & b_{\mathrm{sat}} + (2N)^2 \cdot 
\max_{\begin{array}{c} 
    \scriptscriptstyle{0 \leq i < c_{\mathrm{sat}}} \\[-2mm]
    \scriptscriptstyle{1 \leq s,t \leq 2N} 
\end{array}}
\sum_{j=1}^{L_{st}}\abs{w^{i,j}_{st}(0)} & \text{by Lemma
  \ref{lemma:polynomial-periodicity}} \\ & = & b_{\mathrm{sat}} +
\mathcal{O}(2^{N \log N}) \cdot \max(\mu(\sigma(R^{N^2})),\max_{0 \leq
  i < c_{\mathrm{sat}}}
\mu(\sigma(R_{\mathrm{sat}}^{b_{\mathrm{sat}+i}}))) & \text{by
  (\ref{eq:monomial})}
\end{array}\] \qed}

We have reduced the problem of proving the periodicity of an arbitrary
difference bounds relation $R$ to proving the periodicity of a
saturated difference bounds relation $R_{\mathrm{sat}}$. In the next
two sections, we show that any saturated relation $R$ is periodic
(\ref{sec:zigzag-automata}) and, moreover, that the prefix and the
period of the sequence $\{\sigma(R^k)\}_{k=0}^\infty$ are of the order
of $2^{\mathcal{O}(\bin{R})}$ (\ref{sec:dbm-bounds}). The previous
lemma generalizes these bounds to arbitrary difference bounds
relations. Finally, section \ref{sec:oct-bounds} extends these bounds
to octagonal relations, concluding the proof of
\textsc{Np}-completness for the class of problems $\reachflat{\oct}$. 

\subsection{Zigzag Automata}
\label{sec:zigzag-automata}

In this section we define {\em zigzag automata}, which are an
important tool for reasoning about the powers of difference bounds
relations. Consider an unfolding $\mathcal{G}_R^n$ of the constraint
graph $\mathcal{G}_R$, for some $n>0$. We recall the constraints
(\ref{eq:dbm-min-paths}) which define the power $R^n$, using the
minimal paths in $\mathcal{G}_R^n$ between the vertices in the set
$\xk{0}\cup\xk{n}$. Each such path can be seen as a word over the
finite alphabet of subgraphs of $\mathcal{G}_R$, and the set of paths
between two distinguished vertices is the language of a finite
(weighted) automaton, called \emph{zigzag
  automaton}\cite{Fundamenta09}. Intuitively, a zigzag automaton reads
at step $i$, all edges between $\xk{i}$ and $\xk{i+1}$
simultaneously. The weight of a transition fired by the zigzag
automaton is the sum of the weights of these edges. Each run of length
$n$ in a zigzag automaton recognizes a word consisting of a single
path between two extremal vertices in $\mathcal{G}_R^n$, from the set
$\xk{0}\cup\xk{n}$. Since we are interested in the minimal paths that
occur in the constraints (\ref{eq:dbm-min-paths}), we aim at computing
the minimal weight among all runs of length $n$, as a function of $n$.

Formally, a \emph{weighted automaton}\footnote{We adopt a simplified
  version of the classical definition \cite{Schutzenberger61} that is
  sufficient for our purposes.} \cite{Schutzenberger61} is a tuple $A
= \tuple{\Sigma,\omega,Q,I,F,\Delta}$, where $\Sigma$ is a finite
alphabet, $\omega : \Sigma \rightarrow \zed$ is a function associating
integer weights to alphabet symbols, $Q$, $I$, $F$ are the set of
states, initial and final states, respectively, and $\Delta \subseteq
Q \times \Sigma \times Q$ is a transition relation. The weight of a
non-empty word $w=\sigma_1\ldots\sigma_n\in\Sigma^+$ is defined as
$\omega(w)=\sum_{i=1}^n\omega(\sigma_i)$ and the weight of the empty
word is $\omega(\varepsilon)=0$. When $A$ is clear from the context,
we denote by $q \arrow{\sigma}{} q'$ the fact that $(q,\sigma,q') \in
\Delta$. A \emph{run} of $A$ is a sequence $q_0 \arrow{\sigma_0}{} q_1
\arrow{\sigma_1}{} \ldots \arrow{\sigma_{n-1}}{} q_n$, denoted $q_0
\arrow{\sigma_0\ldots\sigma_{n-1}}{} q_n$. A state $q \in Q$ is
\emph{reachable} if there exists a run from an initial state to it,
and \emph{co-reachable} if there exists a run from it to a final
state. A word $w \in \Sigma^*$ is accepted by $A$ if there exists a
run $q_0 \arrow{w}{} q_n$ such that $q_0\in I$ and $q_n\in F$. We
denote by $\lang{A}$ the set of words accepted by $A$, i.e.\ the
\emph{language} of $A$. Moreover, we define the function $\minw_A(n) =
\min\set{\omega(w) \mid w\in\lang{A}, \len{w}=n}$ yielding, for each
$n \in \nat$, the minimal weight among all words of length $n$
recognized by $A$, or $\infty$ if no such word exists.

\begin{figure}[htb]
\centering{
\begin{tabular}{lcr}
\begin{tabular}{c}
\begin{tabular}{cccccccccccccccccccccc}
  \symbolGone & \symbolGtwo & \symbolGthree & \symbolGfour & \symbolGeps & \symbolGfive & \symbolGsix
  \\
  $\mathcal{G}_1$ & $\mathcal{G}_2$ & $\mathcal{G}_3$ & $\mathcal{G}_4$ & $\mathcal{G}_5$ & $\mathcal{G}_6$ & $\mathcal{G}_7$
\end{tabular}
\\
(a) A subset of the zigzag alphabet
\end{tabular}
&&
\begin{tabular}{c}
\scalebox{0.7}{\includegraphics{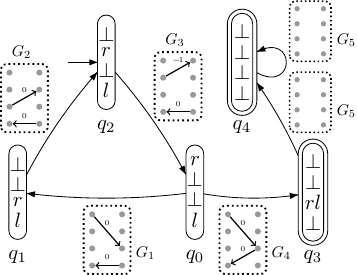}}
\\
(b) The zigzag automaton $\mathcal{A}^{\mathit{ef}}_{24}$
\end{tabular}
\end{tabular}
\vspace*{5mm}
\\
\begin{tabular}{c}
\mbox{\begin{minipage}{8cm}
\mbox{\scalebox{0.7}{\begin{tikzpicture}
  \TermZrBase{1}{4}{4}{0.7}{0.65}{0.5}{1}{3}{1}
  \TermZrBase{5}{8}{4}{0.7}{0.65}{0.5}{4}{3}{1}


  \newarray\aCaption
  \readarray{aCaption}{$q_2$ & $q_0$ & $q_1$ & $q_2$ & $q_0$ & $q_1$ & $q_2$ & $q_0$ & $q_3$}
  \TermZrCapState{0}{8}{4}{0.7}{0.65}{0.5}
  \readarray{aCaption}{$\mathcal{G}_3$ & $\mathcal{G}_1$ & $\mathcal{G}_2$ & $\mathcal{G}_3$}
  \TermZrCapGraph{1}{4}{4}{0.7}{0.65}{0.5}
  \readarray{aCaption}{$\mathcal{G}_1$ & $\mathcal{G}_2$ & $\mathcal{G}_3$ & $\mathcal{G}_4$}
  \TermZrCapGraph{5}{8}{4}{0.7}{0.65}{0.5}

  \TermZrStateElem{0}{1}{\bot}
  \TermZrStateElem{0}{2}{r}
  \TermZrStateElem{0}{3}{\bot}
  \TermZrStateElem{0}{4}{l}

  \TermZrStateElem{1}{1}{r}
  \TermZrStateElem{1}{2}{\bot}
  \TermZrStateElem{1}{3}{\bot}
  \TermZrStateElem{1}{4}{l}

  \TermZrStateElem{2}{1}{\bot}
  \TermZrStateElem{2}{2}{\bot}
  \TermZrStateElem{2}{3}{r}
  \TermZrStateElem{2}{4}{l}

  \TermZrStateElem{3}{1}{\bot}
  \TermZrStateElem{3}{2}{r}
  \TermZrStateElem{3}{3}{\bot}
  \TermZrStateElem{3}{4}{l}

  \TermZrStateElem{4}{1}{r}
  \TermZrStateElem{4}{2}{\bot}
  \TermZrStateElem{4}{3}{\bot}
  \TermZrStateElem{4}{4}{l}

  \TermZrStateElem{5}{1}{\bot}
  \TermZrStateElem{5}{2}{\bot}
  \TermZrStateElem{5}{3}{r}
  \TermZrStateElem{5}{4}{l}

  \TermZrStateElem{6}{1}{\bot}
  \TermZrStateElem{6}{2}{r}
  \TermZrStateElem{6}{3}{\bot}
  \TermZrStateElem{6}{4}{l}

  \TermZrStateElem{7}{1}{r}
  \TermZrStateElem{7}{2}{\bot}
  \TermZrStateElem{7}{3}{\bot}
  \TermZrStateElem{7}{4}{l}

  \TermZrStateElem{8}{1}{\bot}
  \TermZrStateElem{8}{2}{\bot}
  \TermZrStateElem{8}{3}{rl}
  \TermZrStateElem{8}{4}{\bot}

 {\scriptsize
  \TermZrEdgeFWLab{1}{2}{1}{-1}
  \TermZrEdgeBWLab{1}{4}{4}{0}

  \TermZrEdgeFWLab{2}{1}{3}{0}
  \TermZrEdgeBWLab{2}{4}{4}{0}

  \TermZrEdgeFWLab{3}{3}{2}{0}
  \TermZrEdgeBWLab{3}{4}{4}{0}

  \TermZrEdgeFWLab{4}{2}{1}{-1}
  \TermZrEdgeBWLab{4}{4}{4}{0}

  \TermZrEdgeFWLab{5}{1}{3}{0}
  \TermZrEdgeBWLab{5}{4}{4}{0}

  \TermZrEdgeFWLab{6}{3}{2}{0}
  \TermZrEdgeBWLab{6}{4}{4}{0}

  \TermZrEdgeFWLab{7}{2}{1}{-1}
  \TermZrEdgeBWLab{7}{4}{4}{0}

  \TermZrEdgeFWLab{8}{1}{3}{0}
  \TermZrEdgeBWLab{8}{3}{4}{0}
 }

\end{tikzpicture}}}
\end{minipage}}
\\ \\ (c) A~run of $\mathcal{A}^{\mathit{ef}}_{24}$ on the word
$\mathcal{G}_3.(\mathcal{G}_1.\mathcal{G}_2.\mathcal{G}_3)^2.\mathcal{G}_4\in\Sigma_R^+$
encoding an even forward path.
\end{tabular}
}
\caption{Zigzag automaton for the relation $R$ defined by $x_2\mi
  x'_1\leq -1 \wedge x_3\mi x'_2\leq 0 \wedge x_1\mi x'_3\leq 0 \wedge
  x'_4\mi x_4\leq 0 \wedge x'_3\mi x_4\leq 0$. }
\label{fig:rf:zigzag}
\end{figure}


Given a difference bounds relation $R \in \dbm_\x$, where
$\x=\set{x_1,\ldots,x_N}$ is a set of variables, let $\mathcal{G}_R =
\tuple{\x\cup\x',\arrow{}{},w}$ be the constraint graph that defines
$R$. The alphabet $\Sigma_R$ is the set of all subgraphs of
$\mathcal{G}_R$ such that~\begin{inparaenum}[(i)]
\item the in-degree and out-degree of each node are at most $1$, and
\item the difference between the number of edges from $\x$ to $\x'$
  and the number of edges number of edges from $\x'$ to $\x$ is either
  $-1$, $0$ or $1$.
\end{inparaenum}
The weight of a graph symbol $\mathcal{G} \in \Sigma_R$ os the sum of
the weights that occur on its edges $\omega(\mathcal{G}) = \sum_{x
  \arrow{c}{} y} c$.

\begin{example}\label{example:alphabet}
Fig.~\ref{fig:rf:zigzag}\ (a) shows a subset of the zigzag alphabet
$\Sigma_R$ for the difference bounds relation $R \iff x_2\mi x'_1\leq
-1 \wedge x_3\mi x'_2\leq 0 \wedge x_1\mi x'_3\leq 0 \wedge x'_4\mi
x_4\leq 0 \wedge x'_3\mi x_4\leq 0$ from Ex. \ref{ex:unfold-dbm}. The
weights of the symbols in the word are
$\omega(\mathcal{G}_1)\!=\!\omega(\mathcal{G}_2)\!=\!\omega(\mathcal{G}_4)\!=\!0$,
$\omega(\mathcal{G}_3)\!=\!-1$. Fig.~\ref{fig:unfolded}\ (c) shows a
path $\xxki{0}{2} \arrow{}{} \ldots \arrow{}{} \xxki{0}{4}$ from the
unfolding graph $\mathcal{G}^8_R$, encoded by the word
$\mathcal{G}_3.(\mathcal{G}_1.\mathcal{G}_2.\mathcal{G}_3)^2.\mathcal{G}_4$.
\hfill$\blacksquare$
\end{example}


The set of states of the zigzag automaton is \(Q = \set{\ell,r,\ell
  r,r\ell,\bot}^N\), i.e.\ the set of $N$-tuples of symbols
$\ell,r,\ell r,r\ell$ and $\bot$. Intuitively, these symbols capture
the direction of the incoming and outgoing edges of the alphabet
symbols: $\ell$ for a~path traversing from right to left, $r$ for
a~path traversing from left to right, $\ell r$ for a~right incoming
and right outgoing path, $r\ell$ for a~left incoming and left outgoing
path, and $\bot$ when there are no incoming nor outgoing edges from
that node. As a remark, the number of states of a zigzag automaton is
bounded by $5^N$. For example, Figure~\ref{fig:rf:zigzag}\ (c) shows
the use of states in a zigzag automaton.

The transition relation $\Delta \subseteq Q \times \Sigma_R \times Q$
is defined as follows. For all $\vec{q},\vec{q}'\in Q$ and
$\mathcal{G} \in \Sigma_R$, we have $\vec{q} \arrow{\mathcal{G}}{}
\vec{q}'$, if and only if, for all $1 \leq i\leq N$:
\begin{compactitem}
\item $\vec{q}_i = \ell$ iff $\mathcal{G}$ has one edge to $x_i$ and
  no other edge involving $x_i$,
\item $\vec{q'}_i = \ell$ iff $\mathcal{G}$ has one edge from $x_i'$
  and no other edge involving $x_i'$,
\item $\vec{q}_i = r$ iff $\mathcal{G}$ has one edge from $x_i$ and no
  other edge involving $x_i$,
\item $\vec{q'}_i = r$ iff $\mathcal{G}$ has one edge to $x_i'$ and no
  other edge involving $x_i'$,
\item $\vec{q}_i = \ell r$ iff $G$ has exactly two edges involving
  $x_i$, $x^{\scriptscriptstyle (\prime)}_j \arrow{}{} x_i \arrow{}{}
  x^{\scriptscriptstyle (\prime)}_k$ and $j \not\sim_R k$,
\item $\vec{q'}_i = r\ell$ iff $G$ has exactly two edges involving
  $x_i'$, $x^{\scriptscriptstyle (\prime)}_j \arrow{}{} x'_i
  \arrow{}{} x^{\scriptscriptstyle (\prime)}_k$ and $j \not\sim_R k$,
\item $\vec{q'}_i \in \{\ell r, \bot\}$ iff $G$ has no edge involving
  $x_i'$,
\item $\vec{q}_i \in \{r\ell, \bot\}$ iff $G$ has no edge involving
  $x_i$.
\end{compactitem}
Observe that the variables that occur on any path which traverses a
vertex labeled $\ell r$ or $r\ell$ may not belong to the same SCC of
the folded graph $\mathcal{G}_R^f$. As a consequence, every path
recognized by a zigzag automaton is saturated. For example, the path
recognized by the run in Figure \ref{fig:unfolded} (c) goes forward
while crossing the variables $x_1,x_2,x_3$ and, after changing
direction, goes backward while crossing only $x_4$.


We distinguish four types of paths in $\mathcal{G}_R^n$. A path
$\xxki{k}{i} \arrow{*}{} \xxki{\ell}{j}$ is said to be \emph{odd
  forward} if $k=0$ and $\ell=n$, \emph{even forward} if $k=\ell=0$,
\emph{odd backward} if $k=n$ and $\ell=0$, and \emph{even backward} if
$k=\ell=n$. The symbols needed to represent an odd path have an odd
number of edges, while those encoding even paths have even numbers of
edges.

The zigzag automaton for $R$ is a union of four types of
automata. Formally, for each $i,j \in \set{1,\ldots,N}$ and $t \in
\{\mathit{of}, \mathit{ob}, \mathit{ef}, \mathit{eb}\}$ (we use the
abbreviations $\mathit{of}$=odd forward, $\mathit{ob}$=odd backward,
$\mathit{ef}$=even forward and $\mathit{eb}$=even backward), the
weighted automaton $\mathcal{A}_{ij}^t =
\tuple{Q,\omega,I_{ij}^t,F_{ij}^t,\Delta}$ recognizes the saturated
paths $\xxki{p}{i} \arrow{*}{} \xxki{q}{j}$ of type $t$, with
$p,q\in\set{0,n}$. More precisely, we define the initial and final states as follows:
\[\begin{array}{ccl}
I^{\mathit{of}}_{ij} & = & \{ \vec{q} ~|~ \mbox{$\vec{q}_i = r$ and
  $\vec{q}_h \in \{\ell r,\bot\}$, $\forall h \in \{1,\ldots, N\} \setminus \{i\}$} \} \\ 
F^{\mathit{of}}_{ij} & = & \{\vec{q} ~|~ \mbox{$\vec{q}_j = r$ and 
  $\vec{q}_h \in \{r\ell,\bot\}$, $\forall h \in \{1, \ldots, N\} \setminus \{j\}$} \} \\
I^{\mathit{ob}}_{ij} & = & \{ \vec{q} ~|~ \mbox{$\vec{q}_i = \ell$ and
  $\vec{q}_h \in \{\ell r,\bot\}$, $\forall h \in \{1,\ldots,N\} \setminus \{i\}$} \} \\ 
F^{\mathit{ob}}_{ij} & = & \{\vec{q} ~|~ \mbox{$\vec{q}_j = \ell$ and 
  $\vec{q}_h \in \{r\ell,\bot\}$, $\forall h \in \{1, \ldots, N\} \setminus \{j\}$}\} \\
 I^{\mathit{ef}}_{ij} & = & \left\{\begin{array}{ll} \{\vec{q} ~|~
\vec{q}_i = r,~ \vec{q}_j = \ell,~ \vec{q}_h \in \{\ell r, \bot\},~
\forall h \in \{1,\ldots,N\} \setminus \{i,j\}\} & \mbox{if $i \neq j$} \\
\{\vec{q} ~|~ \vec{q}_i = \ell r,~ \vec{q}_h \in \{\ell r,
\bot\},~ \forall h \in \{1,\ldots,N\} \setminus \{i\}\} & \mbox{if
  $i=j$}\end{array}\right.  \\ 
F^{\mathit{ef}}_{ij} & = & \{r\ell, \bot\}^N \\
I^{\mathit{eb}}_{ij} & = & \{\ell r, \bot\}^N \\ 
F^{\mathit{eb}}_{ij} & = & \left\{\begin{array}{ll} \{\vec{q} ~|~
\vec{q}_i = \ell,~ \vec{q}_j = r,~ \vec{q}_h \in \{\ell r, \bot\},~
\forall h \in \{1,\ldots,N\} \setminus \{i,j\}\} & \mbox{if $i \neq
  j$} \\ 
\{\vec{q} ~|~ \vec{q}_i = r\ell,~ \vec{q}_h \in \{\ell r,
\bot\},~ \forall h \in \{1,\ldots,N\} \setminus \{i\}\} & \mbox{if
  $i=j$}\end{array}\right.
\end{array}\]

\begin{example}
Figure~\ref{fig:rf:zigzag}\ (b) shows the zigzag automaton
$\mathcal{A}^{\mathit{ef}}_{24}$ of the difference bounds relation $R
\iff x_2\mi x'_1\leq -1 \wedge x_3\mi x'_2\leq 0 \wedge x_1\mi
x'_3\leq 0 \wedge x'_4\mi x_4\leq 0 \wedge x'_3\mi x_4\leq 0$ from Ex.
\ref{ex:unfold-dbm} and Ex. \ref{example:alphabet}. Note that the
states that are not both reachable and co-reachable are not shown in
this figure, hence the alphabet symbols $\mathcal{G}_6$ and
$\mathcal{G}_7$ are not used. Fig.~\ref{fig:rf:zigzag}\ (c) shows
a~run of $\mathcal{A}^{\mathit{ef}}_{24}$ on the word $\gamma =
\mathcal{G}_3.(\mathcal{G}_1.\mathcal{G}_2.\mathcal{G}_3)^2.\mathcal{G}_4$,
encoding an $\mathit{ef}$-path. \hfill$\blacksquare$
\end{example} 

The following theorem wraps up the above definition, by relating the
language of a zigzag automaton with the weights of the minimal paths
in the unfolding $\mathcal{G}^n_R$ of the constraint graph defining a
difference bounds relation $R$. 

\begin{theorem}\label{thm:zigzag}
  Let $R \in \dbm_\x$ be a $*$-consistent saturated difference bounds
  relation, for $\x=\set{x_1,\ldots,x_N}$. Then, for each $n>0$ and
  all $1 \leq i,j\leq N$, the following hold:
  \begin{compactenum}
  \item\label{it:zigzag1} each word $w \in
    \lang{\mathcal{A}^{\mathit{of}}_{ij}}$ of length $n$ encodes a
    saturated path $\xxki{0}{i} \arrow{*}{} \xxki{n}{j}$ and the
    weight of a minimal such path is
    $\minw_{\mathcal{A}^{\mathit{of}}_{ij}}(n)$,
  \item\label{it:zigzag2} each word $w \in
    \lang{\mathcal{A}^{\mathit{ob}}_{ij}}$ of length $n$ encodes a
    saturated path $\xxki{n}{i} \arrow{*}{} \xxki{0}{j}$ and the
    weight of a minimal such path is
    $\minw_{\mathcal{A}^{\mathit{ob}}_{ij}}(n)$,
  \item\label{it:zigzag3} each word $w \in
    \lang{\mathcal{A}^{\mathit{ef}}_{ij}}$ of length $n$ encodes a
    saturated path $\xxki{0}{i} \arrow{*}{} \xxki{0}{j}$ and the
    weight of a minimal such path is
    $\minw_{\mathcal{A}^{\mathit{ef}}_{ij}}(n)$,
  \item\label{it:zigzag4} each word $w \in
    \lang{\mathcal{A}^{\mathit{eb}}_{ij}}$ of length $n$ encodes a
    saturated path $\xxki{n}{i} \arrow{*}{} \xxki{n}{j}$ and the
    weight of a minimal such path is
    $\minw_{\mathcal{A}^{\mathit{eb}}_{ij}}(n)$.
  \end{compactenum}
\end{theorem}
\proof{We prove that each word $w \in \lang{A}_{ij}^t$, for $t \in
  \set{\mathit{of}, \mathit{ob}, \mathit{ef}, \mathit{eb}}$ encodes a
  saturated path by contradiction. Suppose that $\pi$ is a path in $w =
  \sigma_1\ldots\sigma_n$ which is not saturated. Then there exists
  two adjacent edges $\xxki{k}{s} \arrow{}{} \xxki{\ell}{t} \arrow{}{}
  \xxki{m}{u}$ on $\pi$ belonging to the same alphabet symbol
  $\sigma_i \in \Sigma_R$, for some $1 \leq i \leq n$, $s \sim_R u$
  and either \begin{inparaenum}[(i)]
  \item $\ell=k+1$ and $m \in \set{k,k+1}$, or
  \item $\ell=k-1$ and $m \in \set{k-1,k}$. 
  \end{inparaenum}
  Since $w \in \lang{\mathcal{A}_{ij}^t}$, there exists a run
  $\vec{q}_0 \arrow{\sigma_1}{} \vec{q}_1 \ldots \arrow{\sigma_n}{}
  \vec{q}_n$ in $\mathcal{A}_{ij}^t$. In the first case, we have
  $(\vec{q}_i)_t = r\ell$ and in the second case $(\vec{q}_{i-1})_t =
  \ell r$. In both cases, however, we must have $s \not\sim_R u$, by
  the definition of the transition relation of $\mathcal{A}_{ij}^t$,
  contradiction.

  For the characterization of the weights of minimal paths, the proofs
  of the points (\ref{it:zigzag1}), (\ref{it:zigzag2}),
  (\ref{it:zigzag3}) and (\ref{it:zigzag4}) are based on \cite[Lemmas
    4.6, 4.7, 4.3 and 4.4]{Fundamenta09}, respectively.  \qed}

\subsection{Weighted Graphs and Periodic Powers of Matrices}

This section recalls several results from the theory of weighted
graphs, needed to prove the periodicity of the minimal weight
functions $\minw_{\mathcal{A}}$ of weighted automata. Based on these
facts, we characterize the prefixes and periods of a sequence of
powers of a matrix, which sets the ground for the analysis of the
periodicity of difference bounds relations (Section \ref{sec:dbm-bounds}).

Let $\mathcal{G}=\tuple{V,E,w}$ be a weighted graph for the
rest of this section. A path $\pi$ is \emph{of minimal weight for its
  length} if , for any path $\pi'$ such that
$\mathit{src}(\pi')=\mathit{src}(\pi)$,
$\mathit{dst}(\pi')=\mathit{dst}(\pi)$ and $\len{\pi'}=\len{\pi}$, we
have $w(\pi)\leq w(\pi')$. Two paths $\pi$ and $\pi'$ in $\mathcal{G}$
are \emph{equivalent} if $\mathit{src}(\pi)=\mathit{src}(\pi')$,
$\mathit{dst}(\pi)=\mathit{dst}(\pi')$, $\len{\pi}=\len{\pi'}$ and
$w(\pi)=w(\pi')$. The \emph{average weight} of a path $\pi$ is
$\aw(\pi)=\frac{w(\pi)}{\len{\pi}}$. A cycle is said to be {\em
  critical} if it has minimal average weight among all cycles of
$\mathcal{G}$. The \emph{critical graph} $\mathcal{G}^c$ consists of
those vertices and edges of $\mathcal{G}$ that belong to a critical
cycle. The following theorem states a classical result \cite[Theorem
  3.96]{BaccelliCohenOlsderQuadrat92}:

\begin{theorem}\label{thm:critical-cycle}
  For any weighted graph $\mathcal{G}$, every cycle of the critical
  graph $\mathcal{G}^c$ is critical.
\end{theorem}

If $C$ is a strongly connected component (SCC) of $\mathcal{G}^c$, we
define its \emph{cyclicity} as the greatest common divisor of the
lengths of all its elementary cycles. The cyclicity of $\mathcal{G}^c$
is the least common multiple of the cyclicities of its SCCs, and the
cyclicity of $\mathcal{G}$, denoted $c(\mathcal{G})$, is the cyclicity
of $\mathcal{G}^c$.

Weighted graphs are intimately related with the powers of their
incidence matrices, defined as follows. For two matrices $A,B \in
\zed_\infty^{n \times n}$, let $(A \boxtimes B)_{ij} = \min_{1 \leq k
  \leq n}(A_{ik}+B_{kj})$ and $\idm{n}$ be the matrix
$(\idm{n})_{ii}=0$, for all $1 \leq i \leq n$ and
$(\idm{n})_{ij}=\infty$, for all $1 \leq i,j \leq n$, where $i \neq
j$. The powers of a matrix $M$ are defined as $M^0 = \idm{n}$ and
$M^{k+1} = M \boxtimes M^k$, for all $k\geq0$. If $M$ is the incidence
matrix of a weighted graph $\mathcal{G}$, the coefficient $(M^k)_{ij}$
is the weight of a minimal path of length $k$ between the vertices $i$
and $j$ in $\mathcal{G}$. In this case, we also write $c(M)$ for
$c(\mathcal{G})$. The following theorem provides a basic tool for
proving periodicity of a sequence of relations, in the following.

\begin{theorem}\label{thm:weighted-graph-period}
For a matrix $M \in \zed_\infty^{n \times n}$, the sequence
$\{M^k\}_{k=0}^\infty$ is periodic and its period divides $c(M)$.
\end{theorem}
\proof{See \cite[Theorem 3.3]{DeSchutter00}. \qed}

Despite our best efforts, no estimation of the prefix of a power
sequence of a matrix could be found in the literature, so far. This
gap is filled by the next theorem. We recall that, for a matrix $M$,
$\mu(M)$ stands for the maximum between the absolute values of its
coefficients and one.

\begin{theorem}\label{thm:periodic-matrix-prefix}
  Given a matrix $M \in \zed_\infty^{n \times n}$, the prefix of the
  periodic sequence $\{M^k\}_{k=0}^\infty$ is at most $4\mu(M) \cdot
  n^6$.
\end{theorem}
\proof{See Appendix \ref{app:wg-prefix}. \qed}

\noindent
Observe that the prefix of a sequence of matrix powers
$\{M^k\}_{k=0}^\infty$ depends linearly on the maximal coefficient(s)
and polynomially on the dimension of $M$. On the other hand, its
period (Theorem \ref{thm:weighted-graph-period}) depends
(exponentially) only on the dimension of $M$.

Finally, we state the results of Theorems
\ref{thm:weighted-graph-period} and \ref{thm:periodic-matrix-prefix}
in terms of weighted automata, instead of weighted graphs.  Given a
weighted automaton $A=\tuple{\Sigma,\omega,Q,I,F,\Delta}$, its
underlying weighted graph is defined as $\mathcal{G}(A) =
\tuple{Q,\delta,w}$, where for all $q,q'\in Q$:
\begin{inparaenum}[(i)]
\item $(q,q') \in \delta$ iff there exists $\sigma\in\Sigma$ such that
  $(q,\sigma,q')\in\Delta$, and
\item $w(q,q')=\min\set{\omega(\sigma) \mid \exists \sigma\in\Sigma
  ~.~ (q,\sigma,q')\in\Delta}$.
\end{inparaenum}
We write $c(A)$ and $\mu(A)$ for $c(\mathcal{G}(A))$ and
$\mu(\mathcal{G}(A))$, respectively.

\begin{corollary}\label{cor:weighted-automata-periodic}
  For a weighted automaton $A = \tuple{\Sigma,\omega,Q,I,F,\Delta}$,
  the infinite sequence $\{\minw_A(n)\}_{n=0}^\infty$ is periodic,
  with prefix $b = \mathcal{O}(\mu(A)\cdot c(A)\cdot\card{Q}^{10})$ and
  period $c$ that divides $c(A)$.
\end{corollary}
\proof{Let $I=\set{q_{i_1}, \ldots, q_{i_k}}$, $F=\set{q_{j_1},
    \ldots, q_{j_\ell}}$ be the sets of initial and final states of
  $A$. Clearly we have $k,\ell \leq \card{Q}$. By denoting $m_{st}(n)
  = \min\set{\omega(w) \mid q_{i_s} \arrow{w}{} q_{j_t},~ \len{w}=n}$,
  we have: \[\min\set{\omega(w)\mid w\in\lang{A}, \len{w}=n} =
  \min_{s=1}^k\min_{t=1}^\ell m_{st}(n)\enspace.\] By Theorem
  \ref{thm:weighted-graph-period}, each sequence
  $\set{m_{st}(n)}_{n=0}^\infty$ is periodic, with prefix $b_{st}$ and
  period $c_{st}$ that divides $c(A)$. By Lemma
  \ref{lemma:polynomial-periodicity}, the sequence
  $\min\set{\omega(w)\mid w\in\lang{A}, \len{w}=n}$ is periodic, with
  period $c$ that divides $\lcm_{s=1}^k\lcm_{t=1}^\ell c_{st}$. Since
  each $c_{st}$ divides $c(A)$, we have that $c$ divides $c(A)$ as
  well.  For an upper bound on the prefix $b$ of this sequence, let
  $b_{\mathrm{max}} = \max_{s=1}^k \max_{t=1}^\ell b_{st}$. By Lemma
  \ref{lemma:polynomial-periodicity}, we obtain:
  \[\begin{array}{rcl}
  b & \leq & b_{\mathrm{max}} + k\cdot\ell\cdot\max_{i=0}^{c-1}(\sum_{s=1}^k\sum_{t=1}^\ell\abs{m_{s,t}(b_{\mathrm{max}}+i)}) \\
  b & \leq & b_{\mathrm{max}} + k^2\cdot\ell^2\cdot (b_{\mathrm{max}}+c-1) \cdot \mu(A) \\
  \end{array}\]
  By Thm. \ref{thm:periodic-matrix-prefix}, we have $b_{\mathrm{max}}
  \leq 4\mu(A)\cdot\card{Q}^6$, hence we obtain, after simplifications
  \(b \leq 4\cdot\mu(A)\cdot c(A) \cdot \card{Q}^{10}\). \qed}

\subsection{The Periodicity of Difference Bounds Relations}
\label{sec:dbm-bounds}

We are now ready to prove that the sequence of matrices
$\{\sigma(R^n)\}_{n=0}^\infty$ is periodic, where $R \in \dbm_\x$ is
any difference bounds relation and $\x=\set{x_1,\ldots,x_N}$ is a set
of variables. The coefficients of $\sigma(R^n)$ are the weights of the
minimal paths between extremal vertices from the set
$\xk{0}\cup\xk{n}$ in the unfolding $\mathcal{G}_R^n$ of the
constraint graph $\mathcal{G}_R$ --- see the constraints
(\ref{eq:dbm-min-paths}). By Theorem \ref{thm:zigzag}, these weights
are given by the functions $\minw_{\mathcal{A}^t_{ij}}(n)$, where
$\mathcal{A}^t_{ij} = \tuple{Q,\omega,I_{ij}^t,F_{ij}^t,\Delta}$ are
the zigzag automata for the relation $R$. Since these functions are
periodic, it follows that the sequence $\{\sigma(R^n)\}_{n=0}^\infty$
is periodic (Corollary
\ref{cor:weighted-automata-periodic}). Moreover, the prefix of this
sequence is polynomially bounded by the common cyclicity of zigzag
automata and $\card{Q}$ and its period divides this cyclicity. Since
$\card{Q} = 2^{\mathcal{O}(N)}$ by the construction of zigzag
automata, we are essentially left with bounding the cyclicity of
zigzag automata.

Let us start by proving a structural property of cycles in a zigzag
automaton. A cycle $\vec{q} \arrow{\gamma}{} \vec{q}$ in the
underlying weighted graph $\mathcal{G}(\mathcal{A}^t_{ij})$ of
$\mathcal{A}^t_{ij}$ is \emph{critical} if it is a critical cycle of
$\mathcal{G}(\mathcal{A}^t_{ij})$ and, moreover, $\vec{q}$ is both
reachable and co-reachable in $\mathcal{A}^t_{ij}$.

\begin{lemma}\label{lemma:zigzag-cycle}
  Let $\mathcal{A}$ be a zigzag automaton for a saturated relation $R
  \in \dbm_x$, where $\x=\set{x_1,\ldots,x_N}$ and $\vec{q}
  \arrow{\gamma}{} \vec{q}$ be one of its a critical cycles, for
  $\len{\gamma}>0$. Then $\gamma$ is a set of saturated paths $\{\xi_k
  : \xxki{p_k}{k} \arrow{*}{} \xxki{q_k}{k}\}_{k\in K}$, either
  forward or backward, such that $k \sim_R \ell$ only if $k = \ell$,
  for all $k,\ell\in K$, where $K \subseteq \set{1,\ldots,N}$.
\end{lemma}
\proof{We give the proof for the odd-forward case, the other three
  cases being similar. Since $\vec{q}$ is a reachable and co-reachable
  state of $\mathcal{A}^{\mathit{of}}_{ij}$, there exists a word
  $w=\mu\cdot\gamma\cdot\nu \in \lang{\mathcal{A}^{\mathit{of}}_{ij}}$
  and a run $\vec{q}_0 \arrow{\mu}{} \vec{q} \arrow{\gamma}{} \vec{q}
  \arrow{\nu}{} \vec{q}_f$ in $\mathcal{A}^{\mathit{of}}_{ij}$ for
  some $\vec{q}_0 \in I^t_{ij}$ and $\vec{q}_f \in F^t_{ij}$. By
  Theorem \ref{thm:zigzag} (\ref{it:zigzag1}), $w$ encodes a saturated
  path $\pi : \xxki{0}{i} \arrow{*}{} \xxki{\len{w}}{j}$. We recall
  that $\vec{q}$ is a $N$-tuple of elements from the set
  $\set{\ell,r,\ell r, r\ell, \bot}$ and we denote by $\vec{q}_k$ the
  $k$-th element of $\vec{q}$, for $1 \leq k \leq N$. Let us consider
  first the case $\vec{q}_k = r$ (the case $\vec{q}_k = \ell$ is
  symmetric). Then either: \begin{compactitem}
  \item $\pi$ has a subpath $\xi : \xxki{\len{\mu}}{k} \arrow{*}{}
    \xxki{\len{\mu}+\len{\gamma}}{k}$. Since $\pi$ is saturated and
    all variables that occur on $\xi$ are from the same equivalence
    class of $\sim_R$, all edges in $\xi$ must have the same
    direction, forward or backward. But since $\xi$ has strictly
    positive travel, i.e.\ $\tau(\xi) > 0$, at least one edge on $\xi$
    is forward, thus $\xi$ is forward.
  \item $\pi$ has a subpath $\xi : \xxki{\len{\mu}+\len{\gamma}}{k}
    \arrow{*}{} \xxki{\len{\mu}}{k}$ and all edges of $\xi$ must have
    the same direction, either forward or backward. Since
    $\tau(\xi)<0$, at least one edge on $\xi$ must be backward, thus
    $\xi$ is backward. But since $\vec{q}_k = r$, the only outgoing
    edge from $\xxki{\len{\mu}}{k}$ must be either forward or
    vertical, contradiction.
  \end{compactitem}
  The cases $\vec{q}_k \in \set{\ell r, r\ell}$ both lead to
  contradictions, using similar arguments. We have established that
  $\gamma$ consists of saturated paths that do not change their
  direction. 

  For the second point, assume that there exist two indices $k,\ell
  \in K$, such that $k \sim_R \ell$ and $k \neq \ell$. Then $\pi$ has
  a subpath $\xi : \xxki{\len{\mu}}{k} \arrow{*}{}
  \xxki{\len{\mu}}{\ell}$, and since $k \sim_R \ell$, it follows that
  all variables occuring on $\xi$ must be equivalent. Since $\pi$ is
  saturated, the same holds for $\xi$, thus the edges on $\xi$ are
  either all forward or all backward. But this contradicts the fact
  that the endpoints of $\xi$ are both on the same position
  $\len{\mu}$. \qed}

A path $\xxki{p}{k} \arrow{*}{} \xxki{q}{k}$ is \emph{essential} if
all variables occurring on it are pairwise distinct, except for the
labels of its source and destination vertices. Clearly, the length of
an essential path is bounded by the number $N$ of variables occurring
on this path. An \emph{essential power} is a path $\xi^n$ obtained
from the concatenation of an essential path $\xi$ with itself $n>0$
times.

The following lemma shows that each critical cycle $\vec{q}
\arrow{\gamma}{} \vec{q}$ in a zigzag automaton $\mathcal{A}$ is
necessarily connected to a critical cycle $\vec{q} \arrow{\lambda}{}
\vec{q}$, where $\lambda$ consists of a finite set of essential
powers. This allows us to bound the length of $\lambda$ by a simply
exponential value, which divides $\lcm(1,\ldots,N)$. It follows that
the common cyclicity of these zigzag automata is a divisor of
$\lcm(1,\ldots,N)$. We use the fact that \(\lcm(1,\ldots,N) =
2^{\mathcal{O}(N)}\) \cite{Nair82}, which occurs as a consequence of
the Prime Number Theorem, and bound the cyclicity of zigzag automata
by $2^{\mathcal{O}(N)}$. 

\begin{lemma}\label{lemma:zigzag-essential-cycle}
  Let $\mathcal{A}$ be a zigzag automaton for a saturated relation $R
  \in \dbm_x$, where $\x=\set{x_1,\ldots,x_N}$ and $\vec{q}
  \arrow{\gamma}{} \vec{q}$ be one of its a critical cycles, for
  $\len{\gamma}>0$. Then there exists a critical cycle $\vec{q}
  \arrow{\lambda}{} \vec{q}$ in $\mathcal{A}$ such that $\lambda$ is a
  set of essential powers $\{\pi_k^{n_k}\}_{k \in K}$ and
  $\len{\lambda} = \lcm_{k\in K}\set{\len{\pi_k}}$, for some $K
  \subseteq \set{1,\ldots,N}$.
\end{lemma}
\proof{By Lemma \ref{lemma:zigzag-cycle}, $\gamma$ is a set
  $\{\xi_k\}_{k\in K}$ of forward/backward paths, such that the labels
  of $\xi_k$ and $\xi_\ell$ lie in different equivalence classes of
  the $\sim_R$ relation, for all $k \neq \ell$. We shall build a word
  $\lambda$ as a set of essential powers $\{\pi_k^{n_k}\}_{k\in
    K}$. Clearly, the paths $\pi_k^{n_k}$ and $\pi_\ell^{n_\ell}$ may
  not intersect (because they cannot share labels), for any $k \neq
  \ell$, thus $\vec{q} \arrow{\lambda}{} \vec{q}$ is a valid cycle
  w.r.t. the definition of the transition table of
  $\mathcal{A}$. Before giving the definition of the set
  $\{\pi_k^{n_k}\}_{k\in K}$, we prove the following fact:
  
  \begin{fact}\label{fact:xik-critical}
    For a path $\xi_k : \xxki{p_1}{k} \arrow{\alpha_1}{} \ldots
    \arrow{\alpha_{m-1}}{} \xxki{p_m}{k}$, let $\mathcal{G}_k$ be the
    restriction of $\mathcal{G}_R^f$ to the vertices and edges on
    $\xi_k$. Then $\zeta_k : x_{k} \arrow{\alpha_1}{} \ldots
    \arrow{\alpha_{m-1}}{} x_{k}$ is a critical cycle of
    $\mathcal{G}_k$.
  \end{fact}
   \proof{ Suppose, by contradiction, that there exists $k \in K$ such
     that $\zeta_k$ is not a critical cycle of $\mathcal{G}_k$, hence
     there exists a cycle $\theta$ in $\mathcal{G}_k$, such that
     $\aw(\theta) < \aw(\zeta_k)$. Since $\mathcal{G}_k$ is strongly
     connected, there exist paths $\mu$ and $\nu$ such that $\eta_{st}
     = \mu.\theta^s.\nu.(\mu.\nu)^t$ is a path in $\mathcal{G}_k$,
     with source and destination $x_k$, for all $s,t\geq0$. It is
     sufficient to build a path $\eta_{st}$ such that
     $\len{\eta_{st}}$ is a multiple of $\len{\gamma}$. In this case,
     there exist a cycle $\vec{q} \arrow{\gamma'}{} \vec{q}$, where
     $\gamma'$ is the set consisting of $\eta_{s+K\len{\gamma},t}$,
     for a sufficiently large $K>0$, and powers of $\xi_\ell$, for all
     $\ell \in K \setminus \set{k}$, such that $\aw(\gamma') <
     \aw(\gamma)$, which contradicts the fact that $\vec{q}
     \arrow{\gamma}{} \vec{q}$ is a critical cycle of $\mathcal{A}$.

    In order to find $s$ and $t$ such that $\len{\eta_{st}}$ is a
    multiple of $\len{\gamma}$, let $n_1 = \len{\mu}+\len{\nu}$,
    $n_2=\len{\theta}$, $m_1 = \frac{n_1}{\gcd(n_1,n_2)}$ and $m_2 =
    \frac{n_2}{\gcd(n_1,n_2)}$. Clearly $m_1$ and $m_2$ are coprimes,
    and let $g(n_1,n_2) = n_1n_2 - (n_1 + n_2)$ be their Frobenius
    number. We know that any integer $h > g(n_1,n_2)$ is a conical
    combination $h = k_1m_1 + k_2m_2$, where $k_1,k_2 \geq 0$. Let
    $n_2 = \ell_1m_1 + \ell_2m_2$, for some $\ell_1,\ell_2 \geq0$, be
    the smallest multiple of $\len{\gamma}$ that is greater than
    $g(n_1,n_2)$, and let $u = \gcd(n_1,n_2) \cdot
    \frac{n_2}{\len{\gamma}}$. It is now easy to verify that
    $\len{\eta_{s,t}}=s\cdot n_1+t\cdot n_2 = u\cdot\len{\gamma}$,
    where $s=\ell_1\cdot\gcd(n_1,n_2)$ and
    $t=\ell_2\cdot\gcd(n_1,n_2)$. \qed}
   
   It follows that each graph $\mathcal{G}_k$ is critical,
   i.e.\ $\mathcal{G}_k = \mathcal{G}_k^c$, and we can chose, in
   $\mathcal{G}_k$, an elementary cycle $\rho_k$ with both source and
   destinatin labeled by $x_k$. By Theorem \ref{thm:critical-cycle},
   we have that $\rho_k$ is a critical cycle as well, and consequently
   $\aw(\rho_k) = \aw(\zeta_k)$. Let $\pi_k$ be the path in
   $\mathcal{G}_R^n$ that corresponds to the path $\rho_k$ in
   $\mathcal{G}_R^f$, for each $k \in K$. Since all edges of $\rho_k$
   correspond to edges of $\mathcal{G}_R^n$ that are either all
   forward or all backward, we know that such a path exists.  The path
   $\pi_k$ is essential, because $\rho_k$ is an elementary cycle of
   $\mathcal{G}_R^f$ and, moreover,
   $\aw(\pi_k)=\aw(\rho_k)=\aw(\zeta_k)=\aw(\xi_k)$. The word
   $\lambda$ consists of the essential powers $\set{\pi_k^{n_k}}_{k\in
     K}$, where $n_k = \frac{\lcm_{\ell\in K}
     \len{\pi_\ell}}{\len{\pi_k}}$, for all $k\in K$. We have
   $\len{\lambda} = \lcm_{k \in K} \len{\pi_k}$ by the construction of
   $\lambda$. Moreover, $\vec{q} \arrow{\lambda}{} \vec{q}$ is a cycle
   in $\mathcal{A}$, and since $\frac{w(\lambda)}{\len{\lambda}} =
   \frac{w(\gamma)}{\len{\gamma}}$, it is also a critical cycle. \qed}

The next theorem concludes the proof of periodicity for the class
$\dbm$, providing simply exponential upper bounds on the prefix and
the period of a difference bounds relation. The next section extends
this result to the class $\oct$ and finalizes the proof of Theorem
\ref{thm:flat-cm-np}, which is the main result of this paper. 

\begin{theorem}\label{thm:dbm-period-prefix}
  There exists a constant $d > 0$ such that, for every relation $R \in
  \dbm$, the sequence $\{\sigma(R^k)\}_{k=0}^\infty$ is periodic, with
  prefix $b=2^{\mathcal{O}(\bin{R}^d)}$ and period
  $c=2^{\mathcal{O}(\bin{R}^d)}$.
\end{theorem}
\proof{Let $R \subseteq \zed^\x \times \zed^\x$, where
  $\x=\set{x_1,\ldots,x_N}$ and assume w.l.o.g. that each variable
  $x_i$ occurs in a formula $\phi$ that defines $R$. Then we have $N
  \leq \bin{R}$. Let $\mathcal{G}_R$ be the constraint graph of
  $R$. We also have that $\mu(\mathcal{G}_R) \leq 2^{\mathcal{O}(\bin{R})}$. 

  We distinguish two cases. First, if $R$ is not $*$-consistent, its
  period is $c=1$ and its prefix is $b \leq
  6N^7\cdot\mu(\mathcal{G}_R) = 6\bin{R}^6 \cdot
  2^{\mathcal{O}(\bin{R})} = 2^{\mathcal{O}(\bin{R})}$, by Lemma
  \ref{lemma:spiral} (\ref{it2:spiral}). Second, if $R$ is
  $*$-consistent, there exists a saturated relation $R_{\mathrm{sat}}
  \subseteq R$ such that $R$ is periodic if $R_{\mathrm{sat}}$ is
  periodic, and \begin{compactitem}
    \item $b = b_{\mathrm{sat}} + \mathcal{O}(2^{N\log N}) \cdot
      \max(\mu(\sigma(R^{N^2})),\max_{0 \leq i < c_{\mathrm{sat}}}
      \mu(\sigma(R_{\mathrm{sat}}^{b_{\mathrm{sat}}+i})))$,
    \item $c$ divides $c_{\mathrm{sat}}$,
  \end{compactitem} 
  where $b_{\mathrm{sat}}$ and $c_{\mathrm{sat}}$ are the prefix and
  period of $R_{\mathrm{sat}}$ (Lemma
  \ref{lemma:saturation-prefix-period}). By Theorem \ref{thm:zigzag}
  the sequence $\{\sigma(R_{\mathrm{sat}}^k)\}_{k=0}^\infty$ is
  periodic, and by Corollary \ref{cor:weighted-automata-periodic}, we
  have $b_{\mathrm{sat}} = \mathcal{O}(\mu(\mathcal{A})\cdot
  c(\mathcal{A}) \cdot \card{Q}^{10})$ and $c_{\mathrm{sat}}$ is a
  divisor of $c(\mathcal{A})$, where $\mathcal{A} =
  \tuple{\Sigma_R,\omega,Q,I,F,\Delta}$ is any zigzag automaton for
  $R_{\mathrm{sat}}$. Since $\Sigma_R$ is a set of subgraphs of
  $\mathcal{G}_R$, we have $\mu(\mathcal{A}) \leq \mu(\mathcal{G}_R) =
  2^{\mathcal{O}(\bin{R})}$. Moreover, $\card{Q} =
  2^{\mathcal{O}(\bin{R})}$, by the definition of zigzag automata.

  Observe that the choice of $\mathcal{A}$ does not influence
  $c(\mathcal{A})$, because all zigzag automata share the same
  transition table. It remains to give a bound for the cyclicity of
  $\mathcal{A}$. By Lemma \ref{lemma:zigzag-essential-cycle} each
  critical cycle of $\mathcal{G}(\mathcal{A})$ is connected to a
  critical cycle of length that divides $\lcm(1,\ldots,N)$. Thus the
  cyclicity of each SCC of $\mathcal{A}$ divides $\lcm(1,\ldots,N)$
  and the same holds for $c(\mathcal{A})$, which is the least common
  multiple of the cyclicities of the SCCs of
  $\mathcal{G}(\mathcal{A})$. Since
  $\lcm(1,\ldots,N)=2^{\mathcal{O}(N)}$ \cite{Nair82}, we obtain that
  $c(A)=2^{\mathcal{O}(N)}=2^{\mathcal{O}(\bin{R})}$. Summing up, we
  obtain $b_{\mathrm{sat}} = 2^{\mathcal{O}(\bin{R})}$ and
  $c_{\mathrm{sat}} = 2^{\mathcal{O}(\bin{R})}$. By Lemma
  \ref{lemma:dbm-poly-log}, for all $0 \leq i < c_{\mathrm{sat}}$, we
  have $\bin{R_{\mathrm{sat}}^{b_{\mathrm{sat}+i}}} =
  \mathcal{O}((\bin{R}_{\mathrm{sat}} \cdot \log(b_{\mathrm{sat}} +
  c_{\mathrm{sat}}))^d)$, for a constant $d>0$ that does not depend on
  the choice of $R_{\mathrm{sat}}$. Since $\bin{R_{\mathrm{sat}}} \leq
  N^2 \cdot \bin{R}$, by the definition of $R_{\mathrm{sat}}$ (Lemma
  \ref{lemma:sat-lim}), we can conclude that $\max_{0 \leq i <
    c_{\mathrm{sat}}}
  \mu(\sigma(R_{\mathrm{sat}}^{b_{\mathrm{sat}+i}})) =
  2^{\mathcal{O}(\bin{R}^d)}$. A similar reasoning leads to
  $\mu(\sigma(R^{N^2})) = 2^{\mathcal{O}(\bin{R}^d)}$, and thus
  $b=2^{\mathcal{O}(\bin{R}^d)}$. Since $c \leq c_{\mathrm{sat}}$, we
  also have that $c=2^{\mathcal{O}(\bin{R}^d)}$, which concludes the
  proof. \qed}

\subsection{Finalizing the Proof of Theorem \ref{thm:flat-cm-np}}
\label{sec:oct-bounds}

We have gathered all the elements necessary to prove the second point
of Theorem \ref{thm:flat-cm-np}, namely that the octagonal relations
are periodic, with simply exponential prefixes and periods. As a
consequence, the class of problems $\reachflat(\oct)$ is
\textsc{Np}-complete. The theorem below is a consequence of the
similar result for difference bounds constraints and of the relation
between the powers of octagonal relations and their encodings using
difference bounds constraints (Lemma \ref{lemma:oct-dbm-powers}). We
infer that the class $\oct$ is periodic, because for two periodic
sequences $\{s_k\}_{k=0}^\infty$ and $\{t_k\}_{k=0}^\infty$, with
prefixes $b_s, b_t \geq 0$ and periods $c_s, c_t > 0$, respectively,
the following sequences are periodic: \begin{compactitem}[-]
\item $\{s_k+t_k\}_{k=0}^\infty$ with prefix at most $\max(b_s,b_t)$
  and period which divides $\lcm(c_s,c_t)$ (Lemma
  \ref{lemma:periodic-sum}, Appendix \ref{app:tropical}),
\item $\set{\lfloor \frac{s_k}{2} \rfloor}_{k=0}^\infty$ with prefix
  $b_s$ and period $2c_s$ (Lemma \ref{lemma:periodic-half}, Appendix
  \ref{app:tropical}),
\item $\{\min(s_k,t_k)\}_{k=0}^\infty$ with prefix at most
  $\max(b_s,b_t)+\sum_{i=0}^{\lcm(c_s,c_t)}(\abs{s_i}+\abs{t_i})$ and
  period which divides $\lcm(c_s,c_t)$ (Lemma
  \ref{lemma:periodic-min}, Appendix \ref{app:tropical}).
\end{compactitem} 

\begin{theorem}\label{thm:oct-period-prefix}
  There exists a constant $d > 0$ such that, for every relation $R \in
  \oct$, the sequence $\{\sigma(R^k)\}_{k=0}^\infty$ is periodic, with
  prefix $b=2^{\mathcal{O}(\bin{R}^d)}$ and period
  $c=2^{\mathcal{O}(\bin{R}^d)}$.
\end{theorem}
\proof{ Let $R \in \oct_\x$, where $\x=\set{x_1,\ldots,x_N}$ and
  $\overline{R} \in \dbm_\y$ be the difference bounds relation that
  encodes $R$, for $\y=\set{y_1,\ldots,y_{2N}}$. We have that
  $\bin{\overline{R}} \leq 2\bin{R}$. 

  We consider first the case in which $R$ is $*$-consistent. By Lemma
  \ref{lemma:oct-dbm-powers} the matrix $\sigma(R^k)$ is
  octagonal-consistent, for any $k\geq0$. Then, by Theorem
  \ref{thm:bhz}, $\sigma(\overline{R}^k)$ is consistent, for all
  $k\geq0$, thus $\overline{R}$ is $*$-consistent. Moreover, by
  Theorem \ref{thm:dbm-period-prefix}, the sequence
  $\{\sigma(\overline{R}^k)\}_{k=0}^\infty$ is periodic, with period
  $b$ and prefix $c$ of the order of
  $2^{\mathcal{O}(\bin{\overline{R}}^d)}$, for a constant $d$ which
  does not depend on the choice of $R$. By Lemma
  \ref{lemma:periodic-matrix-sequence}, each of the sequences
  $\{\sigma(\overline{R}^k)_{ij}\}_{k=0}^\infty$ is periodic, with
  prefix $b_{ij} \leq b$ and period $c_{ij}$ that divides $c$. By
  Lemma \ref{lemma:periodic-half}, the sequences
  $\set{\lfloor\frac{\sigma(\overline{R}^k)_{i\bar{\imath}}}{2}\rfloor}_{k=0}^\infty$
  and
  $\set{\lfloor\frac{\sigma(\overline{R}^k)_{\bar{\jmath}j}}{2}\rfloor}_{k=0}^\infty$
  are periodic with prefixes $b_{i\bar\imath}$, $b_{\bar\jmath j}$ and
  periods $2c_{i\bar\imath}$, $2c_{\bar\jmath j}$, respectively. By
  Lemma \ref{lemma:periodic-sum}, the sequence
  $\set{\lfloor\frac{\sigma(\overline{R}^k)_{i\bar{\imath}}}{2}\rfloor
    +
    \lfloor\frac{\sigma(\overline{R}^k)_{\bar{\jmath}j}}{2}\rfloor}_{k=0}^\infty$
  is periodic with prefix at most $\max(b_{i\bar\imath},b_{\bar\jmath
    j})$ and period which divides $2 \lcm(c_{i\bar\imath},
  c_{\bar\jmath j})$, thus also $2c$. Then
  $\{\sigma(R^k)_{ij}\}_{k=0}^\infty$ is periodic with period which
  divides $2c$ and prefix at most:
  \[b_m + \max_{\ell=0}^{2c-1}(\abs{\sigma(\overline{R}^{b_m+\ell})_{ij}} + 
  \abs{\sigma(\overline{R}^{b_m+\ell})_{i\bar\imath}} +
  \abs{\sigma(\overline{R}^{b_m+\ell})_{\bar\jmath j}})\] where
  $b_m=\max(b_{ij},b_{i\bar\imath},b_{\bar\jmath j})$. Since
  $b_{ij},b_{i\bar\imath},b_{\bar\jmath j}$ and $c$ are of the order
  of $2^{\mathcal{O}(\bin{R}^d)}$, for all $\ell=0,\ldots,2c-1$, the
  coefficients $\sigma(\overline{R}^{b_m+\ell})_{ij}$,
  $\sigma(\overline{R}^{b_m+\ell})_{i\bar\imath}$ and
  $\sigma(\overline{R}^{b_m+\ell})_{\bar\jmath j}$ are of the order of
  $2^{\mathcal{O}((\bin{R}\cdot \log (b_m+c))^e)}$, for a constant $e$
  that does not depend on $R$ (Lemma \ref{lemma:oct-poly-log}), thus
  $b$ is of the order of $2^{\mathcal{O}(\bin{R}^{de})}$. Finally, by
  Lemma \ref{lemma:periodic-matrix-sequence}, we obtain that the
  sequence $\{\sigma(R^k)\}_{k=0}^\infty$ is periodic, with prefix and
  period both of the order of $2^{\mathcal{O}(\bin{R}^{de})}$. 

  Second, if $R$ is not $*$-consistent, we have two cases: \begin{compactitem}
  \item if $\overline{R}$ is not $*$-consistent, then its period is
    $1$ and its prefix is of the order of \(\mathcal{O}(\bin{R}^7 \cdot
    2^{\bin{R}})\), by Lemma \ref{lemma:spiral} (\ref{it2:spiral}) and
    the same bounds apply to $R$.
  \item otherwise, $\overline{R}$ is $*$-consistent and there exists
    $b_0 \geq 0$ and $1 \leq i \leq 2N$ such that
    \(\lfloor\frac{\sigma(\overline{R}^m)_{i\bar\imath}}{2} \rfloor+
    \lfloor\frac{\sigma(\overline{R}^m)_{\bar\imath i}}{2} \rfloor<0\)
    for all $m \geq b_0$ (Theorem \ref{thm:bhz}). Because
    $\{\sigma(\overline{R}^m)\}_{m=0}^\infty$ is periodic, with prefix
    $b$ and period $c$ of the order of $2^{\mathcal{O}(\bin{R}^d)}$,
    we obtain:
    \[\lfloor\frac{\sigma(\overline{R}^{b+2kc})_{i\bar\imath}}{2}\rfloor + 
    \lfloor\frac{\sigma(\overline{R}^{b+2kc})_{\bar\imath
        i}}{2}\rfloor \geq
    \lfloor\frac{\sigma(\overline{R}^b)_{i\bar\imath}}{2}\rfloor +
    \lfloor\frac{\sigma(\overline{R}^b)_{\bar\imath i}}{2}\rfloor +
    k\cdot\left(\lfloor\frac{\lambda_{i\bar\imath}}{2}\rfloor +
    \lfloor\frac{\lambda_{\bar\imath i}}{2}\rfloor\right)\] where
    $\lambda_{i\bar\imath}$ and $\lambda_{\bar\imath i}$ are the rates
    of the sequences
    $\{\sigma(\overline{R}^m)_{i\bar\imath}\}_{m=0}^\infty$ and
    $\{\sigma(\overline{R}^m)_{\bar\imath i}\}_{m=0}^\infty$,
    respectively. It must be the case that
    $\lfloor\frac{\lambda_{i\bar\imath}}{2}\rfloor +
    \lfloor\frac{\lambda_{\bar\imath i}}{2}\rfloor < 0$, otherwise we
    could not have
    \(\lfloor\frac{\sigma(\overline{R}^m)_{i\bar\imath}}{2} \rfloor+
    \lfloor\frac{\sigma(\overline{R}^m)_{\bar\imath i}}{2} \rfloor<0\)
    for all $m \geq b_0$. Moreover,
    \(\lfloor\frac{\sigma(\overline{R}^{b+2kc})_{i\bar\imath}}{2}\rfloor
    + \lfloor\frac{\sigma(\overline{R}^{b+2kc})_{\bar\imath
        i}}{2}\rfloor < 0$ for all $k \geq
    \abs{\sigma(\overline{R}^b)_{i\bar\imath}} +
    \abs{\sigma(\overline{R}^b)_{\bar\imath i}} =
    2^{\mathcal{O}(\bin{R}^d)}$.  It is easy to see that $b_0 \leq b +
    2c(\abs{\sigma(\overline{R}^b)_{i\bar\imath}} +
    \abs{\sigma(\overline{R}^b)_{\bar\imath i}}) =
    2^{\mathcal{O}(\bin{R}^d)}$, which gives the bound on the prefix
    of the sequence $\{\sigma(R^k)\}_{k=0}^\infty$ in this case.
  \end{compactitem}
\qed}

\section*{Conclusions}

We prove that the class of reachability problems for flat counter
machines, with octagonal relations labeling the cycles, is
\textsc{NP}-complete. This result is based on the analysis of the
periodic behavior of the matrices that encode the power sequences of
relations. These sequences of matrices have, moreover, simply
exponential prefixes and the periods. The crux of the proof is the
reduction from octagonal to a simpler class of difference bounds
constraints, who are proved to be periodic by the construction of a
weighted automaton. The prefix and period of difference bounds
relations are shown to be simply exponential by a detailed analysis of
this weigthed automaton.

\bibliographystyle{splncs03}
\bibliography{refs}

\appendix
\section{Proof of Theorem \ref{thm:periodic-matrix-prefix}}
\label{app:wg-prefix}

For two paths $\pi$ and $\pi'$, such that
$\mathit{dst}(\pi)=\mathit{src}(\pi')$, we denote by $\pi.\pi'$ their
concatenation. The empty path is denoted $\varepsilon$ and if $\pi$ is
a cycle, we define $\pi^0=\varepsilon$, $\pi^{k+1} = \pi.\pi^k$ and
$\pi^* = \set{\pi^k \mid k \geq 0}$. For two sets of paths $S$ and
$S'$ let $S.S'=\set{\sigma.\sigma' \mid \sigma \in S, \sigma' \in S',
  \mathit{dst}(\sigma) = \mathit{src}(\sigma')}$. If
$\sigma_1,\ldots,\sigma_{k+1}$ are paths and
$\lambda_1,\ldots,\lambda_k$ are pairwise distinct elementary cycles,
such that $\mathit{dst}(\sigma_i) = \mathit{src}(\sigma_{i+1}) =
\mathit{src}(\lambda_i) = \mathit{dst}(\lambda_i)$, for all
$i=1,\ldots,k$, the set $\sigma_1.\lambda_1^*.\sigma_2 \ldots
\sigma_k.\lambda_k^*.\sigma_{k+1}$ is called a \emph{path scheme}.

First, we show that all paths in $\mathcal{G}=\tuple{V,E,w}$, that are
minimal for their length, are captured by path schemes in which the
number of cycles is at most quadratic in the number of vertices.

\begin{proposition}\label{prop:elementary-path-scheme}
For each minimal path $\rho$ there exists an equivalent path $\rho'$
and a path scheme $\theta=\sigma_1.\lambda_1^* \ldots
\sigma_k.\lambda_k^*.\sigma_{k+1}$, such that $k \leq \card{V}^2$ and
$\rho' \in \theta$.
\end{proposition}
\proof{ A similar statement, using a slightly different terminology is
  proved in \cite[Lemma 7.3.2]{LinPhD}. Namely, for each path
  $\rho$ (not necessarily minimal) in $\mathcal{G}$, there exists an
  equivalent path
  $\rho'=\sigma_1.\lambda_1^{n_1}\ldots\sigma_k.\lambda_k^{n_k}.\sigma_{k+1}$
  where $\sigma_1, \ldots, \sigma_{k+1}$ are paths, $\lambda_1,
  \ldots, \lambda_k$ are elementary cycles, $n_1, \ldots, n_k > 0$ and
  $\len{\sigma_1 \ldots \sigma_{k+1}} \leq (\card{V}-1)^2$. \qed }

A path scheme $\sigma_1.\lambda_1^*.\sigma_2 \ldots
\sigma_k.\lambda_k^*.\sigma_{k+1}$ is \emph{bi-quadratic} if
$\len{\sigma_1.\sigma_2 \ldots \sigma_{k+1}} \leq \card{V}^4$. Next we
show that, for every path in the graph, minimal for its length, there
exists an equivalent path which is captured by a bi-quadratic path
scheme with one cycle:

\begin{lemma}\label{lemma:biq-path-scheme}
For each path $\rho$, minimal for its length, there exists an
equivalent path $\rho'$ and a bi-quadratic path scheme
$\sigma.\lambda^*.\sigma'$, such that $\rho' \in
\sigma.\lambda^*.\sigma'$.
\end{lemma}
\proof{By Prop. \ref{prop:elementary-path-scheme}, for any path $\rho$
  there exists a~path scheme $\theta = \sigma_1 . \lambda_1^*
  . \sigma_2 \ldots \sigma_k . \lambda_k^* . \sigma_{k+1}$ such that
  $k \leq \card{V}^2$, and an equivalent path $\rho' = \sigma_1
  . \lambda_1^{n_1} .  \sigma_2 \ldots \sigma_k . \lambda_k^{n_k}
  . \sigma_{k+1}$, for some $n_1,\dots,n_k\geq 0$. Suppose that
  $\lambda_i$ is a~cycle with minimal average weight among all cycles
  in the scheme, i.e.\ $\aw(\lambda_i) =
  \frac{w(\lambda_i)}{\len{\lambda_i}} \leq
  \frac{w(\lambda_j)}{\len{\lambda_j}} = \aw(\lambda_j)$, for all $1
  \leq j \leq k$. For each $n_j$ there exist $p_j \geq 0$ and $0 \leq
  q_j < \len{\lambda_i}$, such that $n_j = p_j \cdot \len{\lambda_i} +
  q_j$.  Let $\rho''$ be the path: $$\sigma_1 . \lambda_1^{q_1} .
  \sigma_2 \ldots \sigma_{i-1} . \lambda_i^{n_i + \sum_{j=1}^{i-1} p_j
    \cdot \len{\lambda_j} + \sum_{j=i+1}^k p_j \cdot \len{\lambda_j}}
  . \sigma_{i+1} . \ldots \sigma_k .  \lambda_k^{q_k} . \sigma_{k+1}$$
  It is easy to check that $\len{\rho''} = \len{\rho'}$ and $w(\rho'')
  = w(\rho')$, since $\rho'$ is minimal for its length. Clearly
  $\rho''$ is captured by the path scheme $\rho_1 .  \lambda_i^*
  . \rho_2$, where $\rho_1 = \sigma_1 .  \lambda_1^{q_1} . \sigma_2
  \ldots \sigma_{i-1}$ and $\rho_2 = \sigma_{i+1} . \ldots \sigma_k
  . \lambda_k^{q_k} .  \sigma_{k+1}$. Since $\sigma_1, \ldots,
  \sigma_k, \sigma_{k+1}$ are acyclic elementary paths, by Prop.
  \ref{prop:elementary-path-scheme}, $\len{\sigma_i} <
  \card{V}$. Also, since $\lambda_1,\ldots,\lambda_k$ are elementary
  cycles, we have $\len{\lambda_i} \leq \card{V}$. Since $q_i <
  \len{\lambda_i} \leq \card{V}$, and $k \leq \card{V}^2$, by Prop.
  \ref{prop:elementary-path-scheme}, we obtain:
  \[\begin{array}{rcl}
  \len{\rho_1 . \rho_2} & \leq & (k+1) \cdot (\card{V}-1) + k \cdot
  (\card{V}) \cdot (\card{V} - 1) \\ & \leq & (\card{V}^2+1) \cdot
  (\card{V}-1) + \card{V}^2 \cdot (\card{V}) \cdot (\card{V} - 1) \\ &
  = & \card{V}^4 - \card{V}^2 + \card{V} - 1 \leq
  \card{V}^4 \end{array}\] Hence $\rho_1 . \lambda_i^* . \rho_2$ is a
  bi-quadratic path scheme. \qed }

For any $\ell \geq 0$ and vertices $u,v \in V$, let
$\mathrm{biq}(u,v,\ell)$ denote the set of all bi-quadratic path
schemes $\sigma.\lambda^*.\sigma'$, for which there exists a path
$\sigma.\lambda^k.\sigma'$ of length $\ell$, between $u$ and $v$ and
\(\mathrm{minbiq}(u,v,\ell) = \set{\sigma.\lambda^*.\sigma' \in
  \mathrm{biq}(u,v,\ell) ~|~ \forall \tau.\eta^*.\tau' \in
  \mathrm{biq}(u,v,\ell) ~.~ \aw(\lambda) \leq \aw(\eta)}\) be the
subset of $\mathrm{biq}(u,v,\ell)$ consisting of bi-quadratic path
schemes of the form $\sigma.\lambda^*.\sigma'$, where $\lambda$ has
minimal average weight. The following lemma shows that, for each
sufficiently long path that is minimal for its length, there exists an
equivalent path following a bi-quadratic path scheme of the form
$\sigma.\lambda^*.\sigma'$, whose cycle $\lambda$ has minimal average
weight, among all path schemes of this form. We recall that
$\mu(\mathcal{G})$ is the maximum between the absolute values of the
weights of $\mathcal{G}$ and $1$.

\begin{lemma}\label{lemma:min-path-scheme}
  For every path $\rho$ that is minimal for its length $\len{\rho} >
  4\mu(\mathcal{G})\cdot\!\!\!\card{V}^6$, there exists an equivalent
  path $\rho'$ and a path scheme $\sigma.\lambda^*.\sigma' \in
  \mathrm{minbiq}(\mathit{src}(\rho),\mathit{dst}(\rho),\len{\rho})$,
  such that $\rho' \in \sigma.\lambda^*.\sigma'$.
\end{lemma}
\proof{Let $u=\mathit{src}(\rho)$ and $v=\mathit{dst}(\rho)$. By Lemma
  \ref{lemma:biq-path-scheme}, for every path $\rho$, minimal for its
  length $L > 0$, there exists an equivalent path $\rho'$ which is
  captured by at least one biquadratic path scheme from
  $\mathrm{biq}(u,v,L)$. We will show that if $L \geq
  4\mu(\mathcal{G}) \cdot \card{V}^6$, the cycle in this path scheme
  must have minimal average weight among the cycles of all path
  schemes in $\mathrm{biq}(u,v,L)$. Let
  $\sigma_i.\lambda_i^*.\sigma'_i, \sigma_j.\lambda_j^*.\sigma'_j \in
  \mathrm{biq}(u,v,L)$ be two path schemes such that $\rho_i =
  \sigma_i .\lambda_i^{b_i}.\sigma'_i$ and $\rho_j =
  \sigma_j.\lambda_j^{b_j}.\sigma'_j$ are two paths of length $L$,
  between the same vertices, for some $b_i, b_j \geq 0$.  First, we
  compute: \[\begin{array}{rclcrcl} b_i & = & \frac{L -
    \len{\sigma_i.\sigma'_i}}{\len{\lambda_i}} & \hspace*{1cm} &
  w(\rho_i) & = & w(\sigma_i.\sigma'_i) + \frac{L -
    \len{\sigma_i.\sigma'_i}}{\len{\lambda_i}} \cdot w(\lambda_i)
  \\ b_j & = & \frac{L - \len{\sigma_j.\sigma'_j}}{\len{\lambda_j}}
  &~& w(\rho_j) & = & w(\sigma_j.\sigma'_j) + \frac{L -
    \len{\sigma_j.\sigma'_j}}{\len{\lambda_j}} \cdot
  w(\lambda_j) \end{array}\] Assume w.l.o.g. that $\aw(\lambda_i) <
  \aw(\lambda_j)$. We compute:
  \begin{equation}\label{eq:len:bnd} \begin{array}{rc}
      w(\rho_i) \leq w(\rho_j) & \iff 
      \\
      w(\sigma_i.\sigma'_i) + \frac{L -
        \len{\sigma_i.\sigma'_i}}{\len{\lambda_i}} \cdot w(\lambda_i)
      \leq w(\sigma_j.\sigma'_j) + \frac{L -
        \len{\sigma_j.\sigma'_j}}{\len{\lambda_j}} \cdot w(\lambda_j) & \iff 
      \\
      \frac{\len{\lambda_i}\len{\lambda_j}(w(\sigma_i.\sigma'_i) -
        w(\sigma_j.\sigma'_j)) + \len{\lambda_i} \cdot
        \len{\sigma_j.\sigma'_j} \cdot w(\lambda_j) - \len{\lambda_j}
        \cdot \len{\sigma_i.\sigma'_i} \cdot w(\lambda_i)}{
        w(\lambda_j) \cdot \len{\lambda_i} - w(\lambda_i) \cdot
        \len{\lambda_j}} \leq L \end{array} \end{equation} Since
  $w(\lambda_j)\cdot\len{\lambda_i} - w(\lambda_i)\cdot\len{\lambda_j}
  > 0$ and since $w(\lambda_i), w(\lambda_j), \len{\lambda_i},
  \len{\lambda_j} \in \zed$, we have that
  $w(\lambda_j)\cdot\len{\lambda_i} - w(\lambda_i)\cdot\len{\lambda_j}
  \geq 1$. By Lemma \ref{lemma:biq-path-scheme}, we have
  $\len{\sigma_i.\sigma'_i}, \len{\sigma_j.\sigma'_j} \leq
  \card{V}^4$, and moreover, for any path $\pi$, $w(\pi) \leq
  \len{\pi}\cdot\mu(\mathcal{G})$. Since $1 \leq \len{\lambda_i},
  \len{\lambda_j} \leq \card{V}$, we compute:
\[\begin{array}{cl}
  & \frac{\len{\lambda_i} \cdot
    \len{\lambda_j} \cdot (w(\sigma_i.\sigma'_i) - w(\sigma_j.\sigma'_j)) +
    \len{\lambda_i} \cdot \len{\sigma_j.\sigma'_j} \cdot w(\lambda_j) -
    \len{\lambda_j} \cdot \len{\sigma_i.\sigma'_i} \cdot
    w(\lambda_i)}{w(\lambda_j) \cdot \len{\lambda_i} - w(\lambda_i) \cdot
    \len{\lambda_j}} \\
  \leq &
    \len{\lambda_i} \cdot
    \len{\lambda_j} \cdot (w(\sigma_i.\sigma'_i) - w(\sigma_j.\sigma'_j))
    + \len{\lambda_i} \cdot \len{\sigma_j.\sigma'_j} \cdot w(\lambda_j) -
    \len{\lambda_j} \cdot \len{\sigma_i.\sigma'_i} \cdot w(\lambda_i) \\
  \leq &
    4\mu(\mathcal{G})\cdot\card{V}^6
\end{array}\]
Combining this with equation \eqref{eq:len:bnd}, we infer that
$\aw(\lambda_i) < \aw(\lambda_j)$ and $L \geq 4\mu(\mathcal{G}) \cdot
\card{V}^6$ only if $w(\rho_i) \leq w(\rho_j)$, or, dually, that
$w(\rho_i) > w(\rho_j)$ and $L \geq 4\mu(\mathcal{G}) \cdot
\card{V}^6$ only if $\aw(\lambda_i) \geq \aw(\lambda_j)$.  Therefore,
a path, minimal for its length, which is greater than
$4\mu(\mathcal{G}) \cdot \card{V}^6$ must belong to a biquadratic path
scheme, whose cycle has minimal average weight, among all path
schemes, to which the path may belong. \qed }

The following lemma shows that the $\mathrm{minbiq}$ sets are
invariant for fixed vertices and lengths that belong to certain
arithmetic progressions.

\begin{lemma}\label{lemma:biqs}
  Given vertices $u$ and $v$, for each arithmetic progression
  $\set{\ell_k}_{k=0}^\infty$ of rate $\lcm(1, \ldots, \card{V})$
  and $\ell_0 \geq \card{V}^4$, $\mathrm{minbiq}(u, v, \ell_i) =
  \mathrm{minbiq}(u, v, \ell_j)$, for all $i,j\geq0$.
\end{lemma}
\proof{ Let $C=\lcm(1, \ldots, \card{V})$. It is sufficient to show 
  $\mathrm{biq}(u, v, \ell_k) = \mathrm{biq}(u, v, \ell_{k+1})$, for
  all $k\geq0$. Let
  $\theta=\sigma.\lambda^*.\sigma'\in\mathrm{biq}(u,v,\ell_0+kC)$ be a
  path scheme. Clearly, \(\ell_0 + kC = \len{\sigma.\sigma'} + p \cdot
  \len{\lambda}\) for some $p\in\nat$. Since $\theta$ is bi-quadratic,
  then $\len{\sigma.\sigma'} \leq \card{V}^4$. Since $\ell_0 \geq
  \card{V}^4$, we obtain that \(\ell_0 \geq \len{\sigma.\sigma'}\). As
  a consequence, $p\cdot\len{\lambda} \geq kC$. Thus, $p \geq
  \frac{kC}{\len{\lambda}}$ and hence $p' = p - \frac{k\cdot
    C}{\len{\lambda}} \geq 0$. Observe that $p'\in\zed$ because
  $\len{\lambda} \in \set{1,\ldots,\card{V}}$ ($\lambda$ is an
  elementary cycle) and $\len{\lambda}$ divides $C$. We can define a
  path $\rho=\sigma.\lambda^{p'}.\sigma'$, such that:
\[\len{\rho} = \len{\sigma.\sigma'} + p' \cdot \len{\lambda} = 
\len{\sigma.\sigma'} + p \cdot \len{\lambda} - k C = \ell_0\enspace.\]
Thus, we have $\theta\in\mathrm{biq}(u,v,\ell_0)$ and since
$\theta\in\mathrm{biq}(u,v,\ell_0+kC)$ was an arbitrary path, we have
$\mathrm{biq}(u,v,\ell_0+kC) \subseteq \mathrm{biq}(u,v,\ell_0)$, for
all $k\in\nat$. The other direction is trivial, by taking $k=0$. \qed}

We denote by $\minw_{\mathcal{G}}(u,v,\ell)$ the minimal weight among
the paths of length $\ell$, between vertices $u$ and $v$ in
$\mathcal{G}$, or $\infty$ is no such path exists. The next lemma
proves that the minimal weights corresponding to a certain arithmetic
progression of lengths form an arithmetic progression as well.

\begin{lemma}\label{lemma:min-weight-periodic}
  Given two vertices $u$ and $v$, for any $\ell_0 >
  4\mu(\mathcal{G})\cdot\!\!\!\card{V}^6$ there exists an arithmetic
  progression $\set{\ell_k}_{k=0}^\infty$ such that the sequence
  $\set{\minw_{\mathcal{G}}(u,v,\ell_k)}_{k=0}^\infty$ forms an
  arithmetic progression.
\end{lemma}
\proof{ It is sufficient to show that, there exists an integer $c > 0$
  such that, for any $\ell_0 > 4\mu(\mathcal{G}) \cdot\!\!\!
  \card{V}^6$, there exists $r \in \zed$ such that
  $\mathrm{minw}(u,v,\ell_0 + (k+1)c) = r + \mathrm{minw}(u,v,\ell_0 +
  kc)$, for all $k \geq 0$. Let $c = \lcm(1,\ldots,\card{V})$. By
  Lemma \ref{lemma:biqs} we have that that
  $\mathrm{minbiq}(u,v,\ell_0) = \mathrm{minbiq}(u,v,\ell_0 + kc)$,
  for all $k \geq 0$.

  We distinguish two cases. First, $\mathrm{minw}(u,v,\ell_0 + kc) =
  \infty$, i.e.\ $\mathrm{minbiq}(\ell+kc,u,v) =
  \mathrm{minbiq}(u,v,\ell_0 + (k+1)c) = \emptyset$, and therefore we
  obtain $\mathrm{minw}(u,v,\ell_0+(k+1)c) = \infty$ as well. Second,
  suppose that $\mathrm{minw}(u,v,\ell_0 + kc) < \infty$. Then there
  exists a path $\rho$ between $u$ and $v$, minimal for its length
  $\len{\rho} = \ell_0 + kc > 4\cdot \mu(\mathcal{G})\!\!\cdot
  \card{V}^6$. By Lemma \ref{lemma:min-path-scheme}, there exists an
  equivalent path $\rho'$ and a biquadratic path scheme
  $\sigma.\lambda^*.\sigma' \in \mathrm{minbiq}(u,v,\ell_0 + kc)$ such
  that $\rho' = \sigma.\lambda^b.\sigma'$ for some $b \geq 0$. Let
  $\rho''$ be the path
  $\sigma.\lambda^{b+\frac{c}{\len{\lambda}}}.\sigma'$. We will show
  that $\rho''$ is minimal for its length. For, if this is the case,
  then $\len{\rho''} = \len{\rho} + c$ and $w(\rho'') = w(\rho) + c
  \cdot \aw(\lambda)$, i.e.\ $\mathrm{minw}(u,v,\ell_0+kc) =
  \mathrm{minw}(u,v,\ell_0+(k+1)c) + c \cdot \aw(\lambda)$. Since
  $\aw(\lambda)$ is the common average weight of all path schemes in
  $\mathrm{minbiq}(u,v,\ell_0+kc) = \mathrm{minbiq}(u,v,\ell_0+k'c)$,
  for any $k,k' \geq 0$, the value of the rate $c \cdot \aw(\lambda)$
  does not depend on the value $k$.

  To show that $\rho''$ is indeed minimal for its length, suppose it
  is not, and let $\pi''$ be a path of length $\len{\rho''} = \ell_0 +
  (k+1)c > 4 \mu(\mathcal{G}) \cdot\!\!\! \card{V}^6$ such that
  $w(\pi'') < w(\rho'')$. By Lemma \ref{lemma:min-path-scheme}, there
  exists an equivalent path $\pi'$ and a biquadratic path scheme
  $\tau.\eta^*.\tau' \in \mathrm{minbiq}(u,v,\ell_0 + (k+1)c) =
  \mathrm{minbiq}(u,v,\ell_0 + kc)$ (by Lemma \ref{lemma:biqs}) such
  that $\pi' = \tau.\eta^d.\tau'$, for some $d \geq 0$. We define the
  path $\pi = \tau.\eta^{d-\frac{c}{\len{\eta}}}.\tau'$, of length
  $\ell_0 + kc$. We have the following relations:
\[\begin{array}{rclcrclcrclcrclrcl} \rho & = & \sigma.\lambda^{b}.\sigma' &  & \rho'' & = &
\sigma.\lambda^{b+\frac{c}{\len{\lambda}}}.\sigma' &  &
w(\rho) & \leq & w(\pi) &  & w(\rho'') & > & w(\pi'')
\\ 
\pi & = & \tau.\eta^{d-\frac{c}{\len{\eta}}}.\tau' && \pi'' & = &
\tau.\eta^{d}.\tau' && \len{\rho} & = & \len{\pi} && \len{\rho''} & =
& \len{\pi''} \end{array}\] 
Since $\aw(\lambda)=\aw(\eta)$, we infer
that \begin{equation}\label{eq:pref}
  w(\rho'')-w(\rho)=w(\lambda)\cdot\frac{c}{\len{\lambda}}=\aw(\lambda)\cdot
  c= \aw(\eta)\cdot
  c=w(\eta)\cdot\frac{c}{\len{\eta}}=w(\pi'')-w(\pi) \end{equation}
Also, $w(\rho)\leq w(\pi)$ and $w(\pi'')<w(\rho'')$ implies that
$w(\rho)+w(\pi'')<w(\pi)+w(\rho'')$ which contradicts equation
\eqref{eq:pref}. \qed
}

\paragraph{\bf Proof of Theorem \ref{thm:periodic-matrix-prefix}}

By Lemma \ref{lemma:periodic-matrix-sequence}, the prefix of the
sequence $\{M^k\}_{k=0}^\infty$ is $\max_{1 \leq i,j \leq n} b_{ij}$
where $b_{ij}$ is the prefix of the sequence
$\{M_{ij}\}_{k=0}^\infty$. Thus it is sufficient to show that $b_{ij}
\leq 4\mu(M) \cdot n^6$, for each pair $1 \leq i,j \leq n$. Let
$\mathcal{G}_M = \tuple{\set{1,\ldots,n},E,w}$ be the weighted graph
whose incidence matrix is $M$ and $i,j \in \set{1,\ldots,n}$ be two
vertices of $\mathcal{G}_M$. By Lemma \ref{lemma:min-weight-periodic},
there exists an arithmetic progression $\{\ell_k\}_{k=0}^n$, where
$\ell_0 = 4\mu(M) \cdot n^6+1$, such that
$\{\minw_{\mathcal{G}_M}(i,j,\ell_k)\}_{k=0}^\infty$ is an arithmetic
progression. Then it must be the case that $b_{ij} \leq 4\mu(M) \cdot
n^6$. \qed

\section{Periodic Sequences of Sums, Minima and Half Terms}
\label{app:tropical}

\begin{lemma}\label{lemma:periodic-sum}
  Let $\set{s_k}_{k=0}^\infty$ and $\set{t_k}_{k=0}^\infty$ be two
  periodic sequences, with prefixes $b_s,b_t$ and periods $c_s,c_t$,
  respectively. Then the sequence $\set{s_k+t_k}_{k=0}^\infty$ is
  periodic, and:
  \[\exists \lambda_0,\ldots,\lambda_{c-1}~\forall k\geq0 ~\forall i\in[c] ~.~ 
  s_{b+(k+1)c+i}+t_{b+(k+1)c+i} = \lambda_i + s_{b+kc+i}+t_{b+kc+i}\]
  where $c=\lcm(c_s,c_t)$ and $b=\max(b_s,b_t)$.  
\end{lemma}
\proof{ Let $\lambda^s_0,\ldots,\lambda^s_{c_s-1}$ be the rates of
  $\set{s_k}_{k=0}^\infty$ and $\lambda^t_0,\ldots,\lambda^t_{c_t-1}$
  be the rates of $\set{t_k}_{k=0}^\infty$, respectively.  For all
  $i\in[c]$, we define $\lambda_i = \frac{c}{c_s} \lambda^s_{i \mod
    c_s} + \frac{c}{c_t} \lambda^t_{i \mod c_t}$. With these
  definitions, the required equality is an easy check. \qed}

\begin{lemma}\label{lemma:periodic-half}
  Let $\set{s_n}_{n=0}^\infty$ be a periodic sequence with prefix
  $b\geq0$ and period $c>0$. Then the sequence
  $\set{\lfloor\frac{s_n}{2}\rfloor}_{n=0}^\infty$ is periodic, and:
  \[\exists \lambda_0,\ldots,\lambda_{2c-1}~\forall k\geq0 ~\forall i\in[2c] ~.~
  \left\lfloor\frac{s_{b+(k+1)\cdot2c+i}}{2}\right\rfloor = \lambda_i + 
  \left\lfloor\frac{s_{b+k\cdot2c+i}}{2}\right\rfloor\enspace.\]
\end{lemma}
\proof{ Let $\lambda^s_0, \ldots, \lambda^s_{c-1}$ be the rates of the
  sequence $\set{s_n}_{n=0}^\infty$. We define
  $\lambda_i=\lambda_{i+c}=\lambda^s_i$, for all $i\in[c]$. Then, we compute: 
  \[\begin{array}{rcl}
  \left\lfloor \frac{s_{b+(k+1)2c + i}}{2} \right\rfloor 
  & = & \left\lfloor \frac{2\lambda^s_i + s_{b+k \cdot 2c + i}}{2} \right\rfloor 
  = \lambda_i + \left\lfloor \frac{s_{b+k \cdot 2c + i}}{2}\right\rfloor \\
  \left\lfloor \frac{s_{b+(k+1)2c + c + i}}{2} \right\rfloor 
  & = & \left\lfloor \frac{2\lambda^s_i + s_{b+k \cdot 2c + c + i}}{2} \right\rfloor 
  = \lambda_i + \left\lfloor \frac{s_{b+k \cdot 2c + c + i}}{2}\right\rfloor\enspace.
  \end{array}\]
  \qed}

\begin{lemma}\label{lemma:periodic-min}
  Let $\{s_k\}_{k=0}^\infty$ and $\{t_k\}_{k=0}^\infty$ be two
  periodic sequences, with prefixes $b_s,b_t$ and periods $c_s,c_t$,
  respectively. Then the sequence $\set{\min(s_k,t_k)}_{k=0}^\infty$
  is periodic, and:
  \[\begin{array}{l}
  \exists b \leq b_{\mathrm{max}} + \max_{i=0}^{c-1}\left(\abs{s_{b_{\mathrm{max}}+(i \xmod c_s)}} + \abs{t_{b_{\mathrm{max}}+(i \xmod c_t)}}\right) \\
  \exists \lambda_0,\ldots,\lambda_{c-1}~\forall k\geq0 ~\forall i\in[c] ~.~ \\
  \min(s_{b+(k+1)c+i},t_{b+(k+1)c+i}) = \lambda_i + \min(s_{b+kc+i},t_{b+kc+i})
  \end{array}\]
  where $c=\lcm(c_s,c_t)$ and $b_{\mathrm{max}}=\max(b_s,b_t)$. 
\end{lemma}
\proof{ We prove first the following facts, for all \[k \geq
  \lceil\frac{\abs{s_{b_{\mathrm{max}} + (i \xmod c_s)}} +
    \abs{t_{b_{\mathrm{max}} + (i \xmod c_t)}}}{c}\rceil\] and each $i\in[c]$:
  \begin{compactenum}
  \item if $\frac{\lambda^s_{i \xmod c_s}}{c_s} < \frac{\lambda^t_{i \xmod c_t}}{c_t}$ then $s_{b_{\mathrm{max}}+kc+i} \leq t_{b_{\mathrm{max}}+kc+i}$, 
  \item if $\frac{\lambda^s_{i \xmod c_s}}{c_s} > \frac{\lambda^t_{i \xmod c_t}}{c_t}$ then $s_{b_{\mathrm{max}}+kc+i} \geq t_{b_{\mathrm{max}}+kc+i}$,
  \item if $\frac{\lambda^s_{i \xmod c_s}}{c_s} = \frac{\lambda^t_{i
      \xmod c_t}}{c_t}$ then \[\begin{array}{rcl}
    s_{b_{\mathrm{max}}+kc+i} - t_{b_{\mathrm{max}}+kc+i} & = & s_{b_{\mathrm{max}} + (i \xmod c_s)} - t_{b_{\mathrm{max}} + (i \xmod c_t)} + \\ 
    && (i \div c_s) \lambda^s_{i \xmod c_s} - (i \div c_t) \lambda^t_{i \xmod c_t}\enspace.
    \end{array}\]
    where $\div$ denotes integer division. 
  \end{compactenum}
  Observe that $s_{b_{\mathrm{max}}+kc+i} = s_{b_{\mathrm{max}}+(i
    \xmod c_s)} + (k \frac{c}{c_s} + i \div c_s) \lambda^s_{i
    \xmod c_s}$ and similar for $t_{b_{\mathrm{max}}+kc+i}$. We have
  thus the following equivalences:
  \[\begin{array}{rcl}
  s_{b_{\mathrm{max}}+kc+i} & \leq & t_{b_{\mathrm{max}}+kc+i} \\
  s_{b_{\mathrm{max}}+(i\xmod c_s)} + (k \frac{c}{c_s} + i \div c_s) \lambda^s_{i\xmod c_s} & \leq & 
  t_{b_{\mathrm{max}}+(i\xmod c_s)} + (k \frac{c}{c_t} + i \div c_t) \lambda^t_{i\xmod c_t} \\
  kc\left(\frac{\lambda^s_{i\xmod c_s}}{c_s} - \frac{\lambda^t_{i\xmod c_t}}{c_t}\right) & \leq & 
  t_{b_{\mathrm{max}}+(i\xmod c_s)} - s_{b_{\mathrm{max}}+(i\xmod c_s)} + \\ 
  && (i \div c_t) \lambda^t_{i\xmod c_r} - (i \div c_s) \lambda^s_{i\xmod c_s}\enspace.
  \end{array}\]
  Under the assumption of this first point, we have:
  \[kc \geq \abs{s_{b_{\mathrm{max}} + (i \xmod c_s)}} + \abs{t_{b_{\mathrm{max}} + (i \xmod c_t)}} \Rightarrow s_{b_{\mathrm{max}}+kc+i} \leq t_{b_{\mathrm{max}}+kc+i}\]
  The second point is symmetric. We obtain the last point by a similar
  argument. The statement of the lemma follows, with the definition
  below. For all $i\in[c]$:
  \[\lambda_i=\left\{\begin{array}{ll}
  c \frac{\lambda^s_{i\xmod c_s}}{c_s} & \mbox{if $\frac{\lambda^s_{i \xmod c_s}}{c_s} < \frac{\lambda^t_{i \xmod c_t}}{c_t}$} \\
  c \frac{\lambda^t_{i\xmod c_t}}{c_t} & \mbox{if $\frac{\lambda^s_{i \xmod c_s}}{c_s} > \frac{\lambda^t_{i \xmod c_t}}{c_t}$} \\
  0 & \mbox{otherwise}
  \end{array}\right.\]
  Observe that, if
  $b=b_{\mathrm{max}}+\max_{i=0}^{c-1}\left(\abs{s_{b_{\mathrm{max}}+(i\xmod
      c_s)}} + \abs{t_{b_{\mathrm{max}}+(i \xmod c_t)}}\right)$, then
  $b+kc+i = b_{\mathrm{max}} +
  \left(\lceil{\frac{\max_{i=0}^{c-1}\left(\abs{s_{b_{\mathrm{max}}+(i\xmod
          c_s)}} + \abs{t_{b_{\mathrm{max}}+(i \xmod
          c_t)}}\right)}{c}}\rceil+k\right)c+i$, and a case split
  based on the above three facts can be applied. \qed}

\begin{lemma}\label{lemma:polynomial-periodicity}
  Let $\set{s^1_k}_{k=0}^\infty, \ldots, \set{s^n_k}_{k=0}^\infty$ be
  periodic sequences with prefixes $b_1,\ldots,b_n$, periods
  $c_1,\ldots,c_n$ and rates $\lambda^1_0, \ldots, \lambda^1_{c_1-1},
  \ldots, \lambda^n_0, \ldots, \lambda^n_{c_n-1}$, respectively. Let
  $m^1_k, \ldots, m^\ell_k$ be linear combinations of $s^1_k, \ldots,
  s^n_k$, respectively. Then the sequence $\{\min(m^1_k, \ldots,
  m^\ell_k)\}_{k=0}^\infty$ is periodic, with prefix at most $b$ and
  period that divides $c$, where: \begin{compactitem}
  \item $b \leq \max_{i=1}^n (b_i) + n \cdot \max_{i=0}^{c-1}(
    \sum_{j=1}^\ell \abs{m_j(s^1_{b_{\mathrm{max}}+i}, \ldots,
      s^n_{b_{\mathrm{max}}+i})})$ and
  \item $c=\lcm_{i=1}^n (c_i)$. 
  \end{compactitem}
\end{lemma}
\proof{ Applying Lemma \ref{lemma:periodic-sum}, we obtain that, each
  sequence $\set{m^i_k}_{k=0}^\infty$, for $i=1,\ldots,\ell$, is
  periodic, with prefix at most $\max_{i=1}^n (b_i)$ and period which
  divides $c$. The upper bound on the prefix and period of
  $\{\min(m^1_k, \ldots, m^\ell_k)\}_{k=0}^\infty$ is obtained by
  applying $n$ times Lemma \ref{lemma:periodic-min}. \qed}

\end{document}